\documentclass[12pt]{article}
\usepackage{amsmath,amsfonts, graphicx, float, caption}
\usepackage{lifecon} 
\usepackage[titletoc,title]{appendix}
\usepackage{color}
\usepackage{xcolor}
\usepackage{algorithm}
\usepackage{algpseudocode}
\usepackage{pifont}
\usepackage[dvips]{epsfig}
\usepackage{subfigure}
\usepackage{caption}
\usepackage{natbib}
\usepackage{eurosym}
\usepackage{natbib}
\usepackage{textcomp}
\usepackage{endnote}
\usepackage{lscape,verbatim}
\usepackage{setspace}
\usepackage{rotating}
\usepackage{rotating,ctable,bm}
\usepackage[latin1]{inputenc}
\usepackage[dvips]{epsfig}
\usepackage{subfigure}
\usepackage{pdflscape}
\usepackage{bm}
\usepackage{calc}
\usepackage{placeins}

\newtheorem{remark}{Remark}[section]

\allowdisplaybreaks

\begin{document}

\title{Cohort effects in mortality modelling: a Bayesian state-space approach} 

\author{Man Chung Fung$^1$, Gareth W. Peters$^{3,4,5}$ \and Pavel V. Shevchenko$^{2}$\\  
{\small{
$^1$ Decision Sciences, Data61, CSIRO, Australia}}\\
{\small{
$^2$ Department of Applied Finance and Actuarial Studies, Macquarie University, Australia}}\\
{\small{
$^3$ Department of Statistical Science, University College London}} \\
{\small{
$^4$ Associate Fellow, Oxford Mann Institute, Oxford University}}\\
{\small{
$^5$ Associate Fellow, Systemic Risk Center, London School of Economics.}}
}

\maketitle

\begin{abstract}
\begin{centering}
Cohort effects are important factors in determining the evolution of human mortality for certain countries. Extensions of dynamic mortality models with cohort features have been proposed in the literature to account for these factors under the generalised linear modelling framework. In this paper we approach the problem of mortality modelling with cohort factors incorporated through a novel formulation under a state-space methodology. In the process we demonstrate that cohort factors can be formulated naturally under the state-space framework, despite the fact that cohort factors are indexed according to year-of-birth rather than year. Bayesian inference for cohort models in a state-space formulation is then developed based on an efficient Markov chain Monte Carlo sampler, allowing for the quantification of parameter uncertainty in cohort models and resulting mortality forecasts that are used for life expectancy and life table constructions. The effectiveness of our approach is examined through comprehensive empirical studies involving male and female populations from various countries. Our results show that cohort patterns are present for certain countries that we studied and the inclusion of cohort factors are crucial in capturing these phenomena, thus highlighting the benefits of introducing cohort models in the state-space framework. Forecasting of cohort models is also discussed in light of the projection of cohort factors.
\end{centering}
\end{abstract}

\begin{centering}
\end{centering}


\maketitle\thispagestyle{empty}

\section{Introduction}\label{sec:intro}
The declining trend of mortality rates is generally observed in many developed countries. It is widely acknowledged among actuaries and demographers that dynamic mortality models are required to account for the uncertainty associated with the projection of mortality for different generations. From governments who are responsible for pension policy design to insures who offer retirement income products, it is important to consider and incorporate different factors that would impact the projections of mortality rates. Arguably one of the most discussed and important factors is the so-called cohort effect, that is, the effect of year-of-birth on mortality rates.

It is perhaps not surprising that people born in different years or generations would undergo different mortality experiences. \cite{Willets04} finds evidence for the existence of cohort trends by examining the average annual mortality improvement rate for males and females in the population of England and Wales. \cite{Murphy09} discusses the ``golden generations" of the British population who were born in early 1930s and have experienced exceptionally rapid improvements in mortality rates.  Possible explanations for this ``golden generations" phenomenon includes changing smoking patterns between generations; better diet and environmental conditions during and after the Second World War; those born in periods of low fertility facing less competition for resources as they age; and benefits from medical advances (\cite{Murphy09}). \cite{Willets04}, \cite{Murphy09} and \cite{Murphy10} provide detailed discussions on cohort effects including their identification from mortality data and competing explanations. The aforementioned studies are not model-based however, but are relying on empirical data analysis and qualitative analysis such as descriptive and graphical representations to conclude the significance of cohort effects on population mortality.

For actuarial applications such as mortality forecasting and longevity risk management (\cite{CairnsBlDo08} and \cite{BarrieuEtAl12}), one is often interested in building dynamic mortality models with stochastic cohort features. \cite{RenshawHa03} extend the well-known Lee-Carter mortality model (\cite{LeeCa92}) by introducing an age-modulated cohort effect. In a similar way, the Cairns-Blake-Dowd mortality model introduced in \cite{CairnsBlDo06b} is extended to incorporate cohort factors in \cite{Cairnsetal09}. In contrast to studying cohort effects via qualitative arguments, mortality models offer a quantitative and statistical approach to identify and analyse cohort patterns exhibited in mortality data.

There are two main approaches for estimating mortality models. The first approach relies on least squares estimation based on singular value decomposition (SVD), pioneered by \cite{LeeCa92} in the mortality context. Other studies using this approach to analyse mortality include \cite{RenshawHa03}, \cite{YangYuHu10} and \cite{ShangBoHy11}. The second approach employs regression-based methods to calibrate mortality models (\cite{BrouhnsDeVe02}). For some recent studies based on this approach, see \cite{OHareLi12}, \cite{vanBerkumAnVe16} and \cite{EnchevKlCa16}. The recent paper \cite{Currie16} provides a comprehensive summary on mortality modelling based on the generalised linear modelling framework. In particular, the paper points out that there are convergence and robustness problems in the regression-based estimation setting that need to be resolved for models with cohort features.

In this paper, we approach the problems of estimating and forecasting mortality with cohort features via state-space methodology, which is a natural extension of the framework recently described in \cite{FungPeSh17} to the cohort model formulation.  In particular we work under the Bayesian paradigm and as a result the important aspect of parameter uncertainty can be accounted for naturally. The main contribution of this paper is to demonstrate how mortality models with cohort effects can be formulated, estimated and forecasted under a Bayesian state-space framework. Other works using the state-space approach for mortality modelling include \cite{Pedroza06}, \cite{deJongTi06}, \cite{KogureKiKu09} and \cite{LiuLi16b}. In our view, the state-space approach has three major advantages.

First, the ability of modelling, estimating and forecasting mortality under a unified framework can avoid the potential pitfalls of the 2-step estimation procedure typically found in the literature. As discussed in details in \cite{FungPeSh17}, a common practice of estimating mortality models consists of two steps:
\begin{description}
  \item[Step 1] Obtain estimates of parameters including period (cohort) effects; the period (cohort) effects are treated as parameters without the assumptions of their dynamics.
  \item[Step 2] Assume a time series model, for example ARIMA models, for the period (cohort) effects; parameters of the time series model are then estimated by fitting the model to period (cohort) effects obtained from Step 1.
\end{description}
In a recent study in \cite{LengPe16}, the authors point out that the least-squares method for the 2-step estimation approach utilised in \cite{LeeCa92} will in general lead to inconsistent estimators unless restrictions are imposed on the possible range of time series models for the underlying dynamics. From a statistical point of view, the two-steps approach is a somewhat ad-hoc procedure. One may argue that it is more satisfactory to perform estimation and forecasting under a single universal, consistent and rigorous framework, and one such example is the state-space methodology which is well-established in the statistics community.

Second, the ability to provide confidence or credible intervals for the estimated parameters, and hence the quantification of parameter risk in mortality projections, is especially important in longevity analysis. The impact of parameter uncertainty on forecasting mortality rates is documented in \cite{CzadoDeDe05}, \cite{KoissiShHo06} and \cite{KleinowRi16}. A Bayesian approach to mortality modelling via credit risk plus methodology is considered in \cite{ShevchenkoHiSc15} and \cite{HirzScSh17}. It is also a significant issue in valuing liabilities in life insurance portfolio and pension schemes. Moreover, the rather short time series data used for calibration purpose typically assumed in the literature\footnote{It is often the case that mortality rates data obtained before 1960 or even 1970 are not used when calibrating mortality models.} further enforce the necessity to account for parameter uncertainty where forecasting intervals are required. The ability to quantify parameter risk will enhance the reliability of the projected mortality rates. As shown in \cite{FungPeSh17}, the state-space approach offers a particularly rich and flexible framework for Bayesian and frequentist estimation, where a range of techniques such as filtering, sequential Monte Carlo and Markov chain Monte Carlo (MCMC) methods can be employed.

Third, the state-space approach allows for a wide range of mortality models to be considered while estimation and forecasting can still be performed efficiently. \cite{FungPeSh17} show that a majority of the popular mortality models can be cast in state-space formulation; in addition, stochastic volatility features can be introduced to dynamic mortality models where numerical filtering techniques are employed for model estimation. State-space formulation of mortality models have been applied to financial problems as well. Annuity pricing via a Bayesian state-space formulation of the Lee-Carter model is considered in \cite{FungPeSh15}. Pricing of longevity instruments based on the maximum entropy principle using a state-space mortality model is studied in \cite{KogureKu10}. The flexibility of the state-space approach is a key element in dealing with diverse issues concerning mortality modelling as well as pricing and risk analysis involving longevity risk, see \cite{LiuLi16a} and \cite{LiuLi16b} for an application of state-space mortality model for longevity hedging .

Despite the advantages of the state-space method, the approach is still under explored in our view. A key element that is as yet missing from this literature, that we aim to address in this paper, is the consideration of cohort effects in a state-space modelling setting. Given the fact that cohort effects are known to be present in certain countries, the possibility of exploiting cohort features under a state-space framework will undoubtedly enhance an actuary's ability to analyse mortality data. The importance of incorporating cohort effects in state-space setting is also emphasized in \cite{LiuLi16b}, where the authors \textit{``acknowledge that cohort effects are significant in certain populations, and that it is not trivial to incorporate cohort effects in a state-space representation in which the vector of hidden states evolve over time rather than year of birth"} (p.66). Therefore, in this paper we focus on addressing this missing piece of model formulation.

The paper is organised as follows. Section~\ref{sec:statespacemortality} provides an overview of different approaches to mortality modelling, leading to the introduction of the state-space methodology. State-space formulation of cohort models is discussed in details in Section~\ref{sec:StateSpaceCohort}. In Section~\ref{sec:bayesian}, we develop Bayesian inference for cohort models under the state-space framework based on efficient MCMC sampling. Empirical studies for male and female population data from various countries using the state-space cohort models are conducted in Section~\ref{sec:empircial}. Finally, Section~\ref{sec:conclusion} concludes. 

\section{State-space approach to mortality modelling}\label{sec:statespacemortality}
In this section we provide an overview of different approaches to modelling mortality as well as their estimation methodologies. Our focus will be on single population mortality modelling, however the essential elements of the approaches and methods discussed can be carried over to multi-population settings, see for example \cite{EnchevKlCa16}.

\subsection{Stochastic mortality models}
The definitions here follow \cite{Dowdetal10}. We use $q_{x,t}$ to denote the true mortality rate, i.e., the probability of death between time $t$ and $t+1$ for individuals aged $x$ at time $t$. The true death rate, denoted by $m_{x,t}$, is related to the true mortality rate via
\begin{equation}
    q_{x,t} = 1-e^{-m_{x,t}},
\end{equation}
where the force of mortality $m_{x,t}$ is assumed to be constant within integers $x$ and $t$. One can use the observed number of deaths $D_{xt}$ and initial exposures $\hat{E}_{xt}$ data\footnote{The initial exposure $\hat{E}_{xt}$ is the population size aged $x$ at the beginning of year $t$.} to obtain the crude mortality rate as $\widetilde{q}_{x,t}=D_{xt}/\hat{E}_{xt}$, which is a crude estimate of $q_{x,t}$. Similarly, the crude death rate $\widetilde{m}_{x,t}$ is defined as the ratio of the observed number of deaths $D_{xt}$ to the average population size $E_{xt}$, known as central exposures, ages $x$ last birthday during year $t$.\footnote{The average population size is often determined approximately as the population size at the middle of the year.} 

The work of \cite{LeeCa92} introduced the so-called Lee-Carter (LC) model
\begin{equation}\label{eqn:LC92}
    \ln{\widetilde{m}_{x,t}} = \alpha_x + \beta_x \kappa_t + \varepsilon_{x,t},
\end{equation}
where the terms $\alpha_x$ and $\beta_x \kappa_t$ aim to capture the age and period effects respectively. Since they proposed to use singular value decomposition (SVD) to calibrate the model to mortality data, the noise term $\epsilon_{x,t}$ is included in addition to the age and period effects; note that the crude death rate $\widetilde{m}_{x,t}$, which is obtained directly using mortality data, is used to obtain estimates of the parameters.

In contrast to the SVD approach where the models are fitted to crude death rates, \cite{BrouhnsDeVe02} considers an alternative approach where mortality models are fitted to the number of deaths instead. Using the number of observed deaths $D_{xt}$ and central exposures $E_{xt}$, the model proposed by Lee and Carter can be reconsidered as
\begin{equation}\label{eqn:PoissonLC}
    D_{xt} \sim \text{Poisson}(E_{xt}\,m_{x,t}), \quad \text{where } m_{x,t}=e^{\alpha_x+\beta_x\kappa_t}.
\end{equation}
In this approach, the number of deaths plays an important role in calibrating the model and consequently the additive error structure in \eqref{eqn:LC92} is replaced by the Poisson error structure. Also notice that confusion may arise if one does not distinguish the true and crude death rate, for example it does not make sense to use $\widetilde{m}_{x,t}$ in \eqref{eqn:PoissonLC}.

Besides having a different statistical assumption on the error structure, the Poisson regression setting, which belongs to the class of models known as generalised linear/non-linear models, has the advantage that the cohort factor can be incorporated while estimation can still be performed without extra difficulty in contrast to the SVD approach. The Lee-Carter model in the Poisson setup can be enriched by adding a cohort factor $\gamma_{t-x}$, where $t-x$ refers to the year-of-birth, as follows:
\begin{equation}\label{eqn:PoissonCohort}
    D_{xt} \sim \text{Poisson}(E_{xt}\,m_{x,t}), \quad \text{where } m_{x,t}=e^{\alpha_x+\beta_x\kappa_t+\beta^{\gamma}_x\gamma_{t-x}}.
\end{equation}
which was proposed in \cite{RenshawHa06}. As $t$ and $x$ take values in $\{t_1,\dots,t_n\}$ and $\{x_1,\dots,x_p\}$ respectively, the cohort index $t-x$ takes values in $\{t_1-x_p,\dots,t_n-x_1\}$. The dimension of the cohort index is thus $n+p-1$ which is different to the dimension of the period index $n$.

Under the Lee-Carter original approach, one might consider modelling the crude death rate with cohort effects as follows:
\begin{equation}\label{SVDcohort}
    \ln{\widetilde{m}_{x,t}} = \alpha_x + \beta_x \kappa_t + \beta^\gamma_x \gamma_{t-x} + \varepsilon_{x,t}.
\end{equation}
However the dimension of the cohort index would cause difficulty for the SVD estimation approach. One of the goals of the present paper is to show that model \eqref{SVDcohort} can be successfully estimated by considering it as a state-space model instead.

\begin{table}[h]
\center \setlength{\tabcolsep}{1em}
\renewcommand{\arraystretch}{1.1}
\scalebox{0.85}{\begin{tabular}{ll}
\hline \hline
Model &  Dynamics \\
\hline
\cite{LeeCa92}   & $\ln{m_{x,t}} = \alpha_x+\beta_x\kappa_t$     \\
\cite{RenshawHa03}   & $\ln{m_{x,t}} = \alpha_x+\sum_{i=1}^k\beta^{(i)}_x\kappa^{(i)}_t$     \\
\cite{RenshawHa06}   & $\ln{m_{x,t}} = \alpha_x+\beta^{(1)}_x\kappa_t+ \beta^{(2)}_x\,\gamma_{t-x}$     \\
\cite{Currie06}   & $\ln{m_{x,t}} = \alpha_x+ \kappa_t+ \gamma_{t-x}$     \\
\cite{CairnsBlDo06b}   & $\text{logit}(q_{x,t}) = \kappa^{(1)}_t+ \kappa^{(2)}_t(x-\bar{x})$     \\
\cite{Cairnsetal09}   & $\text{logit}(q_{x,t}) = \kappa^{(1)}_t+ \kappa^{(2)}_t(x-\bar{x})+ \gamma_{t-x}$     \\
\cite{Plat09}   & $\ln{m_{x,t}} = \alpha_x+ \kappa^{(1)}_t+ \kappa^{(2)}_t(\bar{x}-x)+\kappa^{(3)}_t(\bar{x}-x)^+ + \gamma_{t-x}$     \\
\hline \hline
\end{tabular}}
\center\caption{\label{table:modelsliterature} \small{Examples of dynamic mortality models; here true death/mortality rates are being modelled.}}
\end{table}

Examples of popular mortality models are provided in Table~\ref{table:modelsliterature}. \cite{RenshawHa03} consider a multi-period ($\sum_{i=1}^k\beta^{(i)}_x\kappa^{(i)}_t$) extension of the LC model. \cite{RenshawHa06} introduce a cohort factor ($\gamma_{t-x}$) to LC model. A simplified version of the model in \cite{RenshawHa06} is studied in \cite{Currie06}. \cite{CairnsBlDo06b} propose the so-called CBD model to model $\text{logit}(q_{x,t}):=\ln\left(q_{x,t}/(1-q_{x,t})\right)$, where $\bar{x}$ is the average age of the sample range, i.e. $\bar{x}=\frac{1}{p}\sum^p_{i=1} x_i$, designed to capture the mature-age mortality dynamics. \cite{Cairnsetal09} extend the CBD model by incorporating a cohort factor. \cite{Plat09} introduces a model which combines the desirable features of the previous models and include a term $(\bar{x}-x)^+ := \text{max}(\bar{x}-x,0)$ which aims to capture young-age mortality dynamics.
\newline
\begin{remark}
An alternative approach to modelling cohort mortality is to consider the continuous-time modelling of the mortality intensity (force of mortality) $\mu_{x+t,t}$, with $x$ and $t$ taking continuous values. It is of particular relevance for financial and actuarial applications as pricing formulas can be derived and expressed via the mortality intensity. Continuous-time approaches are discussed in \cite{CairnsBlDo08}, and they have been studied extensively in the literature, see for example \cite{Biffis05}, \cite{DahlMo06}, \cite{LucianoVi08} and \cite{FungIgSh14}.
\end{remark}

\subsection{Generalised linear modelling framework}\label{sec:GLMs}
Estimation of stochastic mortality models such as those presented in Table~\ref{table:modelsliterature} can be performed based on a flexible approach known as generalised linear modelling (GLM) framework (\cite{VillegasMiKa15}, \cite{Currie16}).

Given central exposures $E_{xt}$, it is typical to approximate initial exposures as $\hat{E}_{xt} \approx E_{xt}+\frac{1}{2}D_{xt}$. Under the GLM framework, one is interested in modelling the number of deaths $D_{xt}$ as random variables. Common examples include Poisson error structure
\begin{equation}
D_{xt} \sim \text{Poisson}(E_{xt}\,m_{x,t})
\end{equation}
and Binomial error structure
\begin{equation}
D_{xt} \sim \text{Binomial}(\hat{E}_{xt},\,q_{x,t}).
\end{equation}
Note that we have the expected values $\mathbb{E}\left[D_{xt}/E_{xt}\right]=m_{x,t}$ and $\mathbb{E}[D_{xt}/\hat{E}_{xt}]=q_{x,t}$ for the Poisson and Binomial models respectively. Through the so-called link function $g$, one can associate the mean $\mathbb{E}\left[D_{xt}/E_{xt}\right]$ or $\mathbb{E}[D_{xt}/\hat{E}_{xt}]$ with a predictor $\rho_{x,t}$ as
\begin{equation}
g\left(\mathbb{E}\left[\frac{D_{xt}}{\bar{E}_{xt}}\right]\right)=\rho_{x,t}
\end{equation}
where $\bar{E}_{xt}$ can be initial or central exposures. Typical link functions for the Poisson and Binomial models are the log function and the logit function respectively.

The models in Table~\ref{table:modelsliterature}, from the viewpoint of GLM framework, provide a specification for the predictor. In particular, the last four models in Table~\ref{table:modelsliterature} assume linear predictors while the first three models describe non-linear predictors since multiplicative terms of parameters such as $\beta_x \kappa_t$ are involved in the predictor. The latter models can be referred to as generalised non-linear models (\cite{Currie16}).

We note here that a clear advantage of the GLM framework is that it allows sophisticated error structures such as Poisson and Binomial distributions compared to the SVD approach. However, despite its flexibility, the framework involves a 2-step procedure for estimation which is in contrast to the state-space approach presented in the following sections where a joint estimation of model parameters and latent factors is performed.

\subsection{State-space modelling framework}
In this paper we extend previous work, see for example \cite{Pedroza06}, \cite{KogureKu10} and \cite{FungPeSh17}, on using state-space techniques to model mortality dynamics with cohort features taken into consideration. We first present a brief review of mortality modelling via state-space representation.

A state-space model consists of two equations: the observation equation and the state equation which are given, respectively, by
\begin{subequations}
    \begin{align}
        \bm{z}_t &= a(\bm{\phi}_t, \bm{u}_t), \label{eqn:SSObseqns} \\
        \bm{\phi}_t &= b(\bm{\phi}_{t-1}, \bm{v}_t), \label{eqn:SSStateeqns}
    \end{align}
\end{subequations}
where $\bm{z}_t$ represents an observed multi-dimensional time series and the state $\bm{\phi}_t$ represents a multi-dimensional hidden Markov process. Here, $\bm{u}_t$ and $\bm{v}_t$ are independent random noises and the functions $a$ and $b$ can be nonlinear in general. It is clear that each of the mortality models shown in Table~\ref{table:modelsliterature} specify the observation equation of a state-space model, where the period and cohort factors represent the hidden states. A time series model for the period effects will form the state equation, thus completing the description of a state-space system. As an example, the Lee-Carter model can be reformulated as
\begin{subequations}\label{eqn:LCSS}
\begin{align}
    \boldsymbol{y}_t &= \boldsymbol{\alpha}+\boldsymbol{\beta}\kappa_t+ \boldsymbol{\varepsilon}_t, \quad \boldsymbol{\varepsilon}_t \overset{iid}{\sim} \text{N}(0,\mathbf{1}_p\sigma^2_\varepsilon), \\
    \kappa_t &= \kappa_{t-1}+\theta+\omega_{t}, \quad \omega_t \overset{iid}{\sim} \text{N}(0,\sigma^2_\omega),
\end{align}
\end{subequations}
where $y_{x,t}=\ln{\widetilde{m}_{x,t}}$, $\mathbf{1}_p$ is a $p$-dimensional identity matrix and $\text{N}(a,b)$ denotes the normal distribution with mean $a$ and covariance $b$.

\cite{FungPeSh17} study two generalisations of the state-space Lee-Carter system \eqref{eqn:LCSS} to analyse long term mortality time series for the Danish population. The first generalisation is to incorporate heteroscedasticity into the model, i.e. the homogeneous covariance structure $\mathbf{1}_p\sigma^2_\varepsilon$ in the observation equation in \eqref{eqn:LCSS} which is replaced by a heterogeneous covariance matrix $\mathbf{1}_p\sigma^2_{\varepsilon,x}$. This feature turns out to be a major improvement to model fit for Danish mortality data.

The second generalisation is to consider stochastic volatility for the latent process, that is the period effect $\kappa_t$ in the state equation in \eqref{eqn:LCSS}, to capture the observed characteristics of the long term time series data. Specifically, the following extended LC model with stochastic volatility feature is proposed:
\begin{subequations}\label{eqn:LCSV}
\begin{align}
    \boldsymbol{y}_t &= \boldsymbol{\alpha} + \boldsymbol{\beta} \kappa_t + \boldsymbol{\varepsilon}_t, \quad \boldsymbol{\varepsilon}_t
    \overset{iid}{\sim} \text{N}(\boldsymbol{0},\sigma^2_\varepsilon\boldsymbol{1}_p), \label{eqn:LCSVa} \\
    \kappa_t &= \kappa_{t-1} + \theta + \omega_t, \quad \omega_t|\xi_t \sim
    \text{N}(0,\exp\{\xi_t\}), \label{eqn:LCSVb} \\
    \xi_t &= \lambda_1 \xi_{t-1} + \lambda_2 + \eta_t, \quad \eta_t \overset{iid}{\sim}
    \text{N}(0,\sigma^2_\xi) \label{eqn:LCSVc}
\end{align}
\end{subequations}
where $\xi_t$ captures the stochastic volatility for the period effect. Bayesian inference for the model is developed based on particle MCMC method (\cite{AndrieuDoHo10}).

Despite the fact that cohort models are essential in mortality modelling as shown in Table~\ref{table:modelsliterature}, a formulation of mortality models with cohort factors incorporated is yet to be studied and analysed in the state-space setting in the literature. In the following we will present and demonstrate our approach for dealing with cohort models via state-space methodology.

\section{Cohort effects: state-space formulation}\label{sec:StateSpaceCohort}
This section presents a formulation of cohort models in state-space framework. We first describe how the cohort factor impacts the evolution of the age-specific death rates. The insight will give us a way to derive a state-space representation of the cohort effects. Bayesian inference for cohort models will be developed in Section~\ref{sec:bayesian}.

\subsection{Background}\label{sec:background}
\cite{RenshawHa06}
introduces a cohort factor $\gamma_{t-x}$ to the Lee-Carter model together with an age-modulating coefficient $\beta^{\gamma}_x$ as follows:
\begin{equation}\label{eqn:LCcohortRegression}
    \ln{m_{x,t}} = \alpha_x + \beta_x\kappa_t + \beta^{\gamma}_x\gamma_{t-x}
\end{equation}
where $x\in \{x_1,\dots,x_p\}$ and $t \in\{t_1,\dots,t_n\}$ represent age and calendar-year respectively. Here $t-x$ represents year-of-birth and hence $\gamma_{t-x}$ is a factor created to capture the cohort effect.

Numerical estimation of the cohort model \eqref{eqn:LCcohortRegression}, however, is reported in \cite{HuntVi15} to produce mixed convergence results based on the regression setup. Robustness of the resulting regression-based cohort models is also questioned as the goodness of model fit is reported to be sensitive to the data being used and the fitting algorithm. These robustness and convergence problems are also noticed in \cite{Currie16} where the paper presents a comprehensive approach for mortality modelling based on generalised linear and non-linear models, see also Section~\ref{sec:GLMs} for a brief discussion.

These issues thus provide another motivation to consider cohort models in the state-space framework as the state-space method would potentially eliminate the inefficiency caused by the 2-step estimation procedure required in other approaches, and in addition doesn't seem to suffer from the same sensitivity and poor convergence results identified in the aforementioned literature. We report our empirical findings for the estimation and forecasting of state-space cohort models in Section~\ref{sec:empircial}.

\subsection{State-space formulation}
In this paper we focus on the \textit{full cohort model}
\begin{equation}\label{eqn:CohortSS}
    \ln{\widetilde{m}_{x,t}} = \alpha_x + \beta_x\kappa_t + \beta^{\gamma}_x\gamma_{t-x} + \varepsilon_{x,t},
\end{equation}
where a dynamics for the crude death rate is being modelled and a noise term $\varepsilon_{x,t}$ is included.

To aid in explaining how we derive a state-space representation of cohort models, we consider a matrix of cells where the row and column corresponds to age ($x$) and year ($t$) respectively, see Table~\ref{table:cohorttable}. Here we assume $x=1,\dots,3$ and $t=1,\dots,4$ for illustration. The cohort factor $\gamma_{t-x}$ is indexed by the year-of-birth $t-x$ and its value on each cell is displayed in Table~\ref{table:cohorttable}. We first notice that the value $\gamma_{t-x}$ is constant on the ``cohort direction", that is on the cells $(x,t)$, $(x+1,t+1)$ and so on.

\begin{table}[h]
\center \setlength{\tabcolsep}{1em}
\renewcommand{\arraystretch}{1.1}
\begin{tabular}{c|c|c|c|c}
age/year & $t=1$ & $t=2$ & $t=3$ & $t=4$  \\
\hline
$x=1$ & $\gamma_{0}$ & $\gamma_{1}$ & $\gamma_{2}$ & $\gamma_{3}$    \\
\hline
$x=2$ & $\gamma_{-1}$ & $\gamma_{0}$ & $\gamma_{1}$ & $\gamma_{2}$    \\
\hline
$x=3$ & $\gamma_{-2}$ & $\gamma_{-1}$ & $\gamma_{0}$ & $\gamma_{1}$    \\
\end{tabular}
\center\footnotesize{\caption{\label{table:cohorttable} Values of the cohort factor $\gamma_{t-x}$ on a matrix of cells $(x,t)$.}}
\end{table}

Now consider the cohort model \eqref{eqn:CohortSS} and let $\gamma^x_t := \gamma_{t-x}$. The model can be expressed in matrix form as
\begin{equation}\label{eqn:LCcohortMatrix}
    \begin{pmatrix}
        \ln{\widetilde{m}_{1,t}} \\
        \ln{\widetilde{m}_{2,t}} \\
        \ln{\widetilde{m}_{3,t}}
    \end{pmatrix}
    =
    \begin{pmatrix}
        \alpha_1 \\
        \alpha_2 \\
        \alpha_3
    \end{pmatrix}
    +
    \begin{pmatrix}
        \beta_1 \\
        \beta_2 \\
        \beta_3
    \end{pmatrix} \kappa_t
    +
    \begin{pmatrix}
        \beta^\gamma_1 & 0 & 0  \\
        0 & \beta^\gamma_2 & 0  \\
        0 & 0 & \beta^\gamma_3
    \end{pmatrix}
    \begin{pmatrix}
        \gamma^1_t \\
        \gamma^2_t \\
        \gamma^3_t
    \end{pmatrix}
    +
    \begin{pmatrix}
        \varepsilon_{1,t} \\
        \varepsilon_{2,t} \\
        \varepsilon_{3,t}
    \end{pmatrix}.
\end{equation}
As time flows from $t=1$ to $t=4$, the cohort vector $(\gamma^1_t,\gamma^2_t,\gamma^3_t)^\top$, which represents the cohort factor in matrix form, proceeds as
\begin{equation}\label{eqn:cohortvectorflow}
    \begin{pmatrix}
        \gamma^1_1 (=\gamma_0) \\
        \gamma^2_1 (=\gamma_{-1}) \\
        \gamma^3_1 (=\gamma_{-2})
    \end{pmatrix}
  \rightarrow
    \begin{pmatrix}
        \gamma^1_2 (=\gamma_{1}) \\
        \gamma^2_2 (=\gamma_{0}) \\
        \gamma^3_2 (=\gamma_{-1})
    \end{pmatrix}
  \rightarrow
    \begin{pmatrix}
        \gamma^1_3 (=\gamma_2) \\
        \gamma^2_3 (=\gamma_{1}) \\
        \gamma^3_3 (=\gamma_{0})
    \end{pmatrix}
  \rightarrow
    \begin{pmatrix}
        \gamma^1_4 (=\gamma_3) \\
        \gamma^2_4 (=\gamma_{2}) \\
        \gamma^3_4 (=\gamma_{1})
    \end{pmatrix}.
\end{equation}
The key observation here from \eqref{eqn:cohortvectorflow} is that the first two elements of the cohort vector at time $t-1$ will appear as the bottom two elements of the cohort vector at time $t$. The pattern can also be observed from Table~\ref{table:cohorttable}. Therefore, the evolution of the cohort vector must satisfy
\begin{equation}\label{eqn:evolutioncohort}
\begin{pmatrix}
        \gamma^1_t \\
        \gamma^2_t \\
        \gamma^3_t
    \end{pmatrix}
    =
    \begin{pmatrix}
        * & * & * \\
        1 & 0 & 0 \\
        0 & 1 & 0
    \end{pmatrix}
    \begin{pmatrix}
        \gamma^1_{t-1} \\
        \gamma^2_{t-1} \\
        \gamma^3_{t-1}
    \end{pmatrix}
    + \dots,
\end{equation}
which is in fact a result of the defining property of ``cohort": $\gamma_{t-x}=\gamma_{(t-i)-(x-i)}$. Furthermore, it is obvious from \eqref{eqn:evolutioncohort} that one only needs to model the dynamics of $\gamma^1_t$ but not $\gamma^2_t$ and $\gamma^3_t$. We will use this observation to derive a state-space formulation of cohort models which is presented next.

Let $y_x=\ln{\widetilde{m}_{x,t}}$, in matrix notation we have (recall that $\gamma^x_t := \gamma_{t-x}$)
\begin{equation}\label{eqn:LCcohortObs}
    \begin{pmatrix}
        y_{x_1,t} \\
        y_{x_2,t} \\
        \vdots \\
        y_{x_p,t}
    \end{pmatrix}
    =
    \begin{pmatrix}
        \alpha_{x_1} \\
        \alpha_{x_2} \\
        \vdots \\
        \alpha_{x_p}
    \end{pmatrix}
    +
    \begin{pmatrix}
        \beta_{x_1} & \beta^{\gamma}_{x_1} & 0 & \cdots & 0  \\
        \beta_{x_2} & 0 & \beta^{\gamma}_{x_2} & \cdots & 0  \\
        \vdots & \vdots & \vdots & \ddots & \vdots \\
        \beta_{x_p} & 0 & 0 & \cdots & \beta^{\gamma}_{x_p}
    \end{pmatrix}
    \begin{pmatrix}
        \kappa_t \\
        \gamma^{x_1}_t \\
        \gamma^{x_2}_t \\
        \vdots \\
        \gamma^{x_p}_t
    \end{pmatrix}
    +
    \begin{pmatrix}
        \varepsilon_{x_1,t} \\
        \varepsilon_{x_2,t} \\
        \vdots \\
        \varepsilon_{x_p,t}
    \end{pmatrix}.
\end{equation}
It is clear that, from \eqref{eqn:LCcohortObs}, we have $y_{x_i,t}=\alpha_{x_i}+\beta_{x_i}\kappa_t+\beta^{\gamma}_{x_i}\gamma^{x_i}_t+\varepsilon_{x_i,t}$ which corresponds to \eqref{eqn:CohortSS} for $i \in \{1,\dots,p\}$. Here $(\kappa_t,\gamma^{x_1}_t,\dots,\gamma^{x_p}_t)^\top$ is the $p+1$ dimensional state vector.

From \eqref{eqn:cohortvectorflow}-\eqref{eqn:evolutioncohort}, we can write the state equation in matrix notation as follows:
\begin{equation}\label{eqn:LCcohortState}
    \begin{pmatrix}
        \kappa_{t} \\
        \gamma^{x_1}_{t} \\
        \gamma^{x_2}_{t} \\
        \vdots \\
        \gamma^{x_{p-1}}_{t}\\
        \gamma^{x_p}_{t}
    \end{pmatrix}
    =
    \begin{pmatrix}
        1 & 0 & 0 & \cdots & 0 & 0  \\
        0 & \lambda & 0 & \cdots & 0 & 0  \\
        0 & 1 & 0 &  \cdots & 0 & 0 \\
        0 & 0 & 1 &\cdots & 0 & 0 \\
        \vdots & \vdots & \vdots & \ddots & \vdots & \vdots \\
        0 & 0 & 0 & \cdots & 1 & 0
    \end{pmatrix}
    \begin{pmatrix}
        \kappa_{t-1} \\
        \gamma^{x_1}_{t-1} \\
        \gamma^{x_2}_{t-1} \\
        \vdots \\
        \gamma^{x_{p-1}}_{t-1}\\
        \gamma^{x_p}_{t-1}
    \end{pmatrix}
    +
    \begin{pmatrix}
        \theta \\
        \eta \\
        0 \\
        \vdots \\
        0 \\
        0
    \end{pmatrix}
    +
    \begin{pmatrix}
        \omega^\kappa_t \\
        \omega^{\gamma}_t \\
        0 \\
        \vdots \\
        0 \\
        0
    \end{pmatrix}.
\end{equation}
Here we assume $\kappa_t$ is a random walk with drift process ($\text{ARIMA}(0,1,0)$)
\begin{equation}
    \kappa_t = \kappa_{t-1} + \theta +\omega^\kappa_t, \quad \omega^\kappa_t \overset{iid}{\sim} \text{N}(0,\sigma^2_\omega),
\end{equation}
and the dynamics of $\gamma^{x_1}_t$ is described by a stationary AR(1) process (ARIMA(1,0,0))
\begin{equation}
    \gamma^{x_1}_t = \lambda \gamma^{x_1}_{t-1} + \eta + \omega^{\gamma}_t, \quad \omega^\gamma_t \overset{iid}{\sim} \text{N}(0,\sigma^2_\gamma),
\end{equation}
where $|\lambda|<1$. One may consider other dynamics for $\gamma^{x_1}_t$ by specifying the second row of the $p+1$ by $p+1$ matrix in \eqref{eqn:LCcohortState}. For example, one can consider generally the state equation as
\begin{equation}
    \begin{pmatrix}
        \kappa_{t} \\
        \gamma^{x_1}_{t} \\
        \gamma^{x_2}_{t} \\
        \vdots \\
        \gamma^{x_{p-1}}_{t}\\
        \gamma^{x_p}_{t}
    \end{pmatrix}
    =
    \begin{pmatrix}
        1 & 0 & 0 & \cdots & 0 & 0  \\
        0 & \lambda_1 & \lambda_2 & \cdots & \lambda_{p-1} & \lambda_{p}  \\
        0 & 1 & 0 &  \cdots & 0 & 0 \\
        0 & 0 & 1 &\cdots & 0 & 0 \\
        \vdots & \vdots & \vdots & \ddots & \vdots & \vdots \\
        0 & 0 & 0 & \cdots & 1 & 0
    \end{pmatrix}
    \begin{pmatrix}
        \kappa_{t-1} \\
        \gamma^{x_1}_{t-1} \\
        \gamma^{x_2}_{t-1} \\
        \vdots \\
        \gamma^{x_{p-1}}_{t-1}\\
        \gamma^{x_p}_{t-1}
    \end{pmatrix}
    +
    \begin{pmatrix}
        \theta \\
        \eta \\
        0 \\
        \vdots \\
        0 \\
        0
    \end{pmatrix}
    +
    \begin{pmatrix}
        \omega^\kappa_t \\
        \omega^{\gamma}_t \\
        0 \\
        \vdots \\
        0 \\
        0
    \end{pmatrix},
\end{equation}
where $\gamma^{x_1}_t=\lambda_1 \gamma^{x_1}_{t-1}+\lambda_2\gamma^{x_2}_{t-1} + \dots + \lambda_{p-1}\gamma^{x_{p-1}}_{t-1}+ \lambda_p\gamma^{x_p}_{t-1} + \eta + \omega^\gamma_t$ which is an ARIMA(p,0,0) process since $\gamma^{x_i}_{t-1}=\gamma^{x_1}_{t-i}$, $i=2,\dots,p$.

We can express the matrix form of \eqref{eqn:LCcohortObs}-\eqref{eqn:LCcohortState} succinctly as
\begin{subequations}\label{eqn:SSfullcohort}
    \begin{align}
        \bm{y}_t &= \bm{\alpha} + B \bm{\varphi}_t + \bm{\varepsilon}_t, \quad \bm{\varepsilon}_t \overset{iid}{\sim} \text{N}(0,\sigma^2_\varepsilon \mathsf{1}_p), \\
        \bm{\varphi}_t &= \Lambda\bm{\varphi}_{t-1} + \bm{\Theta} + \bm{\omega}_t, \quad \bm{\omega}_t \overset{iid}{\sim} \text{N}(0,\Upsilon),
    \end{align}
\end{subequations}
where
\begin{equation}
    B=    \begin{pmatrix}
        \beta_{x_1} & \beta^{\gamma}_{x_1} & 0 & \cdots & 0  \\
        \beta_{x_2} & 0 & \beta^{\gamma}_{x_2} & \cdots & 0  \\
        \vdots & \vdots & \vdots & \ddots & \vdots \\
        \beta_{x_p} & 0 & 0 & \cdots & \beta^{\gamma}_{x_p}
    \end{pmatrix}, \quad
    \Lambda =        \begin{pmatrix}
        1 & 0 & 0 & \cdots & 0 & 0  \\
        0 & \lambda & 0 & \cdots & 0 & 0  \\
        0 & 1 & 0 &  \cdots & 0 & 0 \\
        0 & 0 & 1 &\cdots & 0 & 0 \\
        \vdots & \vdots & \vdots & \ddots & \vdots & \vdots \\
        0 & 0 & 0 & \cdots & 1 & 0
    \end{pmatrix},\quad
    \Theta =  \begin{pmatrix}
        \theta \\
        \eta \\
        0 \\
        \vdots \\
        0 \\
        0
    \end{pmatrix},
\end{equation}
and $\bm{\varphi}_t=(\kappa_t,\gamma^{x_1}_t,\dots,\gamma^{x_p}_t)^\top$, $\mathsf{1}_p$ the $p$-dimensional identity matrix and $\Upsilon$ is a $p+1$ by $p+1$ diagonal matrix with diagonal $(\sigma^2_\kappa,\sigma^2_\gamma,0,\dots,0)$. For simplicity we assume homoscedasticity in the observation equation; heteroscedasticity can be considered as developed in \cite{FungPeSh17}.

The full cohort model \eqref{eqn:CohortSS} is suffering from an identification problem since the model is invariant to the following transformation
\begin{equation}
    \left(\alpha_x,\beta_x,\kappa_t,\beta_x^\gamma, \gamma_{t-x}\right) \rightarrow \left(\alpha_x+c_1\beta_x+c_2\beta_x^\gamma, \frac{1}{c_3}\beta_x, c_3\left(\kappa_t-c_1\right), \frac{1}{c_4}\beta_x^\gamma, c_4\left(\gamma_{t-x}-c_2\right)\right),
\end{equation}
where $c_3\neq 0$ and $c_4\neq 0$. The identification problem can be resolved by imposing the following parameter constraints
\begin{equation}\label{eqn:proposedconstraints}
    \sum_{x=x_1}^{x_p} \beta_x=1, \quad \sum_{x=x_1}^{x_p} \beta^\gamma_x=1, \quad \sum_{t=t_1}^{t_n} \kappa_t=0, \quad \sum_{c=t_1-x_p}^{t_n-x_1}\gamma_c=0,
\end{equation}
to ensure a unique model structure is identified. \cite{HuntVi15} and \cite{Currie16} provide further discussions on the identifiability issues for cohort models.
\newline
\begin{remark}
The observation and state equations \eqref{eqn:LCcohortObs}-\eqref{eqn:LCcohortState} imply that the cohort model that we have formulated here belongs to the linear-Gaussian class of state-space models. As a result one can perform efficient maximum-likelihood or Bayesian estimation on fitting the model to data, see \cite{FungPeSh17}. In this paper we focus on Bayesian inference so that mortality forecasts can take into account parameter uncertainty.
\end{remark}

\subsection{A simplified cohort model}
The cohort model \eqref{eqn:LCcohortRegression} assumes that the impact of the cohort factor on age-specific death rates is modulated by the coefficient $\beta^\gamma_x$. Iteration-based estimation of the cohort model is reported in \cite{Cairnsetal09} to be suffering from convergence problems. As a result, \cite{HabermanRe11} consider to simplify the model structure to
\begin{equation}\label{eqn:simplifiedLCcohort}
    \ln{m_{x,t}} = \alpha_x + \beta_x\kappa_t + \gamma_{t-x}.
\end{equation}
That is, the modulating coefficient $\beta^{\gamma}_x$ for the cohort factor is set to be one for all age $x$. It is suggested that the simplified model \eqref{eqn:simplifiedLCcohort} exhibits better estimation convergence behaviour when fitting the model to mortality data.

\begin{remark}
When fitting cohort models to data based on the Poisson or binomial regression setup via iteration-based estimation, it is typical to assume the starting values of the iteration scheme are coming from the estimates of other similar models, for example the LC model or APC model (\cite{HuntVi15}, \cite{Currie16}, \cite{VillegasMiKa15}). Even doing so the convergence is not guaranteed. We will report in Section~\ref{sec:empircial} that our approach based on Bayesian state-space framework do not require such an assumption and is able to successfully perform Bayesian estimation for various countries when cohort patterns are present in the data.
\end{remark}

In the following, we will also consider the model
\begin{subequations}\label{eqn:SSSimpCohort}
    \begin{align}
        \bm{y}_t &= \bm{\alpha} + B^s \bm{\varphi}_t + \bm{\varepsilon}_t, \quad \bm{\varepsilon}_t \overset{iid}{\sim} \text{N}(0,\sigma^2_\varepsilon \mathsf{1}_p),\\
        \bm{\varphi}_t &= \Lambda\bm{\varphi}_{t-1} + \bm{\Theta} + \bm{\omega}_t, \quad \bm{\omega}_t \overset{iid}{\sim} \text{N}(0,\Upsilon),
    \end{align}
\end{subequations}
where
\begin{equation}
    B^s=    \begin{pmatrix}
        \beta_{x_1} & 1 & 0 & \cdots & 0  \\
        \beta_{x_2} & 0 & 1 & \cdots & 0  \\
        \vdots & \vdots & \vdots & \ddots & \vdots \\
        \beta_{x_p} & 0 & 0 & \cdots & 1
    \end{pmatrix}.
\end{equation}
It will be referred to as the \textit{simplified cohort model}. In addition, the following set of parameter constraints
\begin{equation}\label{eqn:proposedconstraints}
    \sum_{x=x_1}^{x_p} \beta_x=1, \quad \sum_{t=t_1}^{t_n} \kappa_t=0, \quad \sum_{c=t_1-x_p}^{t_n-x_1}\gamma_c=0,
\end{equation}
is imposed to ensure the identifiability of the simplified cohort model.

\section{Bayesian inference for cohort models}\label{sec:bayesian}
In this section we begin by detailing the Bayesian estimation of the full cohort model in state-space formulation \eqref{eqn:SSfullcohort}. Nested models including the simplified cohort model \eqref{eqn:SSSimpCohort} and the LC model \eqref{eqn:LCSS} will be discussed in Section~\ref{sec:BayesianNestModels}. We first note that the cohort models and the LC model belong to the class of linear and Gaussian state-space models. As a result one can apply an efficient MCMC estimation algorithm based on Gibbs sampling with conjugate priors combined with forward-backward filtering as described in \cite{FungPeSh17}, this forms a special case of the so-called collapsed Gibbs sampler framework of \cite{vanDykPa08}.

\subsection{Bayesian inference for the full cohort model}\label{sec:BayesianFullCohort}
For the full cohort model \eqref{eqn:SSfullcohort}, the target density is given by
\begin{align}
    \pi\left(\bm{\varphi}_{0:n},\bm{\psi}|\bm{y}_{1:n}\right) &\propto \pi(\bm{y}_{1:n}|\bm{\varphi}_{0:n},\bm{\psi}) \pi(\bm{\varphi}_{0:n}|\bm{\psi}) \pi(\bm{\psi})\\
    &= \prod_{k=1}^n \pi(\bm{y}_{k}|\bm{\varphi}_{0:k},\bm{\psi}) \pi(\bm{\varphi}_{k}|\bm{\varphi}_{k-1},\bm{\psi})\pi(\bm{\varphi}_0) \pi(\bm{\psi})
\end{align}
where $\bm{\varphi}_{0:n}:=(\kappa_{0:n},\gamma^{x_1}_{0:n},\dots,\gamma^{x_p}_{0:n})$ is the $p+1$ dimensional (for each $t$) latent state vector and
$\boldsymbol{\psi} := (\boldsymbol{\beta},\boldsymbol{\beta}^\gamma, \boldsymbol{\alpha}, \theta,\eta,\lambda,\sigma^2_\varepsilon,\sigma^2_\kappa,\sigma^2_\gamma)$ is the $3p+6$ dimensional static parameter vector. In order to simplify the notation, we write $t=1,\dots,n$ instead of $t=t_1,\dots, t_n$. We perform block sampling for the latent state via the so-called forward-filtering-backward-sampling (FFBS) algorithm (\cite{CarterKo94}) and the posterior samples of the static parameters are obtained via conjugate priors. The sampling procedure is described in Algorithm~\ref{GibbsAlgorithm}, where $N$ is the number of MCMC iterations performed. Note also that the notation $\boldsymbol{\psi}_{-\nu}^{(i)}=(\psi_1^{(i)},\dots,\psi_{\nu-1}^{(i)},\psi^{(i-1)}_{\nu+1},\dots,\psi^{(i-1)}_{3p+6})$ is used in Algorithm~\ref{GibbsAlgorithm}.

In Algorithm~\ref{GibbsAlgorithm}, after obtaining $\tilde{\kappa}^{(i)}_{1:n}$ from line 3, one can impose the constraint $\sum_t\kappa_t^{(i)}=0$ by setting
$\kappa^{(i)}_t = \tilde{\kappa}^{(i)}_t - (1/n)\sum_{j=1}^n \tilde{\kappa}^{(i)}_j$ where $t=1,\dots,n$. Similarly, once $\tilde{\gamma}^{(i)}_c$ is obtained from $\bm{\varphi}^{(i)}_{0:n}$ in line 3, setting $\gamma^{(i)}_c = \tilde{\gamma}^{(i)}_c - (1/(n+p-1))\sum^{n-x_1}_{\ell=1-x_p} \tilde{\gamma}^{(i)}_\ell$, where $c=1-x_p,\dots,n-x_1$, will ensure the constraint $\sum_c \gamma^{(i)}_c = 0$ is satisfied. To impose the constraint $\sum_x \beta^{(i)}_x = 1$, we set $\beta^{(i)}_x=\tilde{\beta}^{(i)}_x/\sum_{j=x_1}^{x_p}\tilde{\beta}^{(i)}_j$, where $x=x_1,\dots,x_p$, once $\tilde{\bm{\beta}}^{(i)}$ is obtained from line 5-7. Constraint for $(\bm{\beta}^{\gamma})^{(i)}$ can be imposed similarly.

\begin{algorithm}[H]
\caption{MCMC sampling for $\pi(\bm{\varphi}_{0:n},\bm{\psi}|\bm{y}_{1:n})$}
\label{GibbsAlgorithm}
\begin{algorithmic}[1]
\State{Initialise: $\boldsymbol{\psi}=\boldsymbol{\psi}^{(0)}$.}
\For{$i=1,\dots,N$}
    \State Sample $\bm{\varphi}^{(i)}_{0:n}$ from
        $\pi(\bm{\varphi}_{0:n}|\boldsymbol{\psi}^{(i-1)},\boldsymbol{y}_{1:n})$ via FFBS (Section~\ref{sec:filtering}).
    \State Impose the constraint $\sum_t \kappa^{(i)}_t = 0$ and $\sum_c \gamma^{(i)}_c = 0$.
    \For{$h=1,\dots,p$}
        \State Sample $\beta^{(i)}_{x_h}$ from its posterior $\pi(\beta_{x_h}|\bm{\varphi}^{(i)}_{0:n},\bm{\psi}^{(i)}_{-\beta_{x_h}},\boldsymbol{y}_{1:n})$
    \EndFor
    \State Impose the constraint $\sum_x \beta^{(i)}_x = 1$.
    \For{$h=1,\dots,p$}
        \State Sample $(\beta^{\gamma}_{x_h})^{(i)}$ from its posterior $\pi(\beta^\gamma_{x_h}|\bm{\varphi}^{(i)}_{0:n},\bm{\psi}^{(i)}_{-\beta^\gamma_{x_h}},\boldsymbol{y}_{1:n})$
    \EndFor
    \State Impose the constraint $\sum_x (\beta^{\gamma}_x)^{(i)} = 1$.
    \For{$h=2p+1,\dots,3p+6$}
        \State Sample $\psi^{(i)}_h$ from $\pi(\psi_h|\bm{\varphi}^{(i)}_{0:n},\boldsymbol{\psi}_{-h}^{(i)},\boldsymbol{y}_{1:n})$
    \EndFor
\EndFor
\end{algorithmic}
\end{algorithm}

\subsubsection{Forward-backward filtering for latent state dynamics}\label{sec:filtering}
The FFBS procedure requires to carry out multivariate Kalman filtering forward in time and then sample backwardly using the obtained filtering distributions. For the full cohort model \eqref{eqn:LCcohortObs}-\eqref{eqn:LCcohortState}, the conditional distributions involved in the multivariate Kalman filtering recursions are given by
\begin{subequations}
    \begin{align}
        \bm{\varphi}_{t-1}|\bm{y}_{1:t-1} &\sim \text{N}(\bm{m}_{t-1},C_{t-1}), \label{filterdisttminus1} \\
        \bm{\varphi}_t|\boldsymbol{y}_{1:t-1} &\sim \text{N}(\bm{a}_t, R_t), \\
        \boldsymbol{y}_t|\boldsymbol{y}_{1:t-1} &\sim \text{N}(\bm{f}_t,Q_t), \label{eqn:KFpredictY} \\
        \bm{\varphi}_t|\boldsymbol{y}_{1:t} &\sim \text{N}(\bm{m}_t,C_t), \label{eqn:filterdistt}
    \end{align}
\end{subequations}
where
\begin{subequations}
    \begin{align}
        &\bm{a}_t=\Lambda\bm{m}_{t-1}+\bm{\Theta}, \quad R_t=\Lambda C_{t-1} \Lambda^\top+\Upsilon, \label{eqn:kf1} \\
        &\bm{f}_t=\bm{\alpha}+B\bm{a}_t, \quad Q_t=B R_t B^\top + \sigma^2_\varepsilon \mathsf{1}_p, \\
        &\bm{m}_t=\bm{a}_t + R_t B^\top Q_t^{-1}(\bm{y}_t-\bm{f}_t), \quad
        C_t=R_t-R_t B^\top Q_t^{-1} B R_t. \label{eqn:kf4}
    \end{align}
\end{subequations}
for $t=1,\dots,n$. Since
\begin{equation}\label{eqn:ffbs}
    \pi(\bm{\varphi}_{0:n}|\bm{\psi},\bm{y}_{1:n})=\prod^n_{t=0}\pi(\bm{\varphi}_t|\bm{\varphi}_{t+1:n},\bm{\psi},\bm{y}_{1:n})=
    \prod^n_{t=0}\pi(\bm{\varphi}_t|\bm{\varphi}_{t+1},\bm{\psi},\bm{y}_{1:t}),
\end{equation}
we see that for a block sampling of the latent state, one can first draw $\bm{\varphi}_n$ from $\text{N}(\bm{m}_n,C_n)$ and then, for $t=n-1,\dots,1,0$ (that is backward in time), draws a sample of $\bm{\varphi}_t|_{\bm{\varphi}_{t+1},\bm{\psi},\bm{y}_{1:t}}$ recursively given a sample of $\bm{\varphi}_{t+1}$. It turns out that
$\bm{\varphi}_t|_{\bm{\varphi}_{t+1},\bm{\psi},\bm{y}_{1:t}} \sim \text{N}(\bm{h}_t,H_t)$ where
\begin{subequations}
    \begin{align}
        \bm{h}_t &=\bm{m}_t+C_t\Lambda^\top R^{-1}_{t+1} (\bm{\varphi}_{t+1}-\bm{a}_{t+1}), \\
        H_t &= C_t - C_t\Lambda^\top R^{-1}_{t+1} \Lambda C_t,
    \end{align}
\end{subequations}
based on Kalman smoothing (\cite{CarterKo94}).

\subsubsection{Posteriors for static parameters}\label{sec:posteriors}
To sample the posterior distribution of the static parameters in Algorithm~\ref{GibbsAlgorithm}, we assume the following independent conjugate priors:
\begin{subequations}
    \begin{align}
        & \alpha_x \sim \text{N}(\tilde{\mu}_\alpha,\tilde{\sigma}^2_\alpha), \quad
        \beta_x \sim \text{N}(\tilde{\mu}_\beta,\tilde{\sigma}^2_\beta), \quad
        \beta^\gamma_x \sim \text{N}(\tilde{\mu}_{\beta^\gamma},\tilde{\sigma}^2_{\beta^\gamma}), \\
        &\theta \sim \text{N}(\tilde{\mu}_\theta,\tilde{\sigma}^2_\theta), \quad \eta \sim \text{N}(\tilde{\mu}_\eta,\tilde{\sigma}^2_\eta),\quad
        \lambda \sim \text{N}_{[-1,1]}(\tilde{\mu}_\lambda,\tilde{\sigma}^2_\lambda),\\ &\sigma^2_{\varepsilon} \sim \text{IG}(\tilde{a}_\varepsilon, \tilde{b}_\varepsilon), \quad
        \sigma^2_\kappa \sim \text{IG}(\tilde{a}_\kappa, \tilde{b}_\kappa), \quad \sigma^2_\gamma \sim \text{IG}(\tilde{a}_\gamma, \tilde{b}_\gamma),
    \end{align}
\end{subequations}
where $\text{N}_{[-1,1]}$ denotes a truncated Gaussian with support $[-1,1]$ and $\text{IG}(\tilde{a},\tilde{b})$ denotes an inverse-gamma
distribution with mean $\tilde{b}/(\tilde{a}-1)$ and variance $\tilde{b}^2/((\tilde{a}-1)^2(\tilde{a}-2))$ for $\tilde{a} > 2$. The posteriors of the static parameters are then obtained as follows:\footnote{For simplicity, we denote $\bm{y}=\bm{y}_{1:n}$, $\bm{\varphi}=\bm{\varphi}_{0:n}$ and
$\bm{\psi}_{-h}=(\psi_1,\dots,\psi_{h-1},\psi_{h+1},\dots,\psi_{3p+6})$.}
\begin{align}
\alpha_x|\bm{y},\bm{\varphi},\bm{\psi}_{-\alpha_x} &\sim
        \text{N}\left(\frac{\tilde{\sigma}^2_\alpha \sum_{t=1}^n
           (y_{x,t}-\beta_x\kappa_t-\beta^\gamma_x\gamma^x_t)+\tilde{\mu}_\alpha\sigma^2_\varepsilon}{\tilde{\sigma}^2_\alpha n+\sigma^2_\varepsilon},
            \frac{\tilde{\sigma}^2_\alpha \sigma^2_\varepsilon}{\tilde{\sigma}^2_\alpha n+\sigma^2_\varepsilon}\right), \label{postAlpha}\\
\beta_x|\bm{y},\bm{\varphi}, \bm{\psi}_{-\beta_x} &\sim
            \text{N} \left(\frac{\tilde{\sigma}^2_\beta \sum_{t=1}^n(y_{x,t}-(\alpha_x+\beta^\gamma_x\gamma^x_t))\kappa_t+\tilde{\mu}_\beta \sigma^2_\varepsilon}
            {\tilde{\sigma}^2_\beta \sum_{t=1}^n\kappa^2_t+\sigma^2_\varepsilon},
            \frac{\tilde{\sigma}^2_\beta \sigma^2_\varepsilon}{\tilde{\sigma}^2_\beta \sum_{t=1}^n\kappa^2_t+\sigma^2_\varepsilon}\right), \label{postBeta}\\
\beta^\gamma_x|\bm{y},\bm{\varphi}, \bm{\psi}_{-\beta^\gamma_x} &\sim
            \text{N} \left(\frac{\tilde{\sigma}^2_{\beta^\gamma} \sum_{t=1}^n(y_{x,t}-(\alpha_x+\beta_x\kappa_t))\gamma^x_t+\tilde{\mu}_{\beta^\gamma} \sigma^2_\varepsilon}{\tilde{\sigma}^2_{\beta^\gamma} \sum_{t=1}^n(\gamma^x_t)^2+\sigma^2_\varepsilon},
            \frac{\tilde{\sigma}^2_{\beta^\gamma} \sigma^2_\varepsilon}{\tilde{\sigma}^2_{\beta^\gamma} \sum_{t=1}^n(\gamma^x_t)^2+\sigma^2_\varepsilon}\right), \label{eqn:postBetaG}\\
\theta|\bm{y},\bm{\varphi},\bm{\psi}_{-\theta} &\sim
           \text{N} \left(\frac{\tilde{\sigma}^2_\theta\sum^n_{t=1}(\kappa_t-\kappa_{t-1})+\tilde{\mu}_\theta\sigma^2_\omega}
           {\tilde{\sigma}^2_\theta n+\sigma^2_\omega},
           \frac{\tilde{\sigma}^2_\theta \sigma^2_\omega}{\tilde{\sigma}^2_\theta n+\sigma^2_\omega}\right),\\
\eta|\bm{y},\bm{\varphi},\bm{\psi}_{-\theta} &\sim
           \text{N} \left(\frac{\tilde{\sigma}^2_\eta\sum^n_{t=1}(\gamma^{x_1}_t-\lambda\gamma^{x_1}_{t-1})+\tilde{\mu}_\eta\sigma^2_\gamma}
           {\tilde{\sigma}^2_\eta n+\sigma^2_\gamma},
           \frac{\tilde{\sigma}^2_\eta \sigma^2_\gamma}{\tilde{\sigma}^2_\eta n+\sigma^2_\gamma}\right),\\
\lambda|\bm{y},\bm{\varphi},\bm{\psi}_{-\lambda} &\sim
           \text{N}_{[-1,1]} \left(\frac{\tilde{\sigma}^2_\lambda \sum^n_{t=1}((\gamma^{x_1}_t-\eta)\gamma^{x_1}_{t-1})+\tilde{\mu}_\lambda\sigma^2_\gamma}
           {\tilde{\sigma}^2_\lambda \sum^n_{t=1}(\gamma^{x_1}_{t-1})^2+\sigma^2_\gamma},
           \frac{\tilde{\sigma}^2_\lambda \sigma^2_\gamma}{\tilde{\sigma}^2_\lambda \sum^n_{t=1}(\gamma^{x_1}_{t-1})^2 +\sigma^2_\gamma}\right),\\
\sigma^2_\varepsilon|\bm{y},\bm{\varphi},\bm{\psi}_{-\sigma^2_\varepsilon} &\sim
            \text{IG} \left(\tilde{a}_\varepsilon+\frac{np}{2},\,
            \tilde{b}_\varepsilon+\frac{1}{2}
           \sum_{x=x_1}^{x_p}\sum^n_{t=1}\left(y_{x,t}-\left(\alpha_x+\beta_x\kappa_t+\beta^\gamma_x\gamma^x_t\right)\right)^2\right),\label{postSigmaEps}\\
\sigma^2_\kappa|\bm{y},\bm{\varphi}, \bm{\psi}_{-\sigma^2_\kappa} &\sim
            \text{IG} \left(\tilde{a}_\kappa+\frac{n}{2},\,
           \tilde{b}_\kappa+\frac{1}{2}
           \sum^n_{t=1}\left(\kappa_t-(\kappa_{t-1}+\theta)\right)^2\right), \\
\sigma^2_\gamma|\bm{y},\bm{\varphi}, \bm{\psi}_{-\sigma^2_\gamma} &\sim
            \text{IG} \left(\tilde{a}_\gamma+\frac{n}{2},\,
           \tilde{b}_\gamma+\frac{1}{2}
           \sum^n_{t=1}\left(\gamma^{x_1}_t-\lambda\gamma^{x_1}_{t-1}\right)^2\right).
\end{align}

\subsection{Bayesian inference for nested models}
\label{sec:BayesianNestModels}
The MCMC estimation for the full cohort model presented in Algorithm 1 can be applied to the nested models including the simplified cohort model and the LC model with only small adjustments.

The static parameter vector for the simplified cohort model is given by
\begin{equation}
\boldsymbol{\psi} := (\boldsymbol{\beta}, \boldsymbol{\alpha},\theta,\eta,\lambda,\sigma^2_\varepsilon,\sigma^2_\kappa,\sigma^2_\gamma),
\end{equation}
where the sampling of the age-modulating coefficients $\beta^\gamma_x$ for the cohort factor is not required. Consequently line 9-12 in Algorithm 1 can be removed in this case. Moreover, for the simplified cohort model, we set $\beta^\gamma_x=1$ in the posterior distributions of $\alpha_x$, $\beta_x$ and $\sigma^2_\varepsilon$ in \eqref{postAlpha}, \eqref{postBeta} and \eqref{postSigmaEps} respectively.

The LC model can be viewed as a further nested model of the simplified cohort model with the cohort factor $\gamma_{t-x}= 0$. As a result the state equation is one-dimensional with $\bm{\varphi}_{0:n}:=\kappa_{0:n}$ and the static parameter vector consists of
\begin{equation}
\boldsymbol{\psi} := (\boldsymbol{\beta}, \boldsymbol{\alpha},\theta,\sigma^2_\varepsilon,\sigma^2_\kappa).
\end{equation}
Identification constraint for the LC model is given by
\begin{equation}
     \sum_{x=x_1}^{x_p} \beta_x=1, \quad \sum_{t=t_1}^{t_n} \kappa_t=0.
\end{equation}
Hence the constraint $\sum_c \gamma^{(i)}_c = 0$ in line 4 and the sampling of $\beta^\gamma_x$ in line 9-12 in Algorithm 1 are not required. We also set $\gamma^x_t=0$ in the posterior distributions for $\boldsymbol{\beta}$, $\boldsymbol{\alpha}$ and $\sigma^2_\varepsilon$ in \eqref{postAlpha}, \eqref{postBeta} and \eqref{postSigmaEps} respectively. Further details for a Bayesian estimation of the LC model can be found in \cite{FungPeSh17}.

\section{Empirical Studies}\label{sec:empircial}
We analyse several set of mortality data from different countries based on the cohort models formulated in the state-space framework. We consider both the full cohort model and the simplified cohort model. In addition, we compare the cohort models against the LC model for model fitting as well as their forecasting properties. The countries that we consider includes England and Wales (UK), United States (US) and Italy (ITA). We perform our analysis on both male and female mortality data to investigate whether cohort effect within a country is shared for both genders. The data is obtained from the Human Mortality Database\footnote{www.mortality.org}. The year range is from year 1970 to 2010, and we restrict our attention to the age range 65-95. Consequently the range for the year-of-birth is 1875-1945.

\subsection{Model estimation}\label{sec:ModelEstimation}
We run the Markov chain sampler described in Section~\ref{sec:bayesian} for 30,000 iterations and the burn-in period is set to be 15,000 iterations to ensure that the chain has arrived to the stationary state; thus we are left with 15,000 posterior samples. For all Gaussian priors $\text{N}(\tilde{\mu},\tilde{\sigma}^2)$, including the truncated Gaussian $\text{N}_{[-1,1]}(\tilde{\mu},\tilde{\sigma}^2)$, we assume $\tilde{\mu}=0$ and $\tilde{\sigma}^2=10$; while for the inverse-gamma priors $\text{IG}(\tilde{a},\tilde{b})$ we set $\tilde{a}=2.01$ and $\tilde{b}=0.01$. The hyperparameters are chosen and tested to ensure the priors are sufficiently vague.  To start the Kalman filter, we assume $\bm{m}_0=\bm{0}$ and $C_0$ is a diagonal covariance matrix with diagonal elements all equal to $10$ (see \eqref{filterdisttminus1}).  We refer $x=\{x_1,\dots, x_p\}$ to $\text{age}=\{65,\dots, 95\}$ and $t=\{t_1,\dots,t_n\}$ to $\text{year}=\{1970,\dots,2010\}$ (i.e. $p=31$ and $T=41$).

As noted previously, no particular special initialization is required with this methodology, it seems relatively robust to the choice of starting points. To start the chain for the full cohort model, initial values of the static parameters are set as follows: $\alpha^{(0)}_x=(1/n)\sum_t y_{x,t}$, $\beta^{(0)}_x=(\beta^\gamma_x)^{(0)}=1/p$, $\theta^{(0)}=\eta^{(0)}=-0.1$, $(\sigma^2_\varepsilon)^{(0)}=(\sigma^2_\omega)^{(0)}=(\sigma^2_\gamma)^{(0)}=0.01$ and $\lambda^{(0)}=0.5$ where $x \in \{x_1,\dots,x_p\}$. The same set of values is also used for the simplified cohort model except that sampling of $\beta^\gamma_x$ is not required.

\subsubsection{Full cohort model}
\begin{figure}[h] 
\begin{center}
\includegraphics[width=5.5cm, height=20cm]{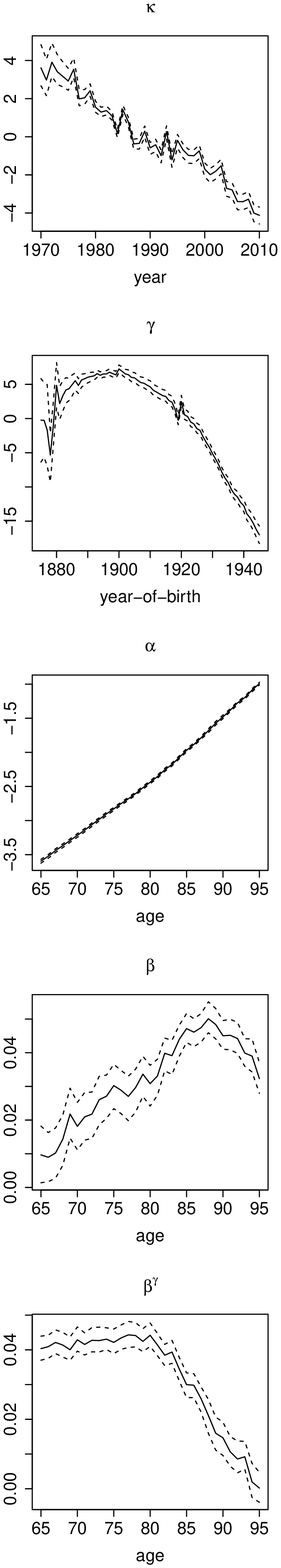}\includegraphics[width=5.5cm, height=20cm]{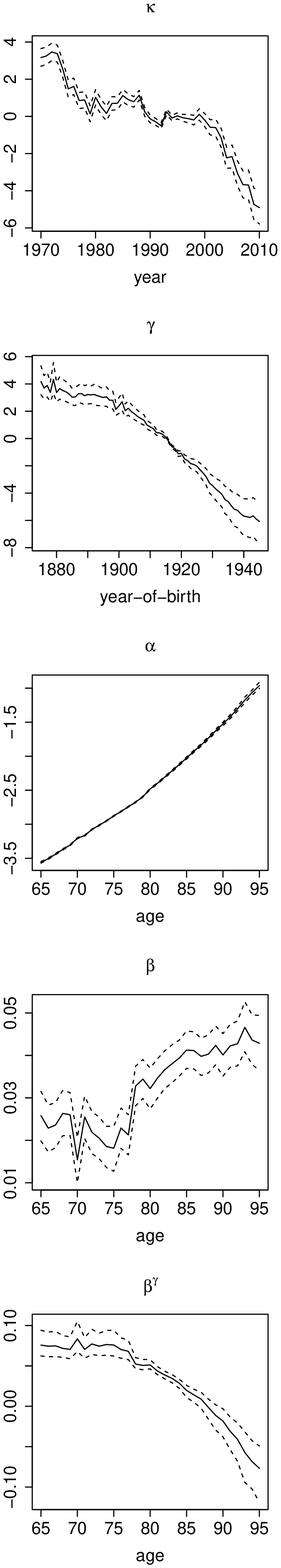}\includegraphics[width=5.5cm, height=20cm]{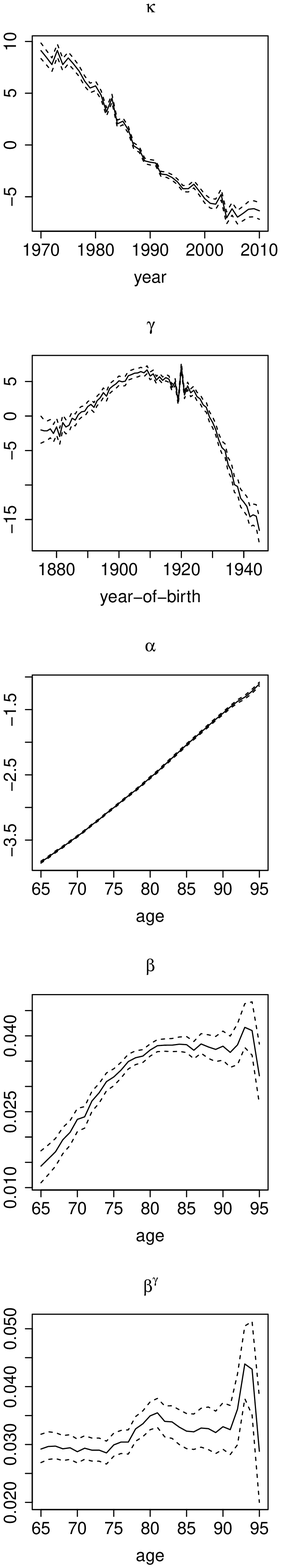}
\caption{\small{Full cohort model: estimated $\bm{\kappa}$, $\bm{\gamma}$, $\bm{\alpha}$, $\bm{\beta}$ and $\bm{\beta}^\gamma$ for the (left column) UK, (middle column) US and (right column) Italy male populations. Mean (solid line) and 95\% credible interval (dash lines) of the posterior distributions are plotted.}}
\label{fig:FullCohortEstMales}
\end{center}
\end{figure}

For the full cohort model, the posterior mean and 95\% credible intervals for the parameters $\bm{\kappa}$, $\bm{\gamma}$, $\bm{\alpha}$, $\bm{\beta}$ and $\bm{\beta}^\gamma$ for the UK, US and Italy male populations are shown in Figure~\ref{fig:FullCohortEstMales}.

Given the linear trend observed, the assumption that the period effect follows a random walk with drift process seems to be reasonable for the UK and Italy male populations, but is less appropriate for the US male population where structural changes seem to be present for the considered age range (see also \cite{LiChCh11} and \cite{vanBerkumAnVe16}). The estimated cohort factor shows a clear kink around the year-of-birth at 1920 for the UK and Italy data. It corresponds to the young generation (aged around 20) where the Second World War was taking place centrally in Europe around 1940. A mild kink appears around the generation born in 1900 for the US population.  The implication for these sudden changes of the cohort factor on model fit will be discussed in more details in Section~\ref{sec:modelComparison} where a comparison of the cohort models with the LC model through residuals is presented. The heightened volatility surrounding the first several years of the cohort factor can be attributed to having a limited data that are used to infer the cohort factor for the very first birth years, thus creating greater uncertainty.

\begin{figure}[!htb] 
\begin{center}
\includegraphics[width=5.5cm, height=20cm]{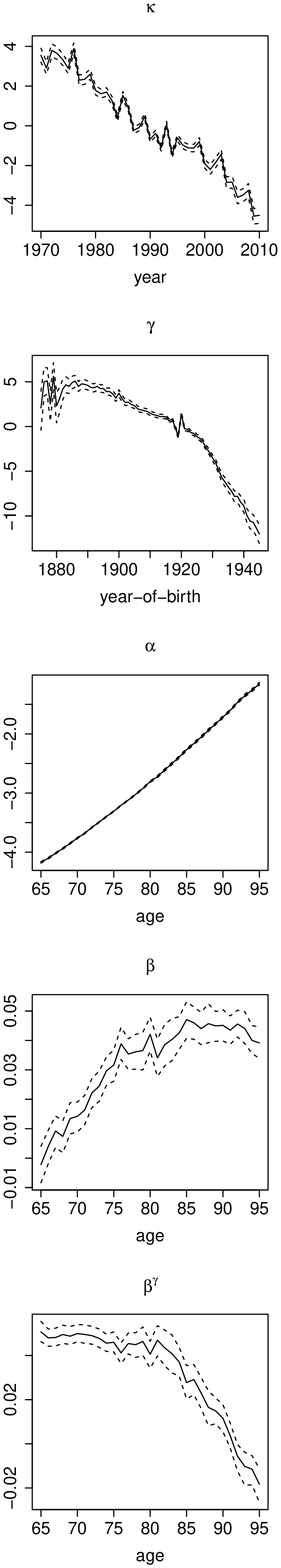}\includegraphics[width=5.5cm, height=20cm]{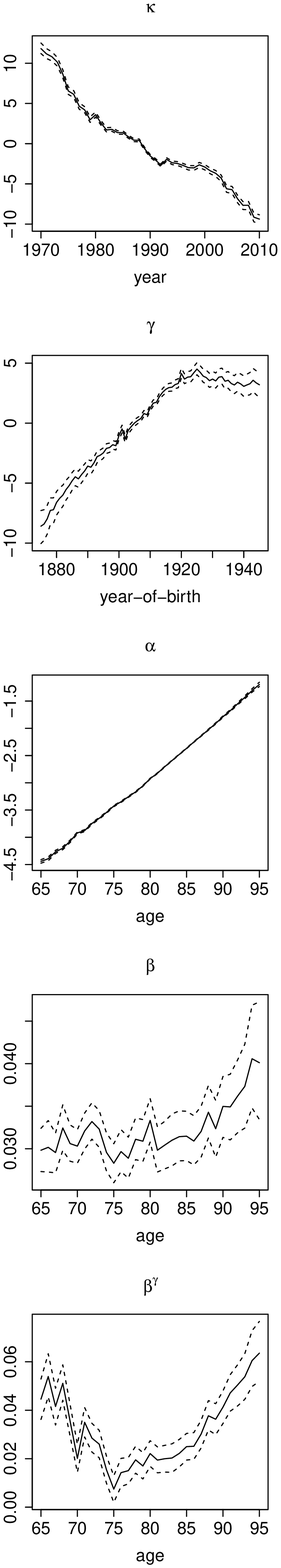}\includegraphics[width=5.5cm, height=20cm]{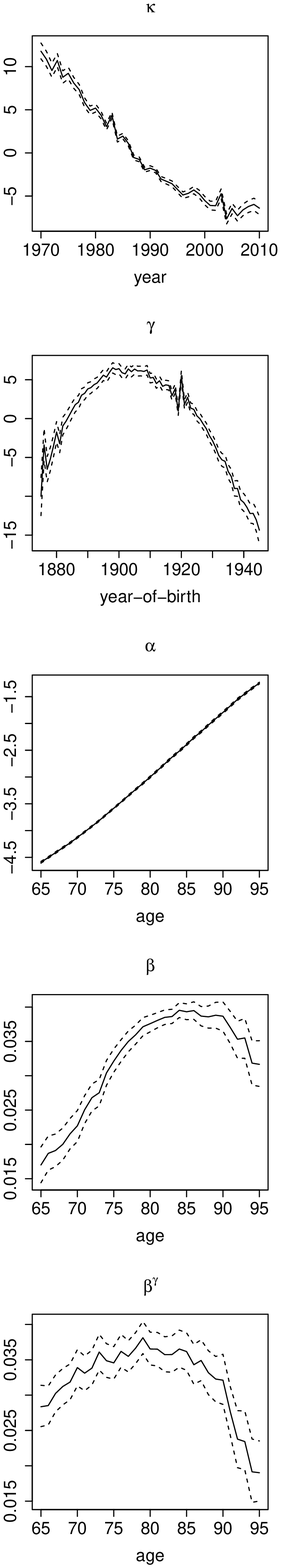}
\caption{\small{Full cohort model: estimated $\bm{\kappa}$, $\bm{\gamma}$, $\bm{\alpha}$, $\bm{\beta}$ and $\bm{\beta}^\gamma$ for the (left column) UK, (middle column) US and (right column) Italy female populations. Mean (solid line) and 95\% credible interval (dash lines) of the posterior distributions are plotted.}}
\label{fig:FullCohortEstFemales}
\end{center}
\end{figure}

Figure~\ref{fig:FullCohortEstFemales} displays the corresponding results for female populations. We also observe similar patterns where the kinks in the cohort factors for the female populations appear at almost exactly the same generations as in the male populations. It suggests that, at least for the countries that we presented here, cohort effect is not gender-specific but is a common phenomenon for people who belong to a particular generation. Such a finding in this case can perhaps be attributed to shared mortality experience in both genders arising from a significant global effect for Europe, corresponding to the Second World War.

Posterior statistics for the other parameters are displayed in Table~\ref{table:ParasFullCohort}.


\begin{table}[h]
\center \setlength{\tabcolsep}{1em}
\renewcommand{\arraystretch}{1.1}
\scalebox{0.9}{\begin{tabular}{l|ccc}
\hline \hline
  & UK  & US & Italy    \\
\hline
 & & Males & \\
\hline
$\theta$                & -0.18 [-0.40, 0.02]     & -0.20 [-0.35, -0.04]    & -0.37 [-0.63, -0.10]  \\
$\eta$                  & -0.57 [-0.79, -0.36]    & -0.21 [-0.29, -0.14]    & -0.55 [-0.98, -0.12]  \\
$\lambda$               & 0.993 [0.977, 0.999]       & 0.990 [0.975, 0.999]       & 0.98 [0.94, 0.99]  \\
$\sigma^2_\varepsilon$  & 0.00028 [0.00026, 0.00030] & 0.00020 [0.00019, 0.00022] & 0.00032 [0.00030, 0.00035]  \\
$\sigma^2_\omega$       & 0.46 [0.29, 0.72]       & 0.23 [0.14, 0.36]       & 0.71 [0.44, 1.10]  \\
$\sigma^2_\gamma$       & 0.46 [0.28, 0.72]       & 0.019 [0.008, 0.03]      & 1.93 [1.22, 3.00]  \\
\hline
 & & Females & \\
\hline
$\theta$                & -0.19 [-0.41, 0.02]     & -0.51 [-0.70, -0.33]    & -0.42 [-0.73, -0.11]  \\
$\eta$                  & -0.37 [-0.56, -0.19]    & 0.38 [0.17, 0.61]    & -0.51 [-0.88, -0.14]  \\
$\lambda$               & 0.990 [0.966, 0.999]       & 0.89 [0.81, 0.96]       & 0.98 [0.94, 0.99]  \\
$\sigma^2_\varepsilon$  & 0.00023 [0.00021, 0.00025] & 0.00022 [0.00020, 0.00024] & 0.00029 [0.00027, 0.00032]  \\
$\sigma^2_\omega$       & 0.52 [0.33, 0.81]       & 0.34 [0.22, 0.54]       & 0.98 [0.61, 1.54]  \\
$\sigma^2_\gamma$       & 0.35 [0.22, 0.56]       & 0.07 [0.04, 0.13]      & 1.41 [0.90, 2.19]  \\
\hline \hline
\end{tabular}}
\center\caption{\label{table:ParasFullCohort} \small{Estimated posterior mean of the static parameters for the full cohort model on male and female population data. $[.,.]$ next to the estimates represents $95\%$ posterior credible interval.}}
\end{table}

\subsubsection{Simplified cohort model}
\begin{figure}[!htb] 
\begin{center}
\includegraphics[width=5.5cm, height=20cm]{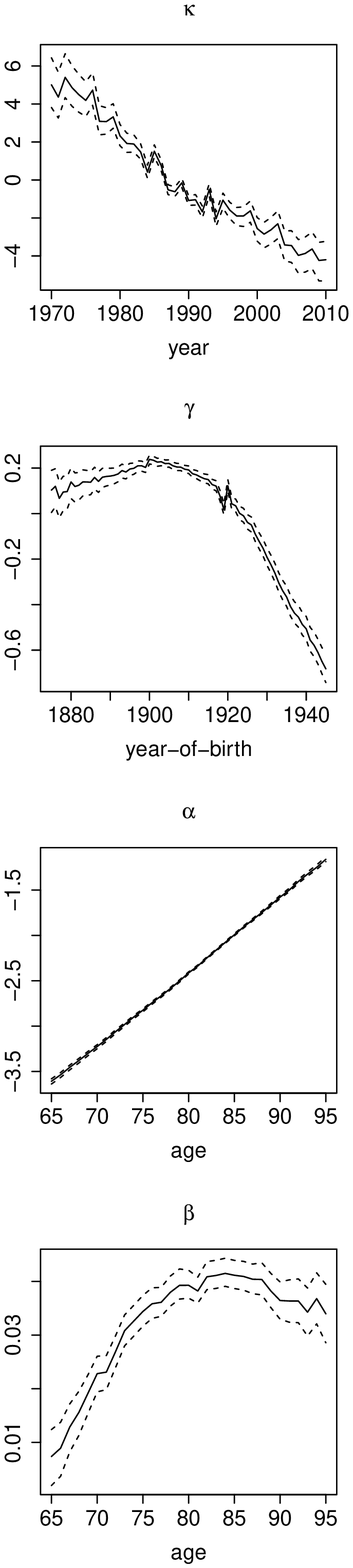}\includegraphics[width=5.5cm, height=20cm]{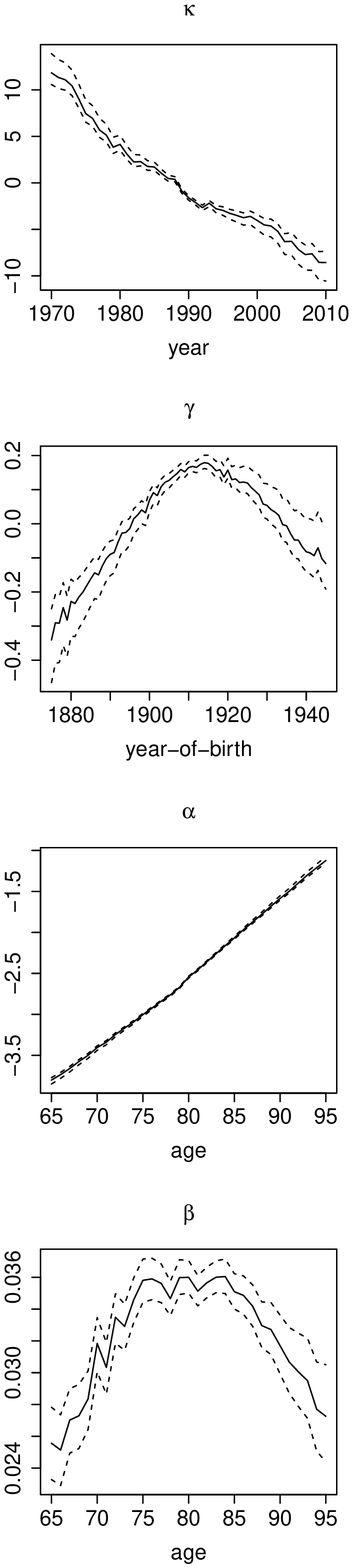}\includegraphics[width=5.5cm, height=20cm]{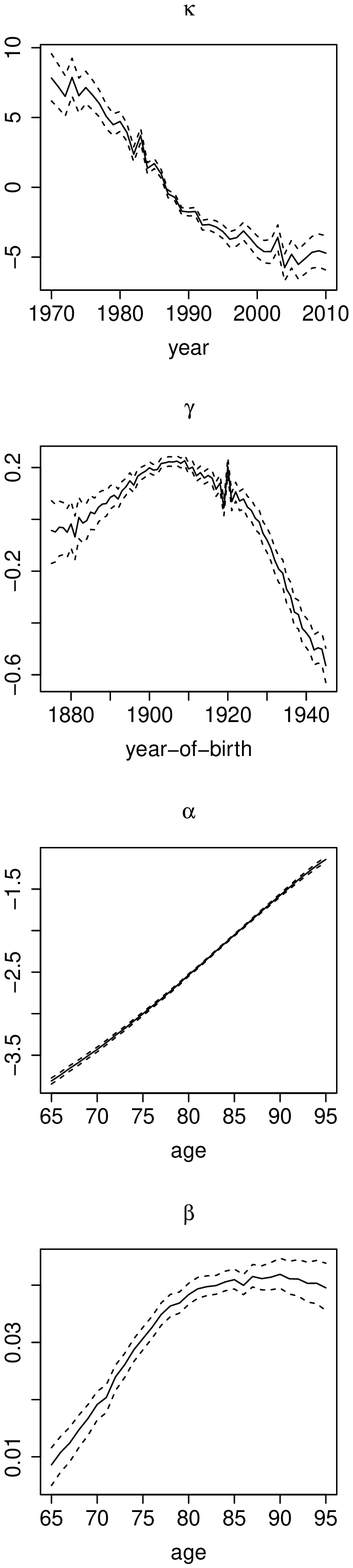}
\caption{\small{Simplified cohort model: estimated $\bm{\kappa}$, $\bm{\gamma}$, $\bm{\alpha}$ and $\bm{\beta}$ for the (left column) UK, (middle column) US and (right column) Italy male populations. Mean (solid line) and 95\% credible interval (dash lines) of the posterior distributions are plotted.}}
\label{fig:SimpCohortEstMales}
\end{center}
\end{figure}

Figure~\ref{fig:SimpCohortEstMales} shows the posterior mean and 95\% posterior credible intervals for the estimated parameters $\bm{\kappa}$, $\bm{\gamma}$, $\bm{\alpha}$, $\bm{\beta}$ where UK, US and Italy males mortality data are used to fit the simplified cohort model where it is assumed that $\beta^\gamma_x=1$ for all age $x$.

The estimated figures from the simplified cohort model share very similar patterns to those obtained from the full cohort model for the UK and Italy data. Interestingly, the same may not be said for the US data, as the estimated $\bm{\kappa}$ and $\bm{\gamma}$ are quite different to the corresponding ones obtained from the full cohort model. As the only difference between the simplified and the full cohort model is that $\beta^\gamma_x$ is allowed to be flexible in the full cohort model, the rather substantial difference for the estimated $\bm{\kappa}$ produced by the two models may lead one to question whether incorporating a cohort factor is appropriate for the US mortality data; in order words, it could be possible that cohort effects may not be significant for the past US mortality experience. We will further examine this in Section~\ref{sec:modelComparison}. Note also that the estimated values of $\bm{\gamma}$ for the simplified model are significantly smaller than the full model. The reason for this is that the constraint $\sum_x \beta^\gamma_x=1$ for the full model dictates that $\beta^\gamma_x$ will take value in the order of $1/p$ which is substantially smaller than $1$, where $p=31$ is the number of ages considered in the data; while $\beta^\gamma_x=1$ is assumed for the simplified model. One may consider fixing $\beta^\gamma_x=1/p$ in the simplified model so that the estimated values of $\bm{\gamma}$ will be in the same order of magnitude for both the simplified and full model. However this should not affect the fitting and forecasting of the resulting model.

\begin{figure}[!htb] 
\begin{center}
\includegraphics[width=5.5cm, height=20cm]{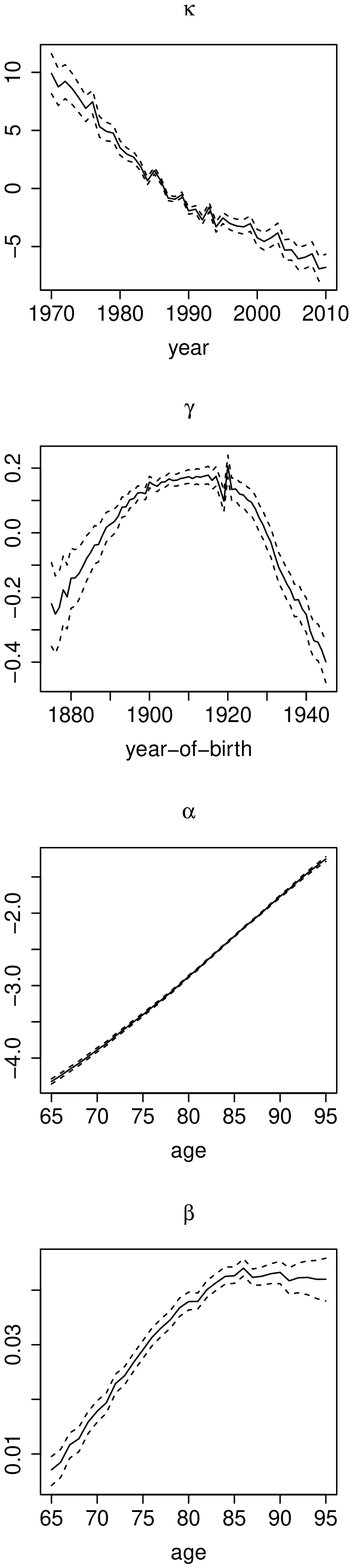}\includegraphics[width=5.5cm, height=20cm]{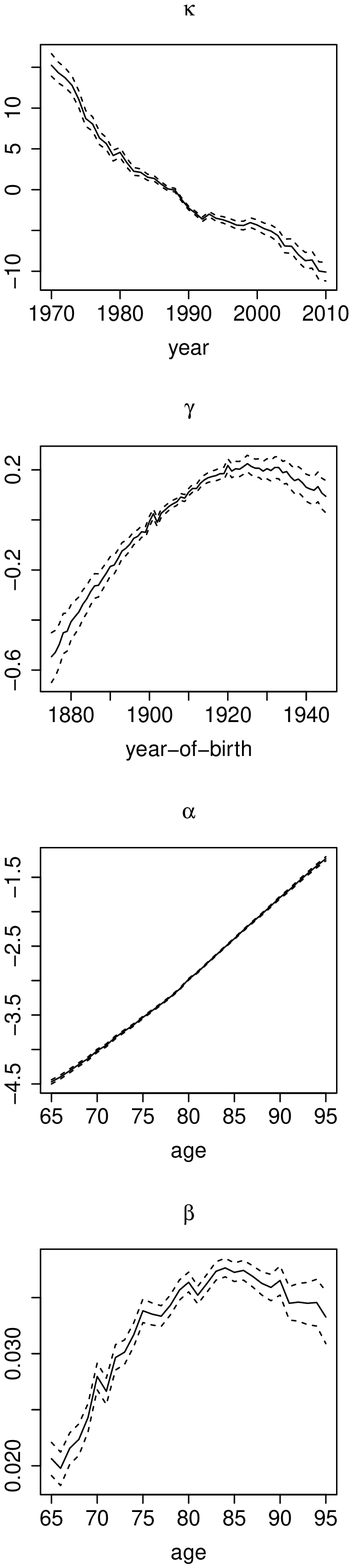}\includegraphics[width=5.5cm, height=20cm]{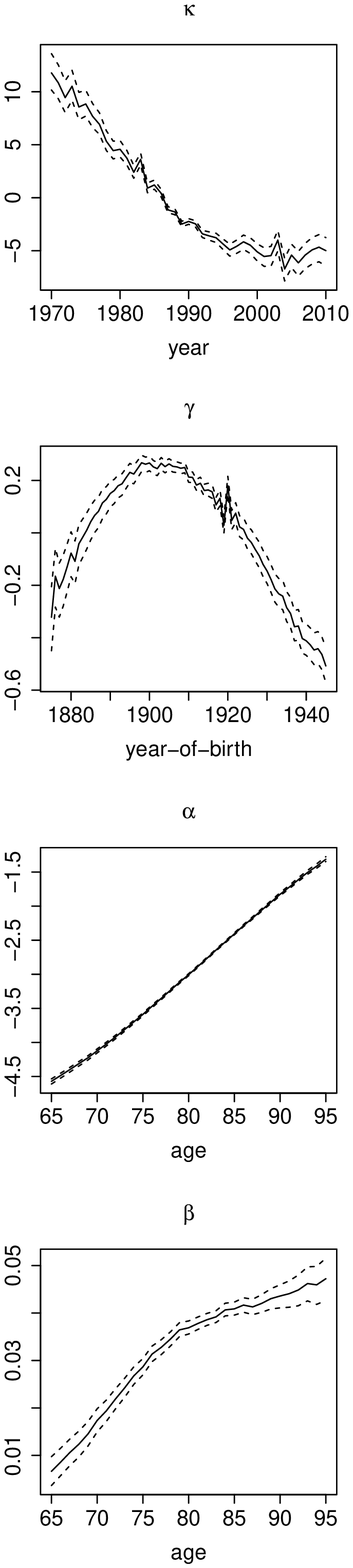}
\caption{\small{Simplified cohort model: estimated $\bm{\kappa}$, $\bm{\gamma}$, $\bm{\alpha}$ and $\bm{\beta}$ for the (left column) UK, (middle column) US and (right column) Italy females populations. Mean (solid line) and 95\% credible interval (dash lines) of the posterior distributions are plotted.}}
\label{fig:SimpCohortEstFemales}
\end{center}
\end{figure}

The corresponding estimation results for female populations are shown in Figure~\ref{fig:SimpCohortEstFemales}. We again observe that the estimates show similarity between the male and female populations for all the three countries, thus suggesting that mortality experiences for both genders share major characteristics.

Posterior estimates for other static parameters for the simplified cohort model on male and female mortality data are reported in Table~\ref{table:ParasSimpCohort}.

\begin{table}[h]
\center \setlength{\tabcolsep}{1em}
\renewcommand{\arraystretch}{1.1}
\scalebox{0.9}{\begin{tabular}{l|ccc}
\hline \hline
  & UK  & US & Italy    \\
\hline
 & & Males & \\
\hline
$\theta$                & -0.22 [-0.44, 0.007]     & -0.50 [-0.67, -0.33]    & -0.30 [-0.56, -0.03]  \\
$\eta$                  & -0.022 [-0.034, -0.011]    & -0.003 [-0.013, -0.007]    & -0.019 [-0.034, -0.004]  \\
$\lambda$               & 0.991 [0.970, 0.999]       & 0.973 [0.909, 0.999]       & 0.982 [0.944, 0.999]  \\
$\sigma^2_\varepsilon$  & 0.00035 [0.00032, 0.00038] & 0.00025 [0.00023, 0.00027] & 0.00033 [0.00031, 0.00036]  \\
$\sigma^2_\omega$       & 0.46 [0.29, 0.73]       & 0.23 [0.14, 0.37]       & 0.68 [0.43, 1.06]  \\
$\sigma^2_\gamma$       & 0.0012 [0.0008, 0.0019]       & 0.0007 [0.0004, 0.0010]      & 0.0023 [0.0015, 0.0036]  \\
\hline
 & & Females & \\
\hline
$\theta$                & -0.40 [-0.64, -0.14]     & -0.61 [-0.81, -0.41]    & -0.39 [-0.70, -0.08]  \\
$\eta$                  & -0.013 [-0.024, -0.002]    & 0.018 [0.0005, 0.043]    & -0.019 [-0.033, -0.005]  \\
$\lambda$               & 0.98 [0.95, 0.99]       & 0.89 [0.74, 0.99]       & 0.97 [0.93, 0.99]  \\
$\sigma^2_\varepsilon$  & 0.00025 [0.00023, 0.00028] & 0.00025 [0.00023, 0.00027] & 0.00032 [0.00029, 0.00035]  \\
$\sigma^2_\omega$       & 0.60 [0.38, 0.94]       & 0.38 [0.24, 0.60]       & 0.95 [0.60, 1.50]  \\
$\sigma^2_\gamma$       & 0.0012 [0.0008, 0.0019]       & 0.0006 [0.0004, 0.0010]      & 0.002 [0.001, 0.003]  \\
\hline \hline
\end{tabular}}
\center\caption{\label{table:ParasSimpCohort} \small{Estimated posterior mean of the static parameters for the simplified cohort model on male and female population data. $[.,.]$ represents $95\%$ posterior credible interval.}}
\end{table}




\subsection{Model fitting: comparison with LC model}\label{sec:modelComparison}
We examine the fitting of the cohort models via residual heatmap as well as model ranking via deviance information criterion (DIC). We emphasize the importance of a comparison for the cohort models with the LC model since if cohort patterns are present in mortality data, the ordinary LC model should not be able to capture this phenomenon but cohort models are designed to accomplish this.

\subsubsection{Residual heatmap}

Residuals under a state-space model are defined as the difference between the observed data $\bm{y}_t$ and the mean of the in-sample one-step-ahead model forecast given by (see \eqref{eqn:KFpredictY})
\begin{equation}
    \mathbb{E}\left[\boldsymbol{y}_t|\bm{\psi},\boldsymbol{y}_{1:t-1}\right]=\bm{f}_t,
\end{equation}
where $t=1,\dots,n$. Explicitly, we have
\begin{equation}
 \bm{e}_t := \bm{y}_t -\bm{f}_t
\end{equation}
where $\bm{e}_t$ is the vector of residuals at time $t$. Using the posterior mean as point estimator for the static parameters, we obtain the mean of the in-sample one-step-ahead forecast $\bm{f}_t$ via Kalman filtering for the cohort models and the LC model.

\begin{figure}[h]
\begin{center}
\includegraphics[width=5.5cm, height=4.5cm]{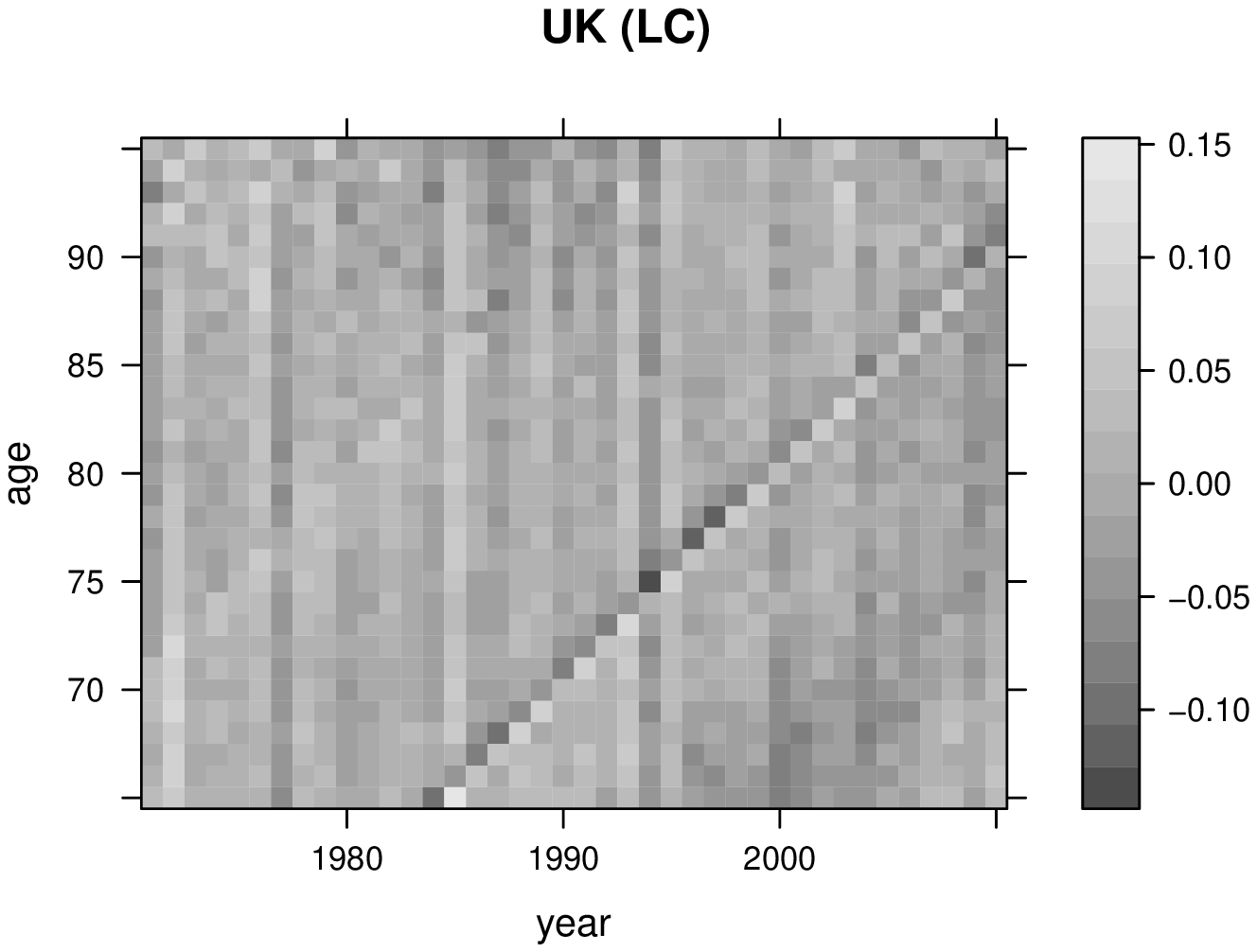}\includegraphics[width=5.5cm, height=4.5cm]{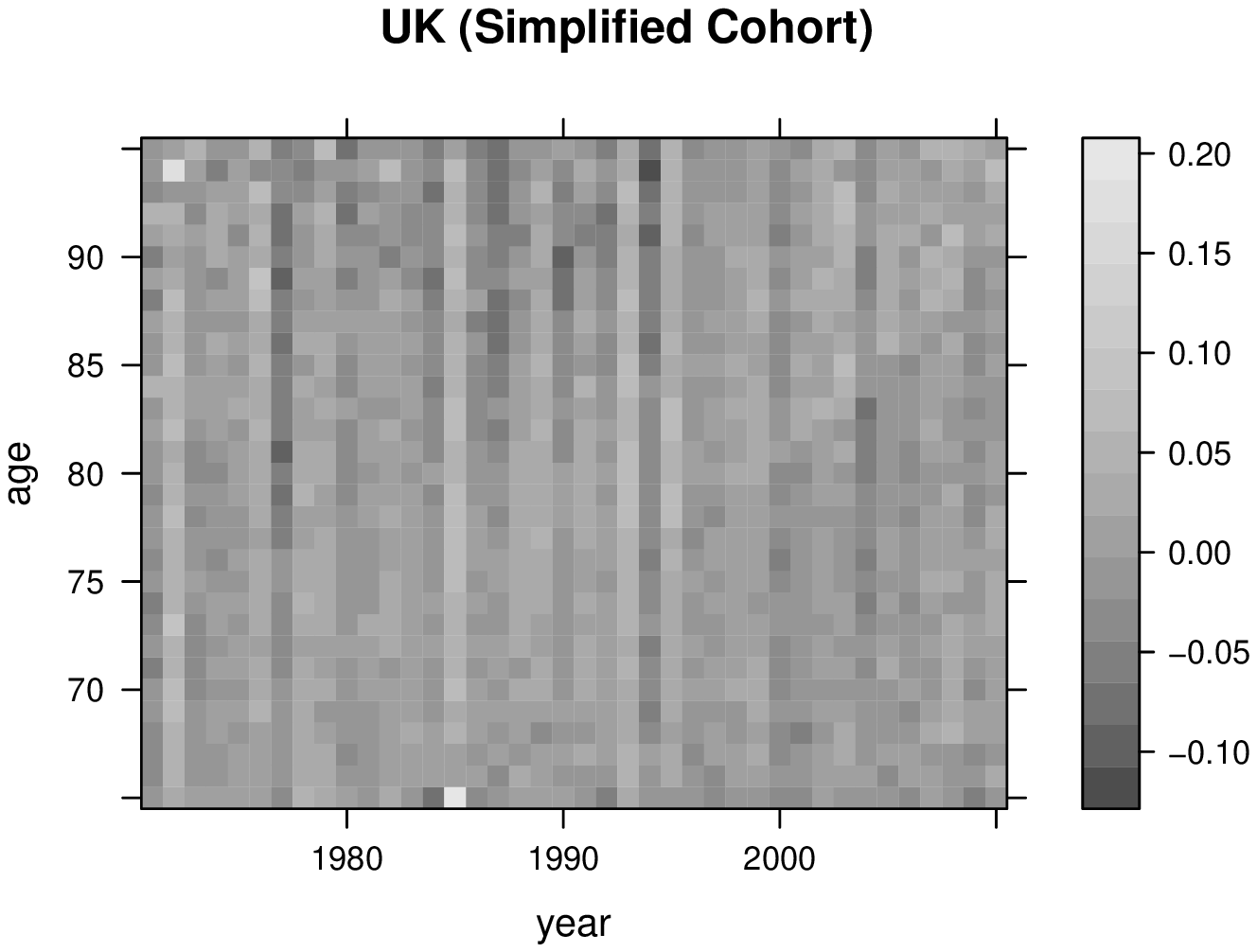}\includegraphics[width=5.5cm, height=4.5cm]{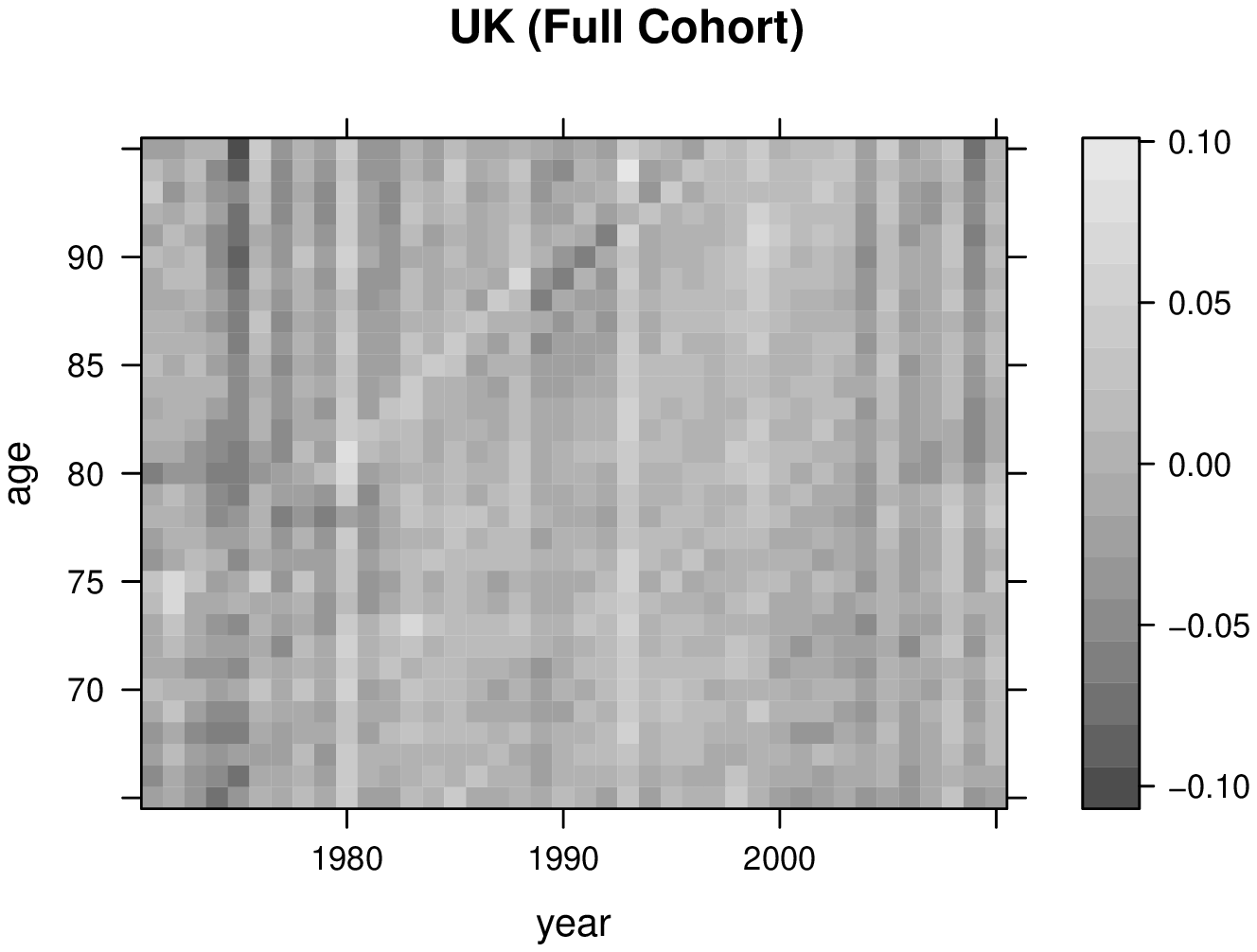}

\includegraphics[width=5.5cm, height=4.5cm]{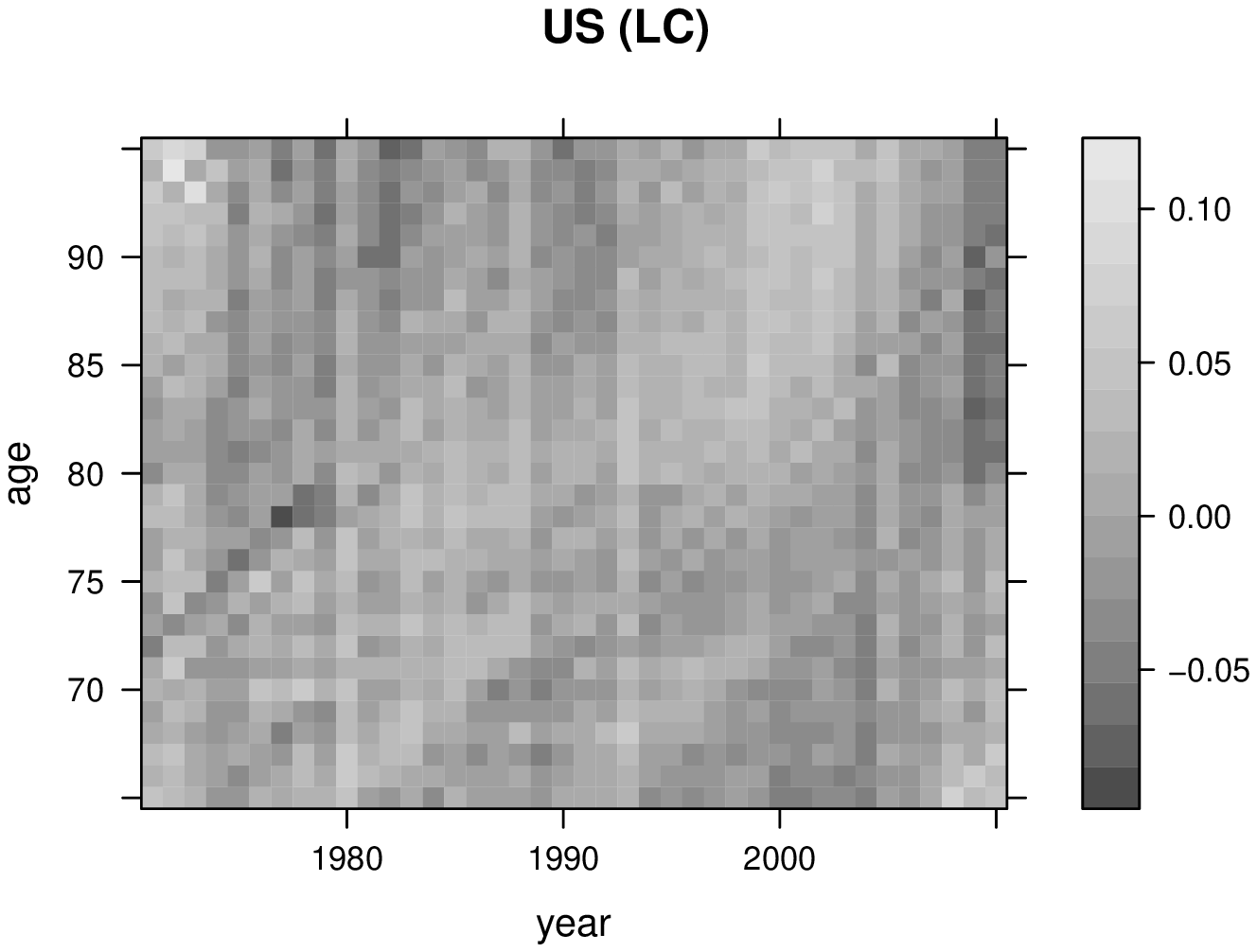}\includegraphics[width=5.5cm, height=4.5cm]{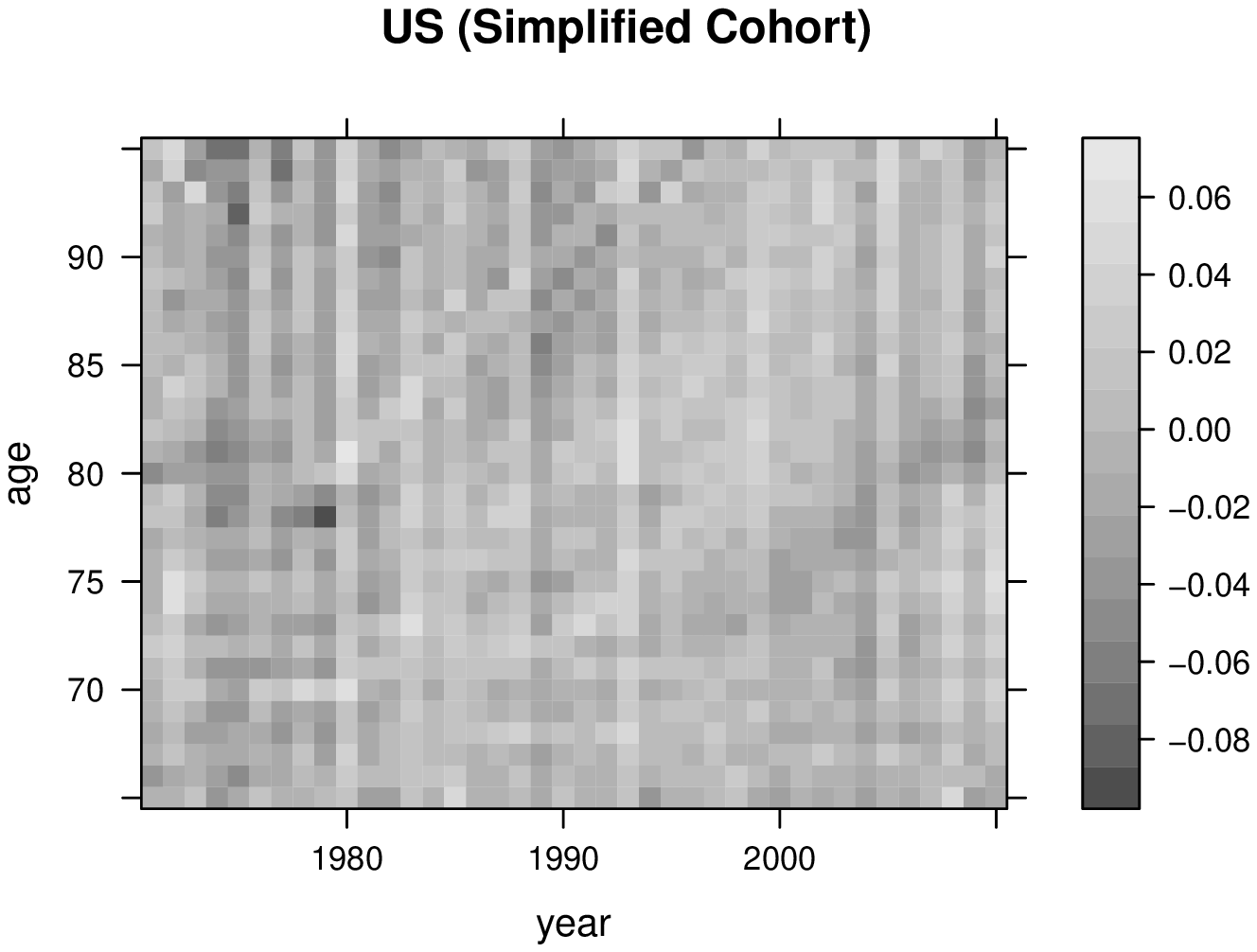}\includegraphics[width=5.5cm, height=4.5cm]{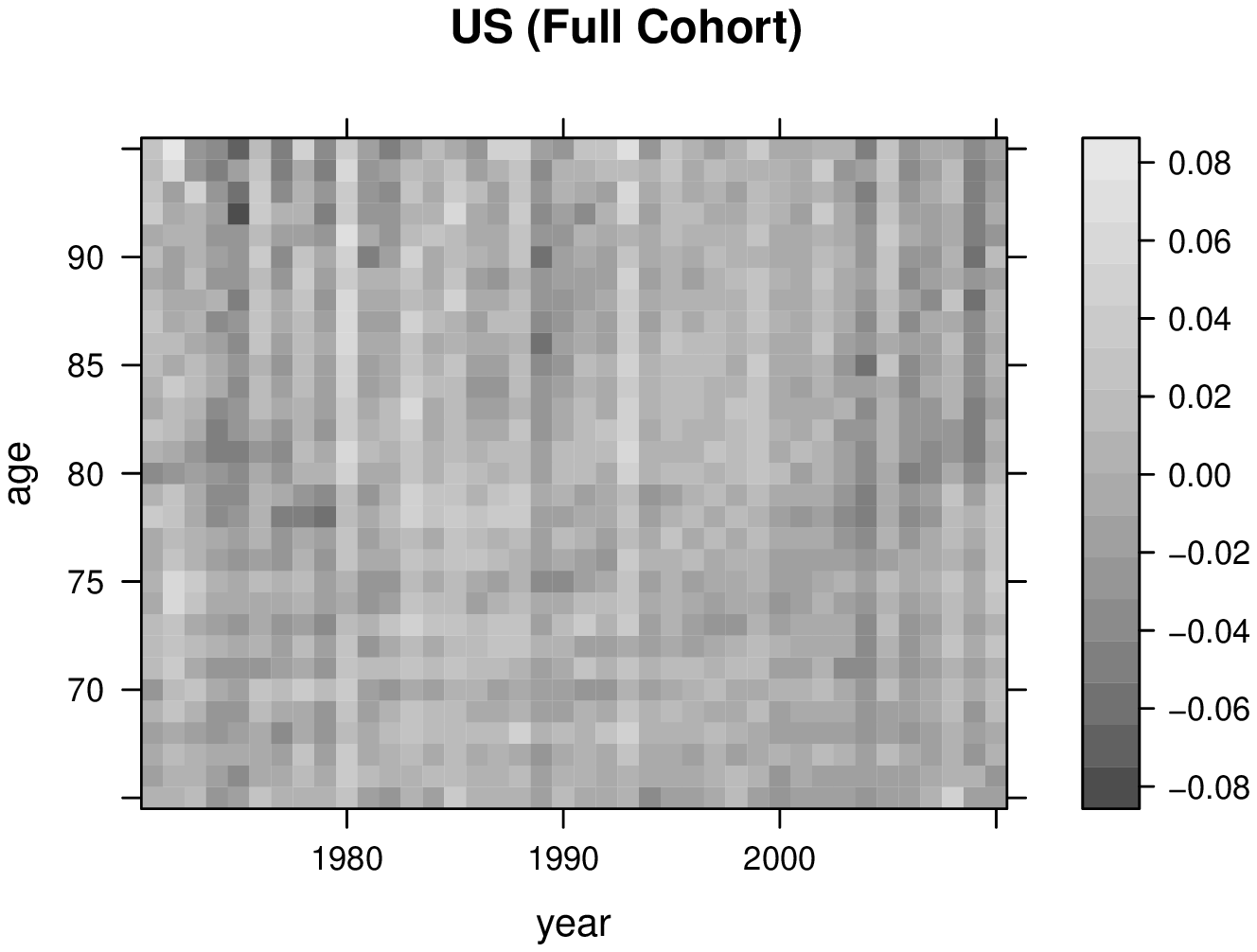}

\includegraphics[width=5.5cm, height=4.5cm]{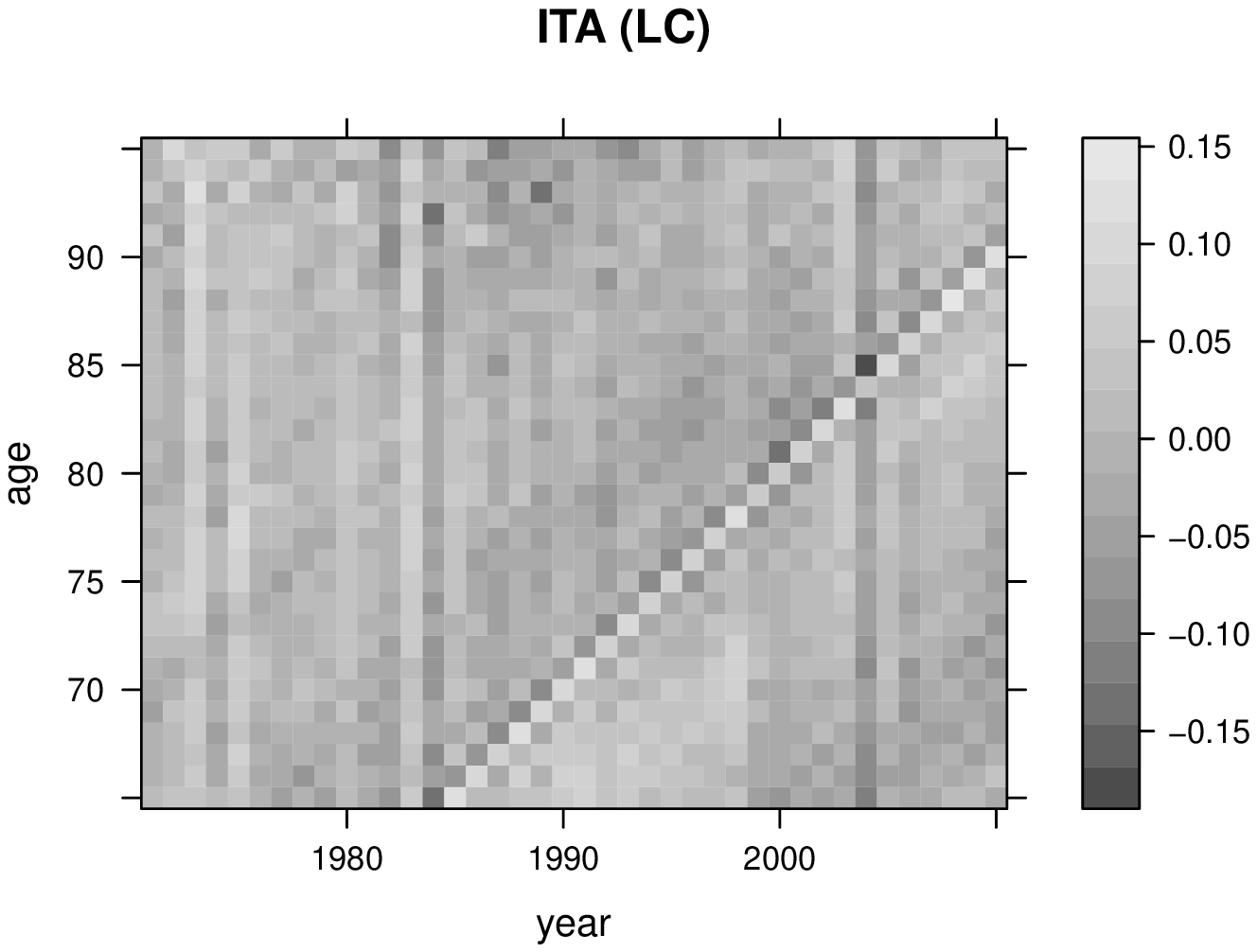}\includegraphics[width=5.5cm, height=4.5cm]{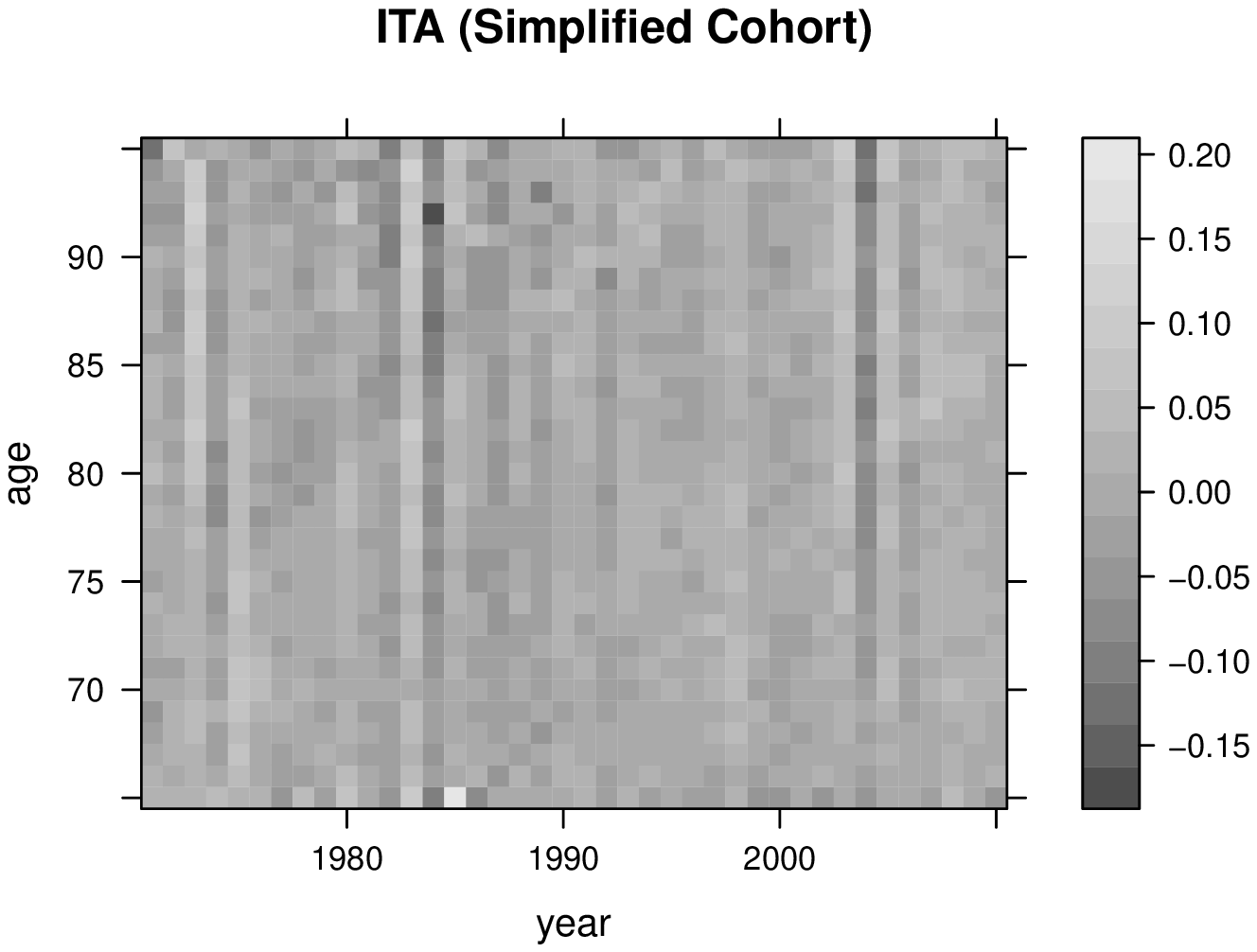}\includegraphics[width=5.5cm, height=4.5cm]{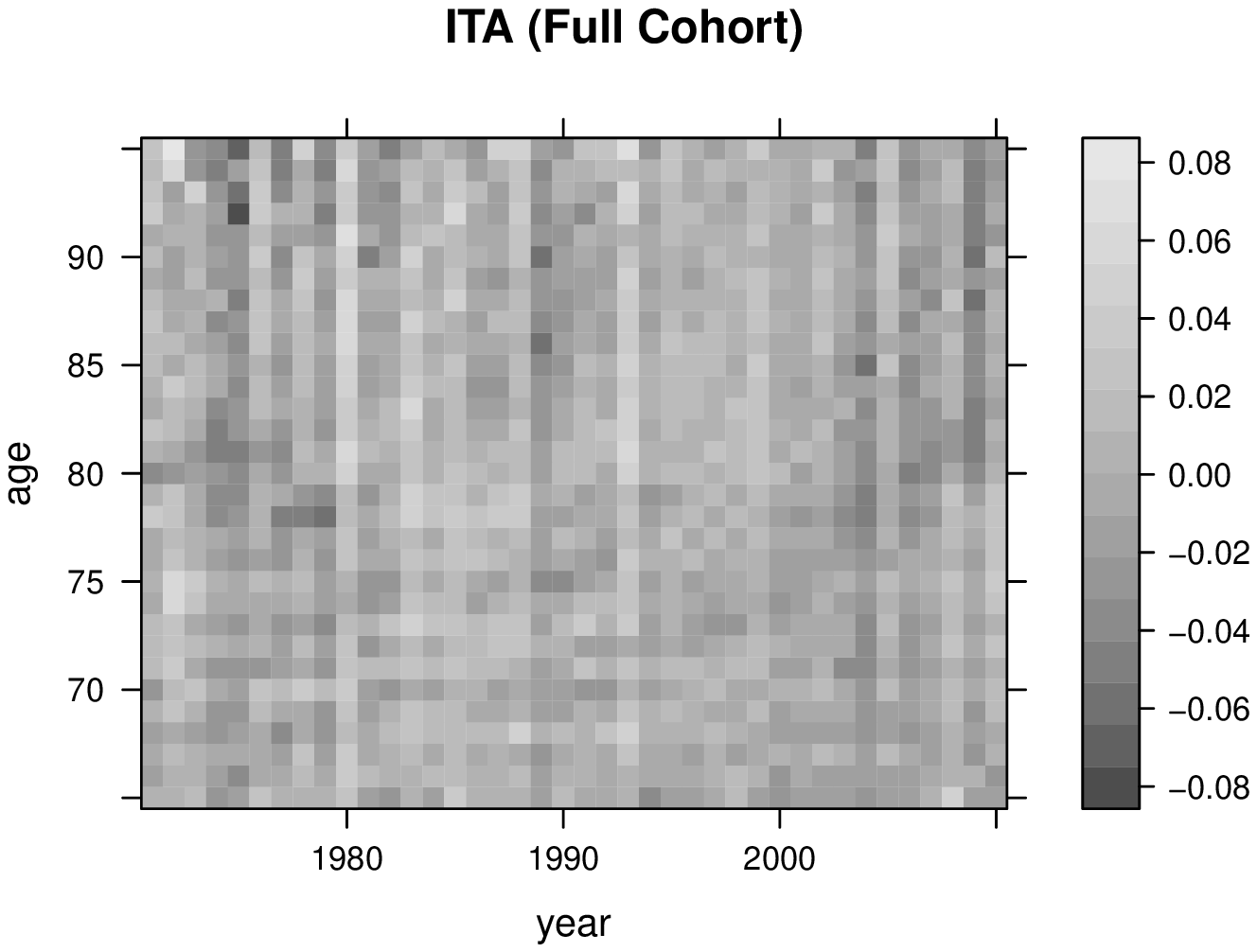}
\caption{\small{Residual heatmap produced from the LC model, the simplified cohort model and the full cohort model for the UK, US and Italy male populations.}}
\label{fig:heatmapMales}
\end{center}
\end{figure}

Figure~\ref{fig:heatmapMales} shows residual heatmaps produced from the LC model, the simplified cohort model and the full cohort model for the UK, US and Italy male populations. The distinctive diagonal bands observed for the fitting of the LC model for the UK and Italy population, which correspond to the generations born around 1920, clearly suggest that cohort effects are strongly present in these countries and the LC model fails to account for these patterns. In contrast, the cohort models considered are capable of capturing these effects which is supported by the observation that the diagonal bands are removed from the residual plots. These results are consistent with the estimated cohort factors shown in Figure~\ref{fig:FullCohortEstMales} and \ref{fig:FullCohortEstFemales} where an apparent irregularity appears around the cohort born in 1920.

Residual heatmap from the LC model for the US male population, on the other hand, shows only small traces of diagonal bands around the cohorts 1900 and 1920 which are barely noticeable. The corresponding plots from the cohort models show that these bands are diminished even further. The very faint occurrence of the diagonal bands for the US males data is compatible with the remarks suggested in Section~\ref{sec:ModelEstimation} that there is no clear evidence that the US male population exhibits certain cohort patterns.

\begin{figure}[h]
\begin{center}
\includegraphics[width=5.5cm, height=4.5cm]{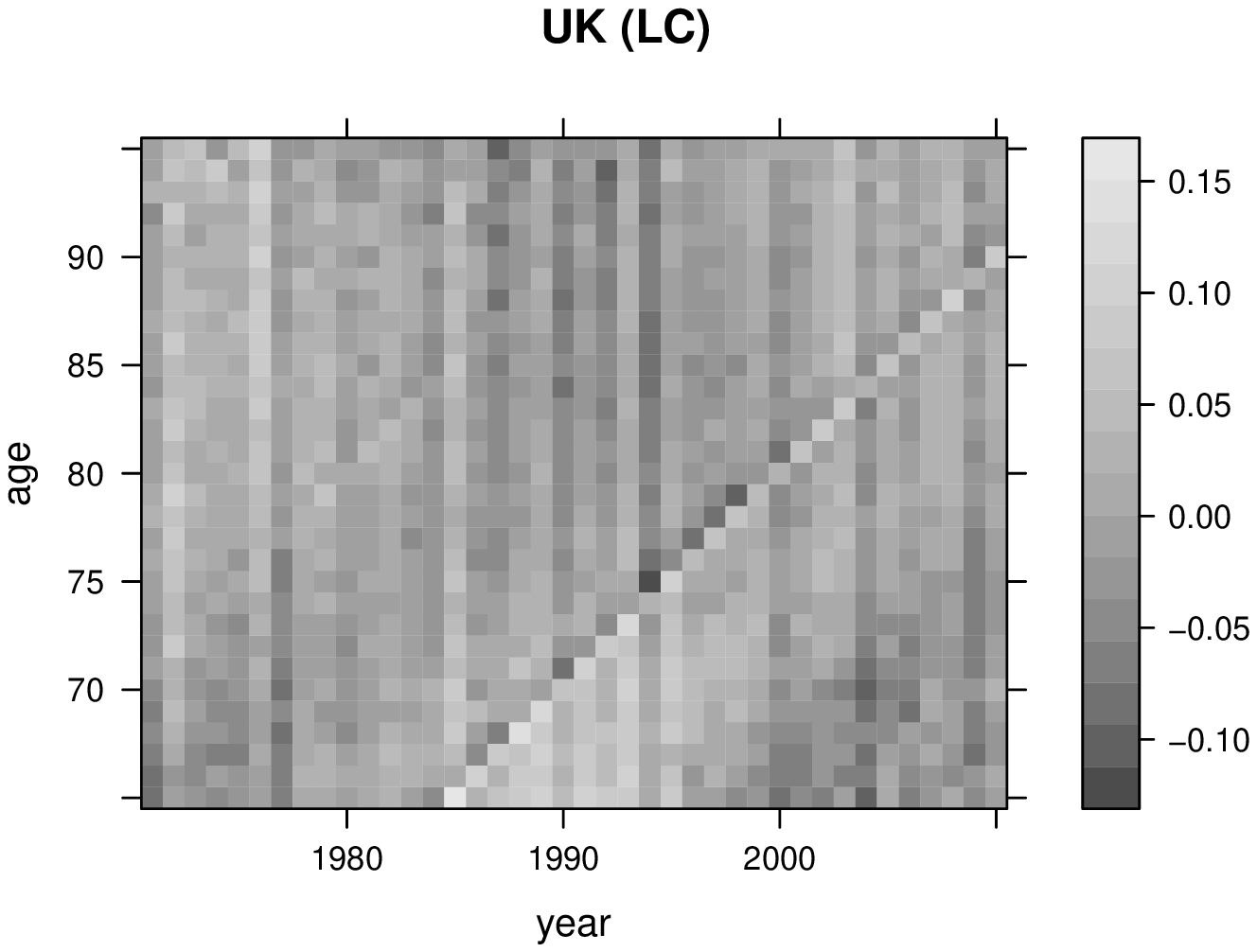}\includegraphics[width=5.5cm, height=4.5cm]{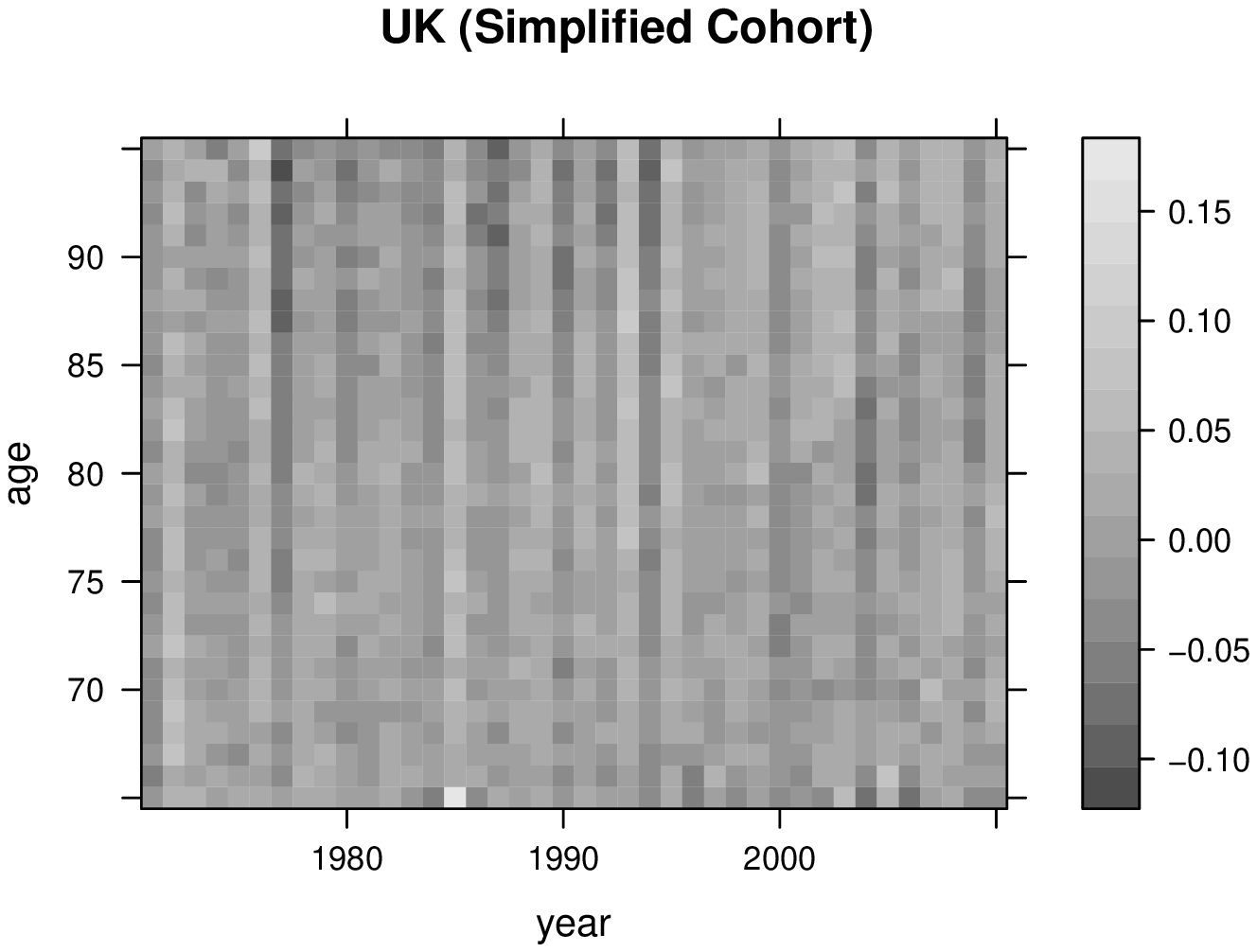}\includegraphics[width=5.5cm, height=4.5cm]{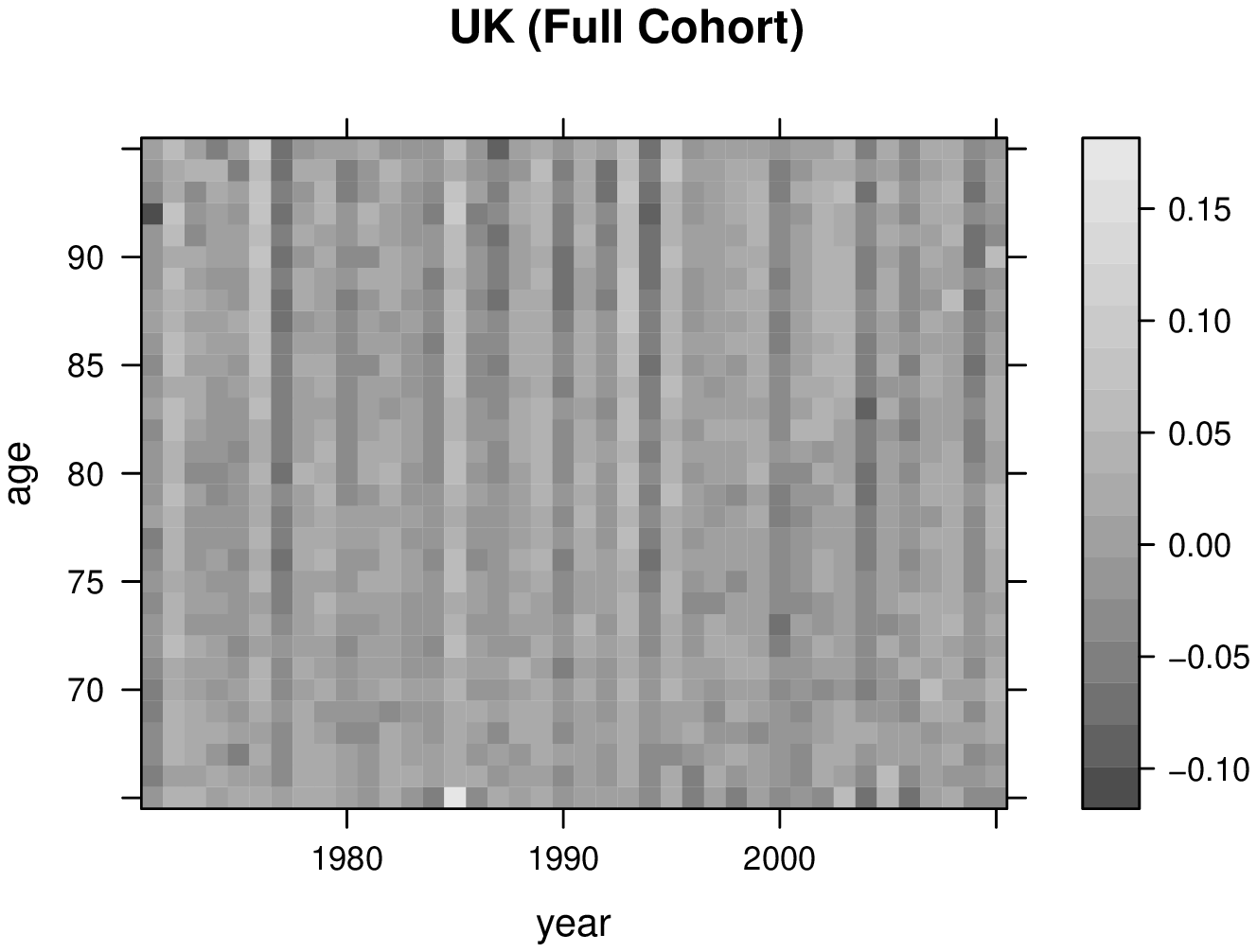}

\includegraphics[width=5.5cm, height=4.5cm]{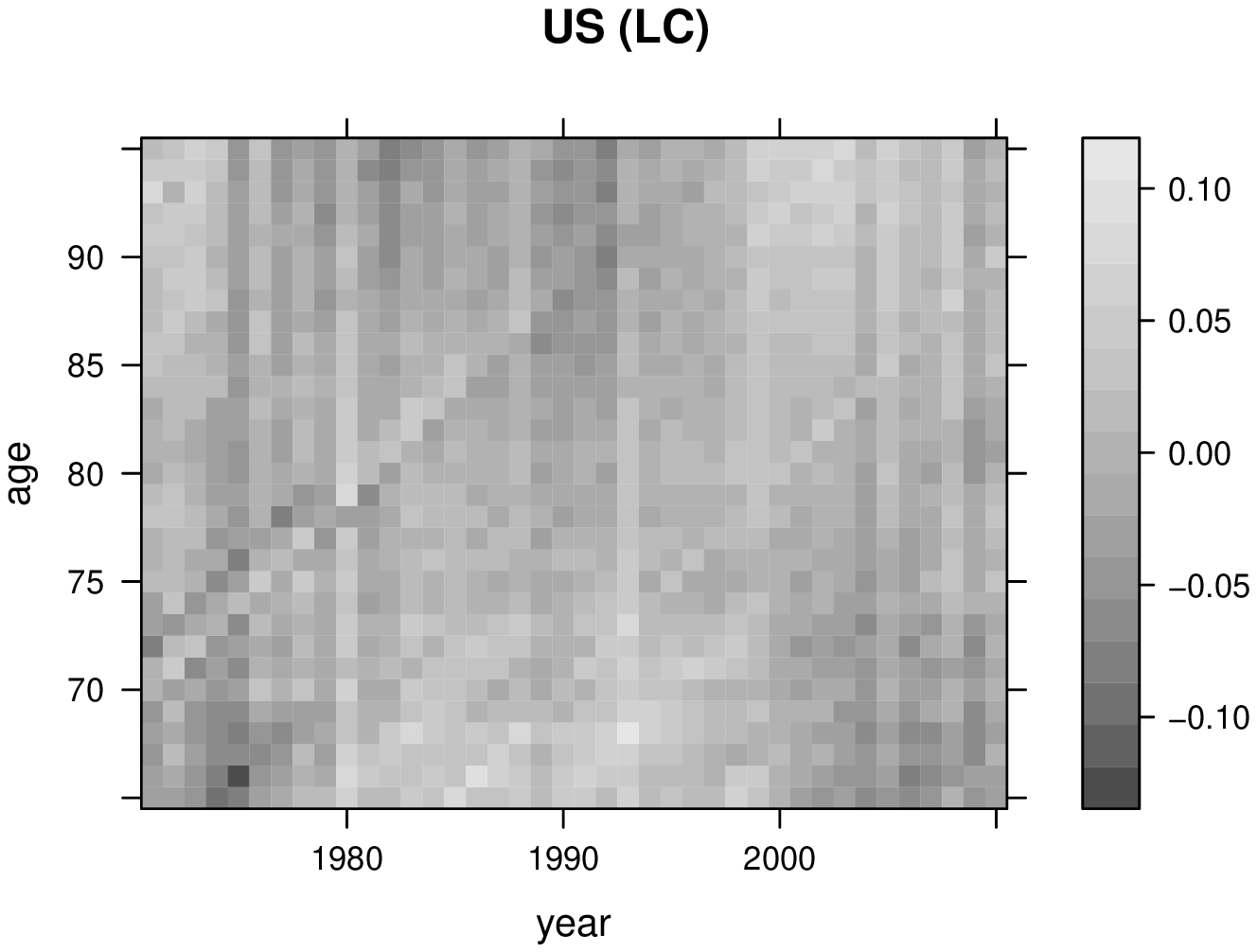}\includegraphics[width=5.5cm, height=4.5cm]{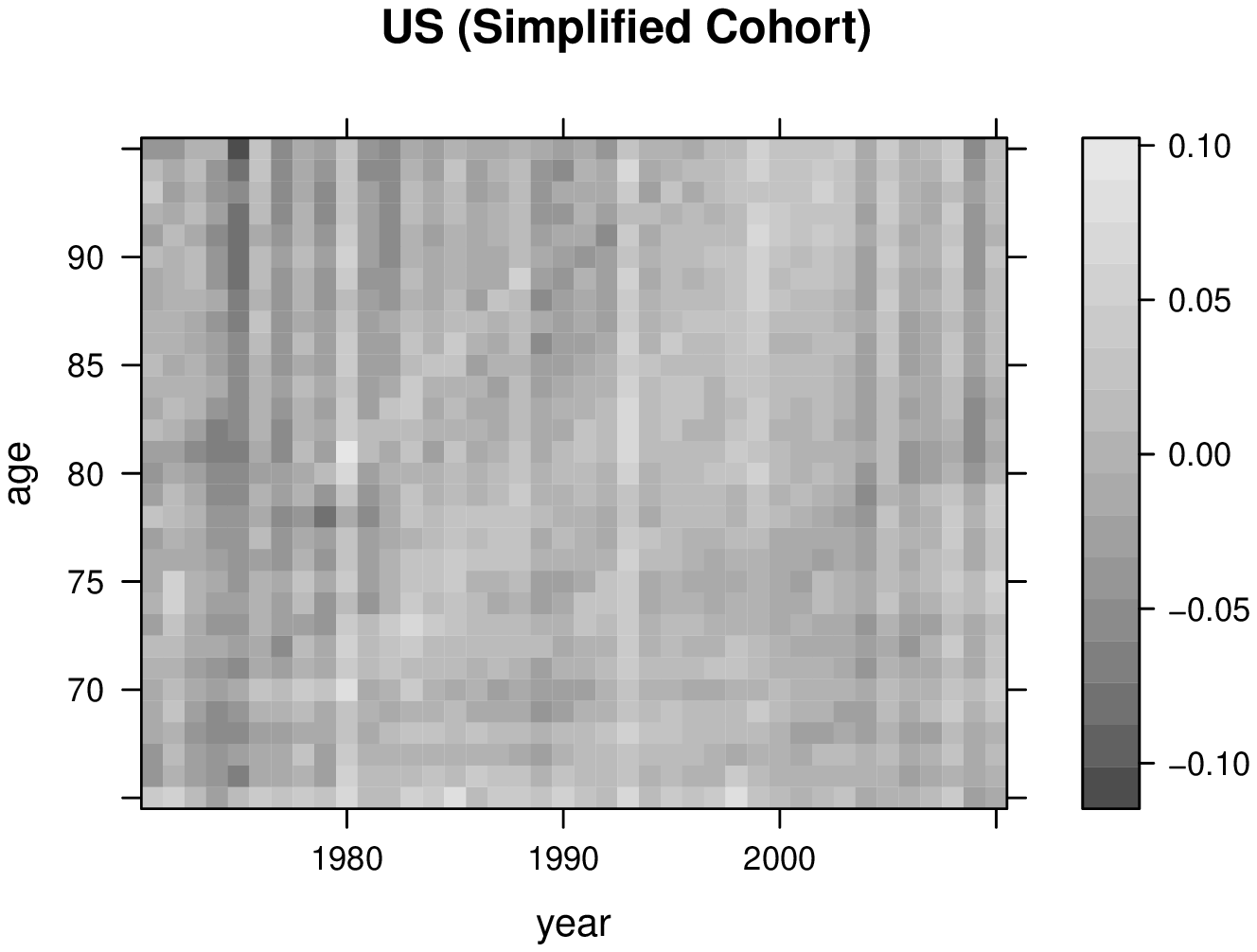}\includegraphics[width=5.5cm, height=4.5cm]{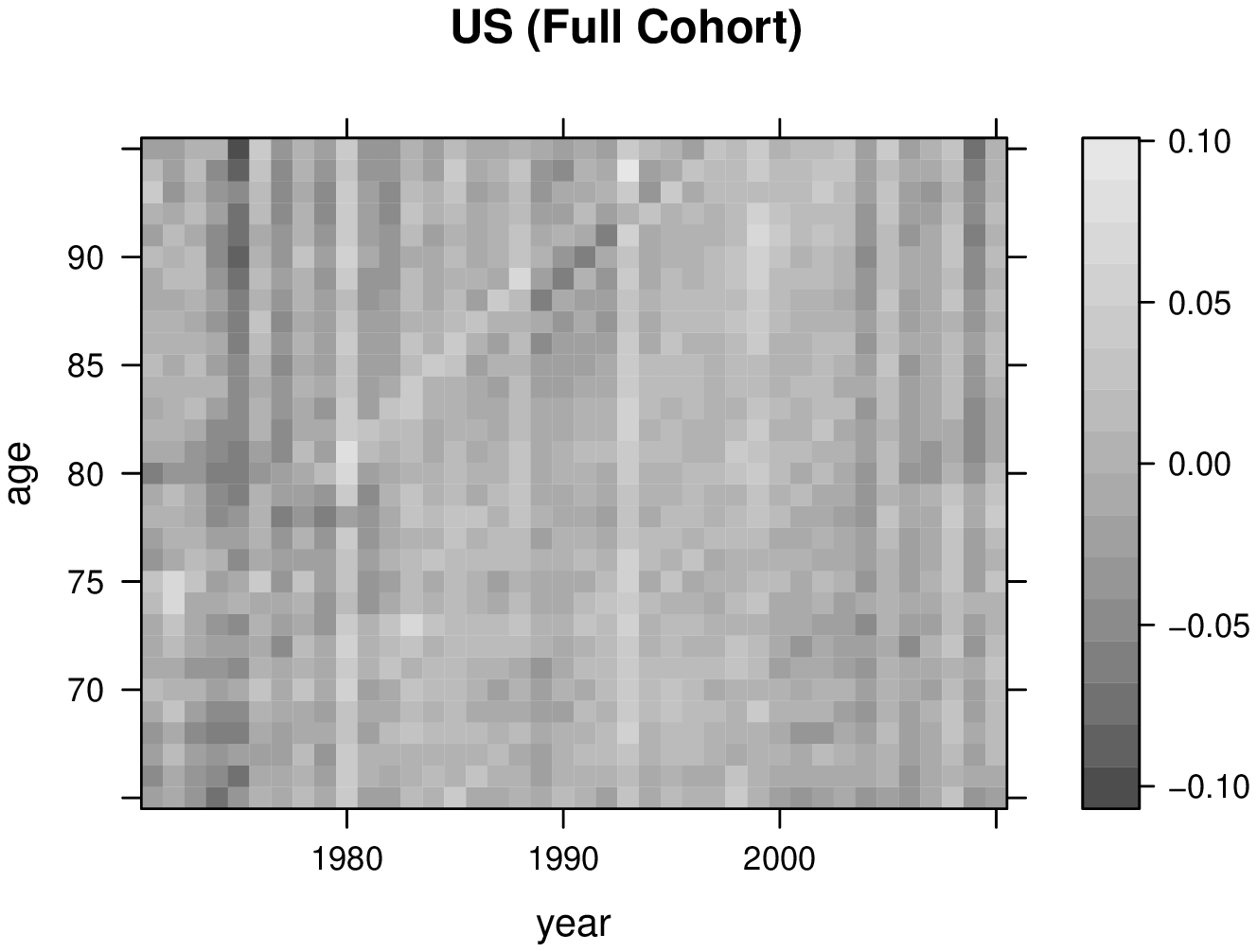}

\includegraphics[width=5.5cm, height=4.5cm]{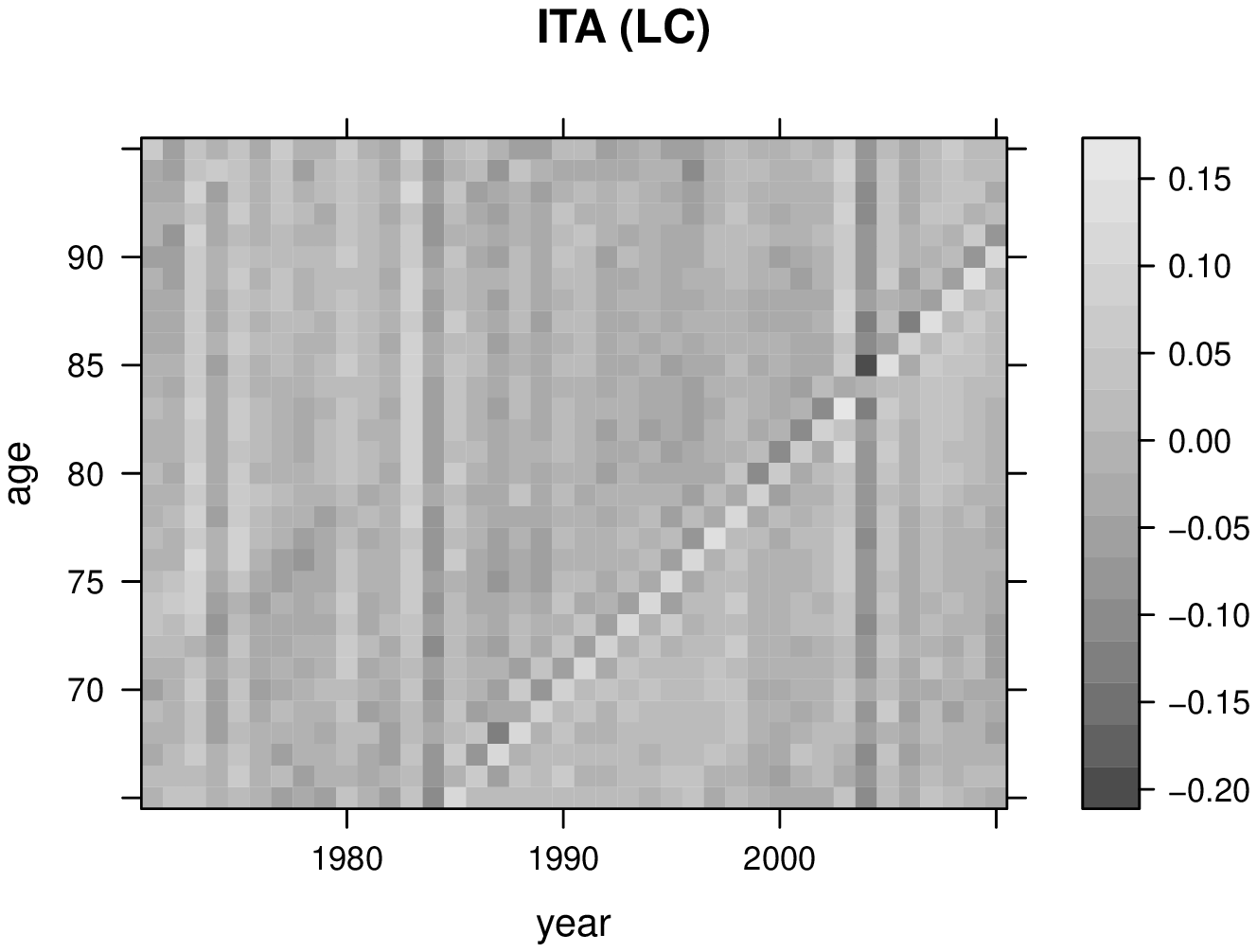}\includegraphics[width=5.5cm, height=4.5cm]{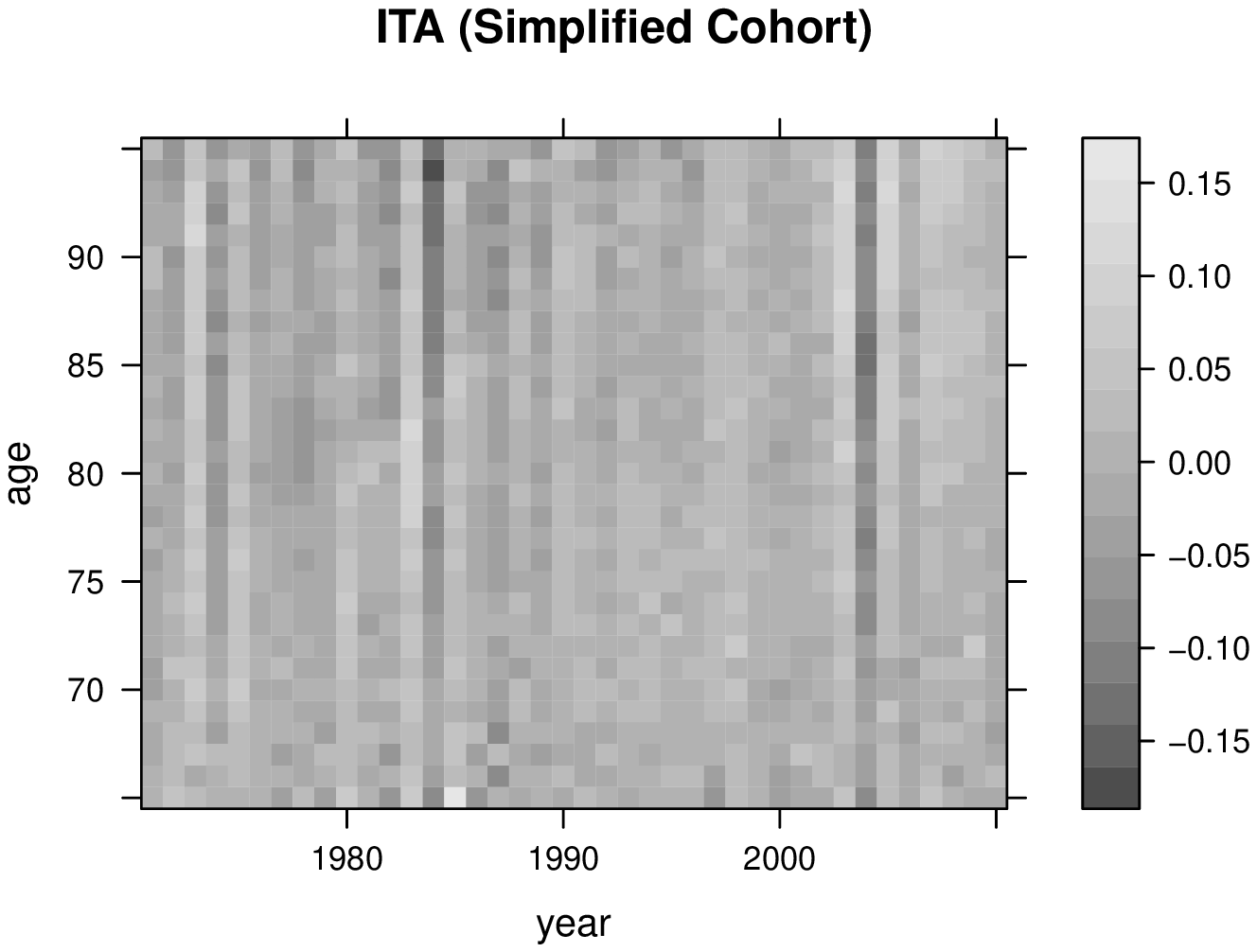}\includegraphics[width=5.5cm, height=4.5cm]{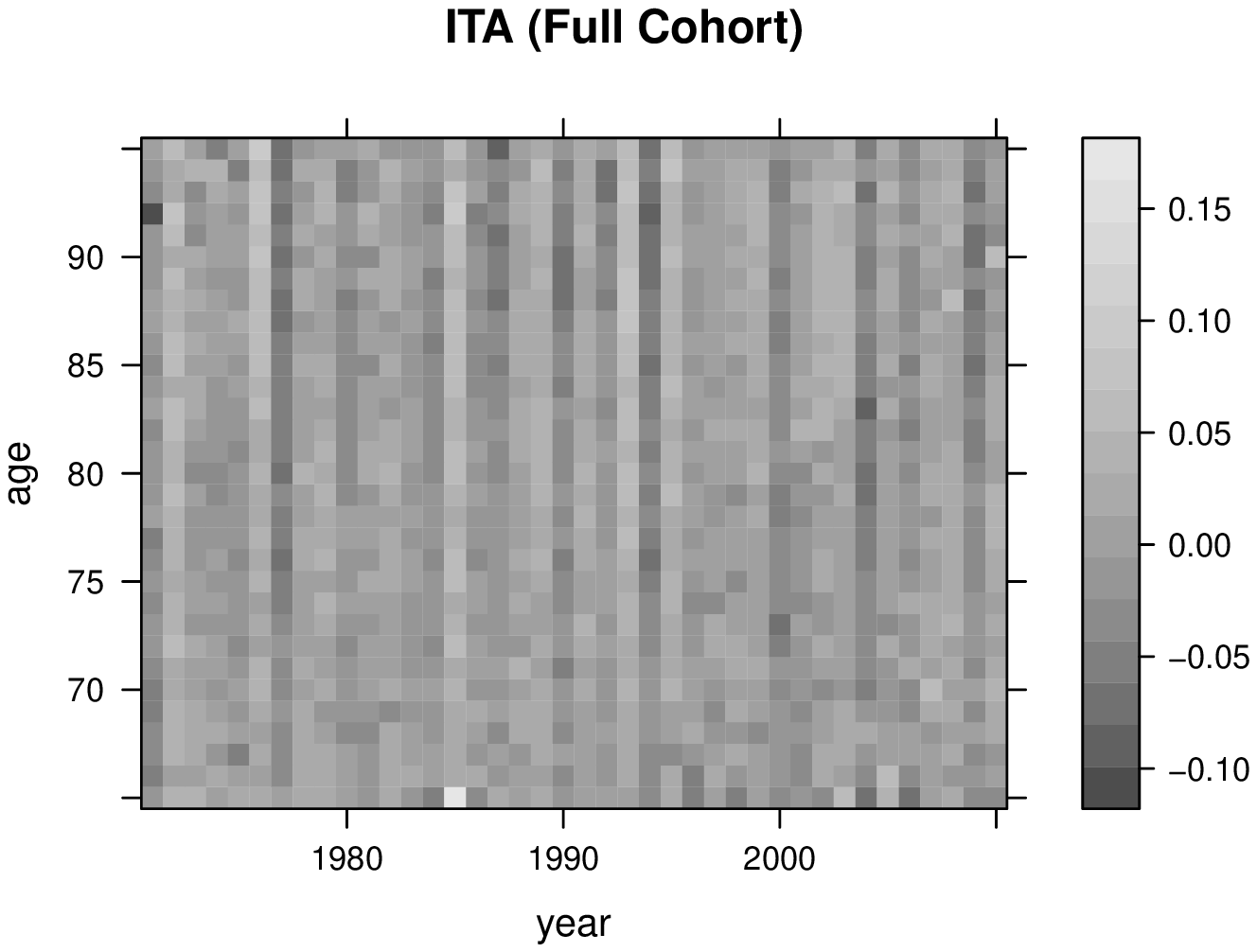}
\caption{\small{Residual heatmap produced from the LC model, the simplified cohort model and the full cohort model for the UK, US and Italy female populations.}}
\label{fig:heatmapFemales}
\end{center}
\end{figure}

Corresponding residual heatmaps for the female populations are displayed in Figure~\ref{fig:heatmapFemales}. The results follow very closely to the discussion for the male populations. It provides further evidence that major mortality characteristics including cohort effects are not gender-specific in the considered countries.

The residual plots shown here also suggest that the deciding factor for the presence of cohort patterns in a specific country depend on whether there are any abrupt irregularities observed in the estimated cohort factor; such non-smooth irregularities indicate that strong cohort patterns exist in the data where the LC model is not able to capture. On the other hand, if the estimated cohort factor is reasonably smooth as for the US data, cohort factors in mortality models may be not be required.

\subsubsection{Deviance information criterion}
We perform model ranking using deviance information criterion (DIC) which is designed specifically for Bayesian models taking into account the trade-off between model fits and complexity (\cite{Spiegelhalteretal02}). There are several versions of DIC and here we will focus on the so-called conditional DIC where the latent states are treated as parameters when calculating the conditional likelihood (\cite{CeleuxFoRoTi06}, \cite{ChanGr16}).

Specifically, the conditional DIC utilises the conditional log-likelihood which is given by
\begin{equation}\label{eqn:DIClikelihood}
    \ln \pi(\bm{y}_{1:n}|\bm{\psi},\bm{\varphi}_{0:n}) = -\frac{1}{2}\sum_{x=x_1}^{x_p} \sum_{t=t_1}^{t_n} \left(\ln 2\pi \sigma^2_{\varepsilon}+ \left(\frac{y_{x,t}-(\alpha_x+\beta_x \kappa_t+\beta^\gamma_x \gamma_{t-x})}{\sigma_{\varepsilon}}\right)^2 \right),
\end{equation}
for the full cohort model; conditional log-likelihoods for the simplified cohort model and the LC model can be obtained similarly. Denote $\bm{\Psi}:=(\bm{\psi},\bm{\varphi}_{0:n})$, one defines
\begin{equation}
    D(\boldsymbol{\Psi}) = -2\ln\pi(\boldsymbol{y}_{1:n}|\boldsymbol{\Psi}) + 2\ln h(\boldsymbol{y}_{1:n}),
\end{equation}
as the deviance of the model. As $h(\boldsymbol{y}_{1:n})$ is independent to the models considered, it is typical to assume $h(\boldsymbol{y}_{1:n})=1$. The effective number of parameters $p_D$ is defined to be
\begin{equation}
    p_{D} = \bar{D}(\boldsymbol{\Psi})- D(\bar{\boldsymbol{\Psi}}),
\end{equation}
where $\bar{D}(\boldsymbol{\Psi})$ is the mean of $D(\boldsymbol{\Psi})$ while $\bar{\boldsymbol{\Psi}}$ is the posterior mean of $\bm{\Psi}$. One then defines the conditional DIC as
\begin{equation}
    \text{DIC} := \bar{D}(\boldsymbol{\Psi}) + p_{D} = 2\bar{D}(\boldsymbol{\Psi}) - D(\bar{\boldsymbol{\Psi}}),
\end{equation}
which can be calculated using the MCMC samples obtained as described in Section~\ref{sec:bayesian}. Note that models with smaller DIC values are ranked higher then models with larger DIC values.

\begin{table}[h]
\center \setlength{\tabcolsep}{1em}
\renewcommand{\arraystretch}{1.1}
\begin{tabular}{c|ccc}
\hline \hline
& LC model & Simplified cohort model & Full cohort model \\
\hline
 & & Males & \\
 \hline
UK   & -5418  & -6376  & \textbf{-6666} \\
US   & -5575  & -6836  & \textbf{-7111} \\
ITA  & -4758  & -6433  & \textbf{-6474}   \\
\hline
& & Females & \\
\hline
UK   & -5053  & -6794  & \textbf{-6910} \\
US   & -5395  & -6824  & \textbf{-6993} \\
ITA  & -5098  & -6485  & \textbf{-6607}  \\
\hline \hline
\end{tabular}}
\center\footnotesize{\caption{\label{table:DIC} DIC values for the considered models on male and female population data.}
\end{table}


Table~\ref{table:DIC} reports the estimated conditional DIC values obtained for the considered models on male and female population data. We observe that the improved fits are more than compensated for the complexity arising from the inclusion of cohort factors for the considered countries based on the estimated DIC values. The improvement of model fits is more significant for the simplified cohort model over the LC model than the full cohort model over the simplified cohort model. Interestingly, there is a distinct improvement in using cohort models over the LC model on the US mortality data, despite the lack of cohort patterns for the US data shown in Figure~\ref{fig:heatmapMales} and \ref{fig:heatmapFemales}. It suggests that multi-factor models are preferred for the US population over a single period factor model such as the LC model. Whether a multi-period model or a period-cohort model is preferred for the US data is an interesting question which is, however, beyond the scope of the current paper.

\subsection{Forecasting from cohort models}
Forecasting of death rates for the cohort models and LC model is studied here. One of the advantages of considering Bayesian inference via MCMC sampling for state-space mortality models is that the forecasting distributions of death rates can be derived rigourously and the samples of the forecasting distributions can be easily generated from the samples obtained from the MCMC estimation step.

Explicitly, we can express the forecasting distribution for the cohort models as
\begin{equation}\label{eqn:forecastCohort}
    \pi(\bm{y}_{n+k}|\bm{y}_{1:n})=\int \pi(\bm{y}_{n+k}|\bm{\varphi}_{n+k},\bm{\psi})\pi(\bm{\varphi}_{n+k}|\bm{\varphi}_{n+k-1},\bm{\psi})
    \dots \pi(\bm{\varphi}_n,\bm{\psi}|\bm{y}_{1:n})\,
    d\boldsymbol{\psi}d\bm{\varphi}_{n:n+k},
\end{equation}
where $\pi(\bm{y}_{n+k}|\bm{y}_{1:n})$ is the $k$-step ahead forecasting posterior predictive distribution. It means that one can simulate the forecasting samples of the dynamic factors to obtain recursively the forecasting samples of death rates as follow
\begin{subequations}\label{eqn:forecastCohortsample}
\begin{align}
    \bm{\varphi}_{n+k}^{(\ell)} &\sim \text{N}\left(\Lambda^{(\ell)}\bm{\varphi}^{(\ell)}_{n+k-1}+\Theta^{(\ell)},\Upsilon^{(\ell)}\right), \\
    \bm{y}^{(\ell)}_{n+k} &\sim \text{N}\left(\bm{\alpha}^{(\ell)}+B^{(\ell)}\bm{\varphi}^{(\ell)}_{n+k},\left(\sigma^2_\varepsilon\right)^{(\ell)}\mathsf{1}_p\right),
\end{align}
\end{subequations}
for the full cohort model, where $\ell=1,\dots,L$ and $L$ is the number of MCMC samples after burn-in. For the simplified cohort model one simply replaces $B$ by $B^s$. Forecasting for the LC model can be carried out by setting $\bm{\varphi}=\kappa$ in \eqref{eqn:forecastCohort} and forecasting samples of death rates are obtained recursively as follow
\begin{subequations}\label{eqn:forecastLCsample}
\begin{align}
    \kappa_{n+k}^{(\ell)} &\sim \text{N}\left(\kappa^{(\ell)}_{n+k-1}+\theta^{(\ell)},\left(\sigma^2_\omega\right)^{(\ell)}\right), \\
    \bm{y}^{(\ell)}_{n+k} &\sim \text{N}\left(\bm{\alpha}^{(\ell)}+\bm{\beta}^{(\ell)}\kappa^{(\ell)}_{n+k},\left(\sigma^2_\varepsilon\right)^{(\ell)}\mathsf{1}_p\right),
\end{align}
\end{subequations}
which is a special case of \eqref{eqn:forecastCohortsample}.

\subsubsection{Projection of death rates}\label{sec:projectDeathrates}

\begin{figure}[h]
\begin{center}
\includegraphics[width=5.5cm, height=20cm]{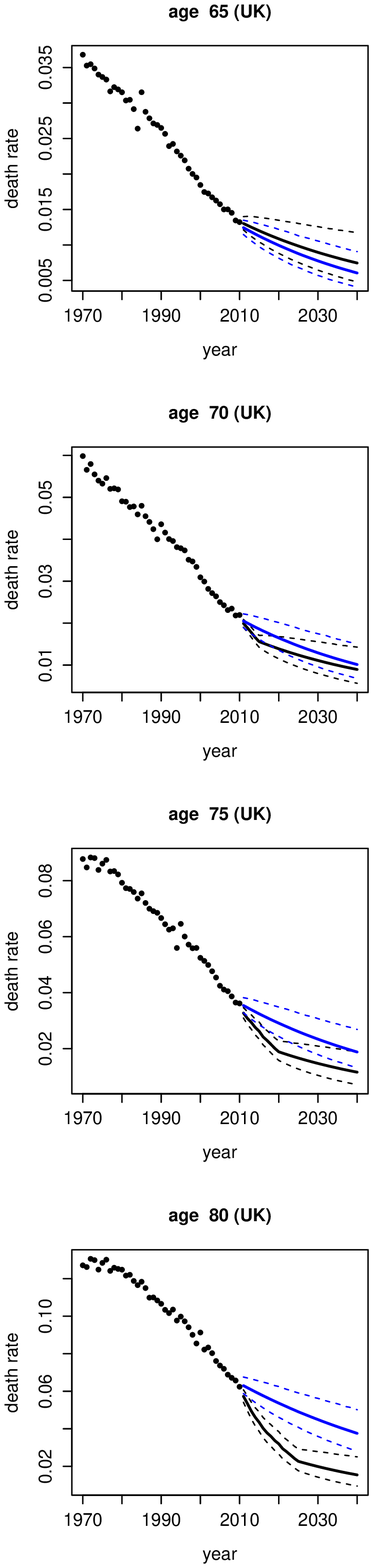}\includegraphics[width=5.5cm, height=20cm]{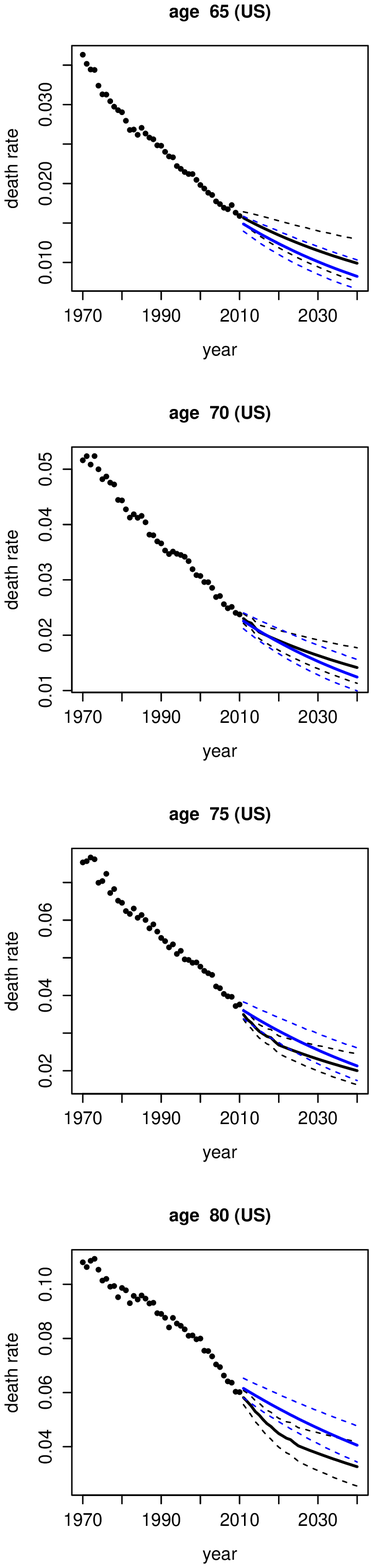}\includegraphics[width=5.5cm, height=20cm]{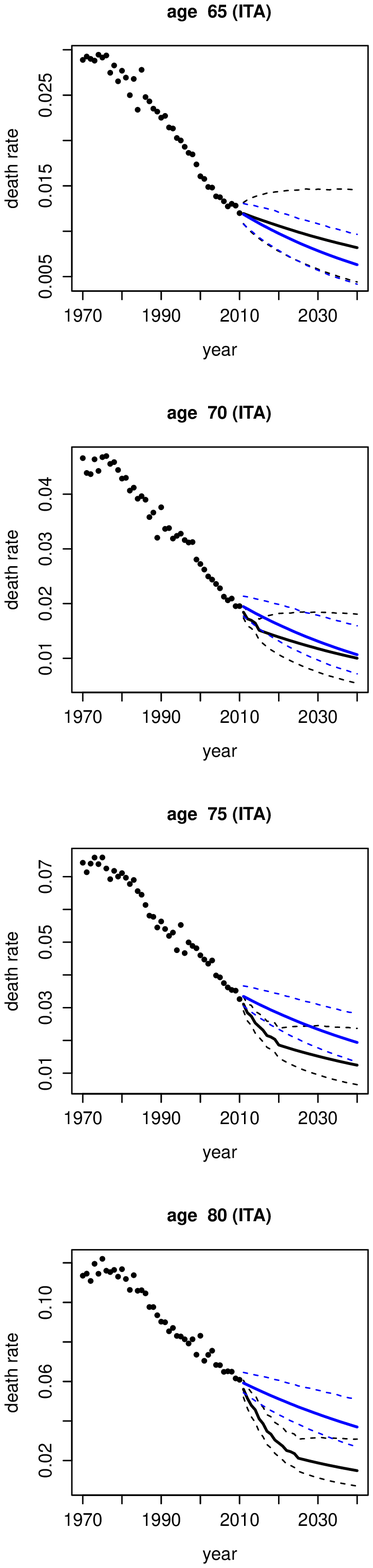}
\caption{\small{Forecasted death rates for the UK, US and Italy male populations from the full cohort model (black lines) and the LC model (blue lines). Solid circles: observed data; solid lines: posterior mean; dash lines: 95\% forecasting interval.}}
\label{fig:forecastDeathRatesFullCohortLCmales}
\end{center}
\end{figure}

Figure~\ref{fig:forecastDeathRatesFullCohortLCmales} displays the mean and 95\% posterior predictive forecast interval of the forecasted death rates from the full cohort model and LC model based on UK, US and Italy male mortality data for selected ages 65, 70, 75 and 80.

We observe that forecasts from the full cohort model show substantial difference to the forecasts from the LC model for the UK and Italy populations, while the difference is noticeably smaller for the US data. It is consistent with our results found in Section~\ref{sec:ModelEstimation} and \ref{sec:modelComparison} that fitting of the cohort models suggest that cohort effect is strong for the UK and Italy populations but is weak for the US mortality data. As a result, it is expected that, from the forecasting perspective, cohort models will show greater difference to the LC model for the UK and Italy data but will be less so for the US data, which is confirmed by the forecasting results shown here.

Another important observation is that there is a clear change of trend for the forecasted death rates from the full cohort model. A closer inspection suggests that the change appears at the generation born in 1945. For example, the change is located in year 2020 for the forecasted death rates for age 75, which corresponds to the cohort with year-of-birth at 1945; the same conclusion can be drawn from the forecasted death rates for other ages. Interestingly, the most recent generation considered in our data is exactly the year-of-birth at 1945.\footnote{Age range and year range for our data are 65-95 and 1970-2010 respectively; hence the most recent generation considered is the cohort born in $2010-65=1945$.} This observation indicates that this change of trend behaviour is a consequence of the projection of the cohort factor which starts at 1945.\footnote{Using the R package StMoMo (\cite{VillegasMiKa15}), which perform estimation and forecasting of mortality models based on the GLM framework, we also observe the change of trend behaviour in the forecasted death rates from the cohort models.} We provide further discussion on this issue in details in Section~\ref{sec:projectCohortFactors}.

\begin{figure}[h]
\begin{center}
\includegraphics[width=5.5cm, height=20cm]{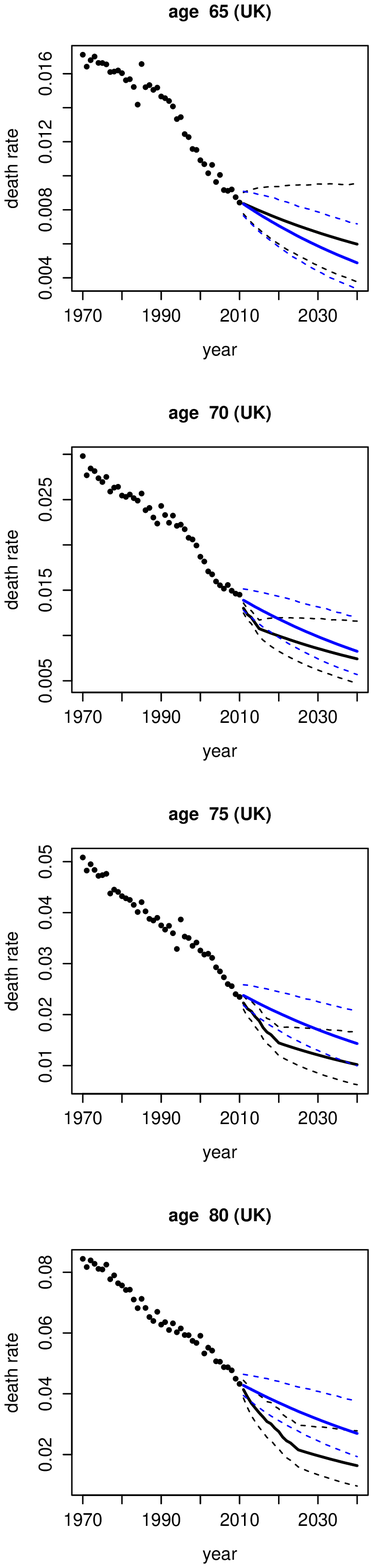}\includegraphics[width=5.5cm, height=20cm]{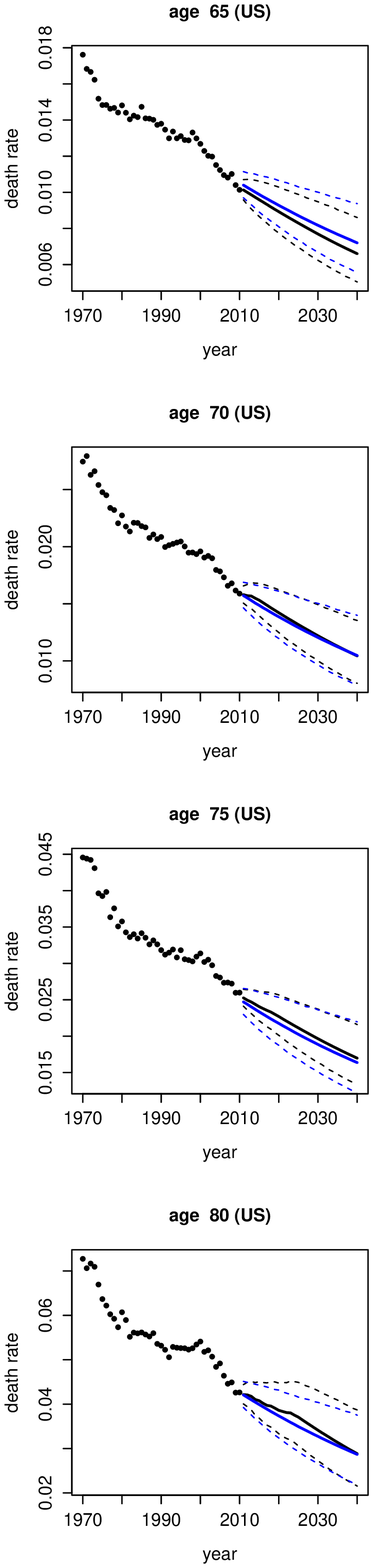}\includegraphics[width=5.5cm, height=20cm]{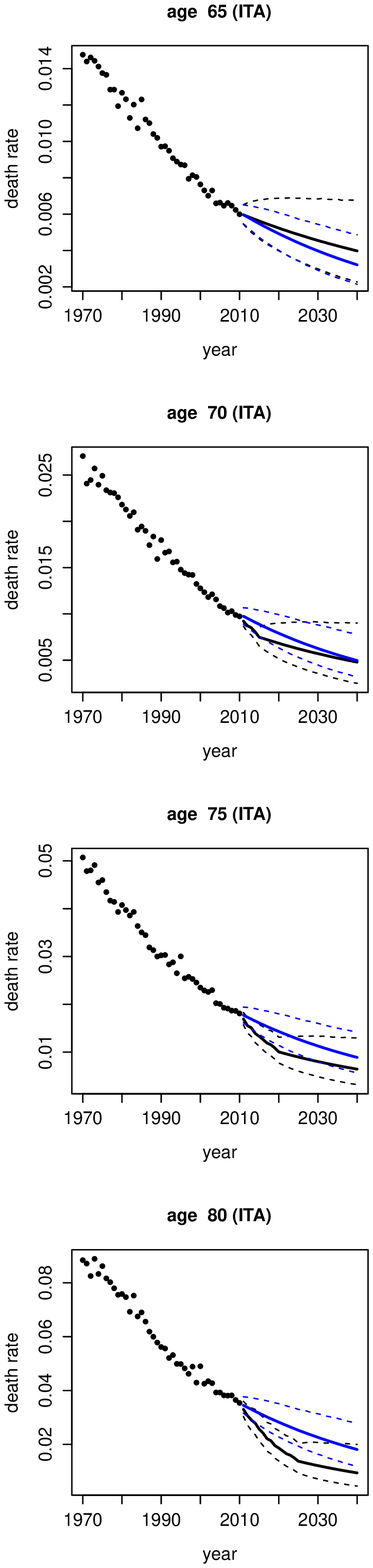}
\caption{\small{Forecasted death rates for the UK, US and Italy female populations from the full cohort model (black lines) and the LC model (blue lines). Solid circles: observed data; solid lines: posterior mean; dash lines: 95\% forecasting interval.}}
\label{fig:forecastDeathRatesFullCohortLCfemales}
\end{center}
\end{figure}

The corresponding forecasting distributions based on female mortality data are shown in Figure~\ref{fig:forecastDeathRatesFullCohortLCfemales}. The observations remarked above can also be applied in this case. As discussed in Section~\ref{sec:ModelEstimation} and \ref{sec:modelComparison}, we find that male and female populations in the studied countries show similar mortality features including cohort patterns in terms of model fitting. The result here signals that the same can be said from the forecasting standpoint.

\begin{figure}[h]
\begin{center}
\includegraphics[width=5.5cm, height=20cm]{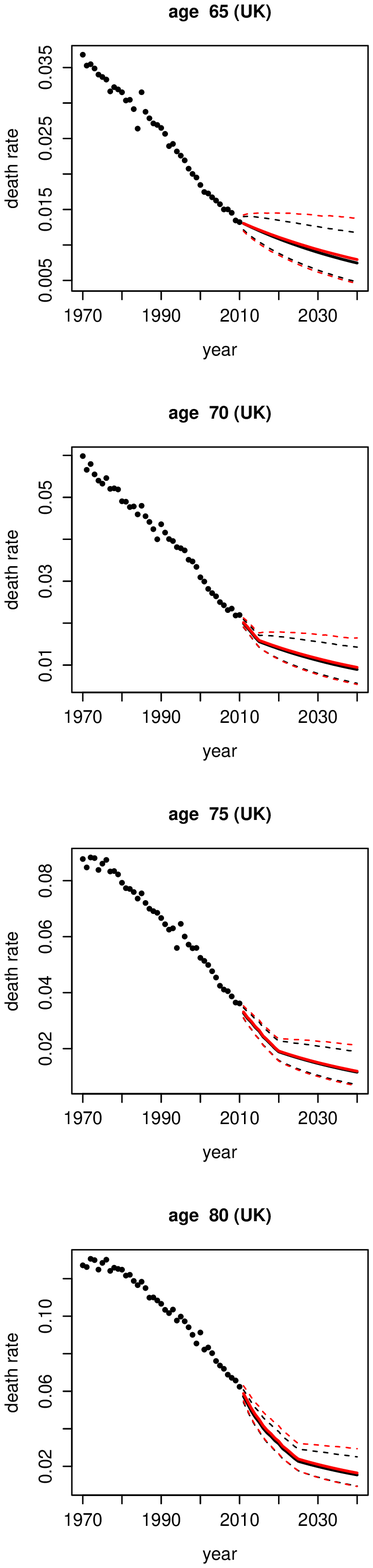}\includegraphics[width=5.5cm, height=20cm]{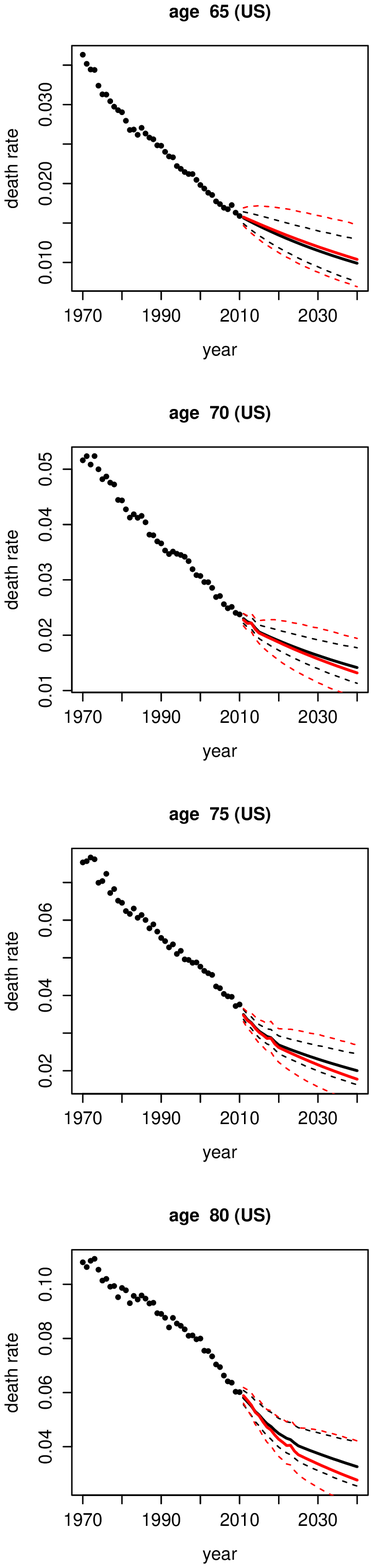}\includegraphics[width=5.5cm, height=20cm]{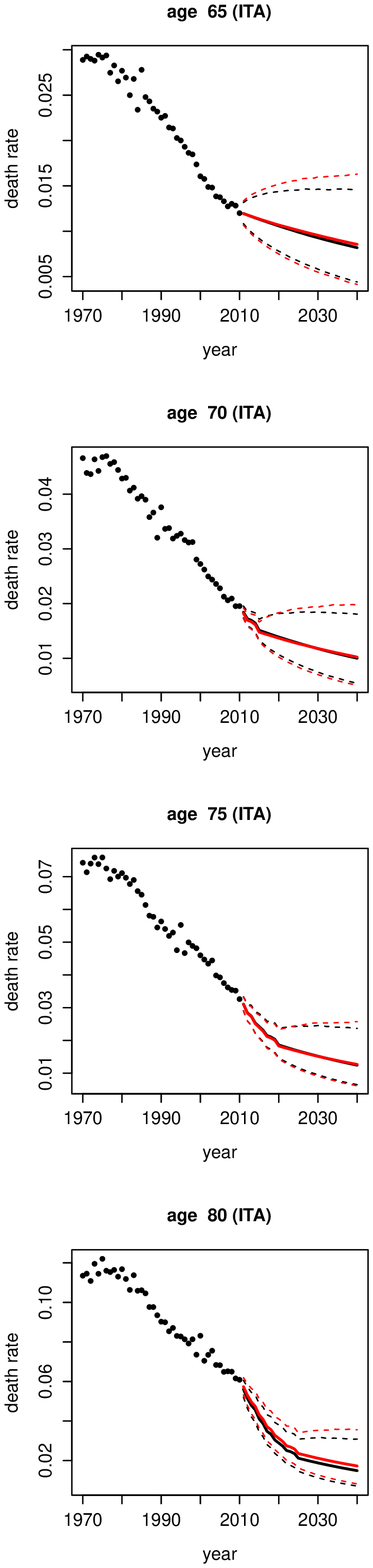}
\caption{\small{Forecasted death rates for the UK, US and Italy male populations from the full cohort model (black lines) and the simplified cohort model (red lines). Solid circles: observed data; solid lines: posterior mean; dash lines: 95\% forecasting interval.}}
\label{fig:forecastDeathRatesFullSimpCohortmales}
\end{center}
\end{figure}

\begin{figure}[h]
\begin{center}
\includegraphics[width=5.5cm, height=20cm]{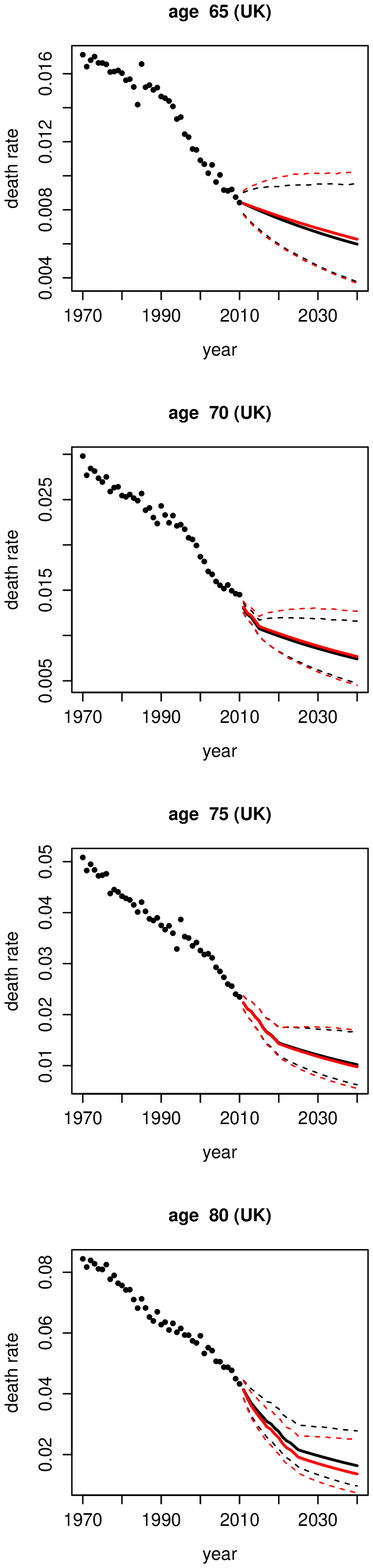}\includegraphics[width=5.5cm, height=20cm]{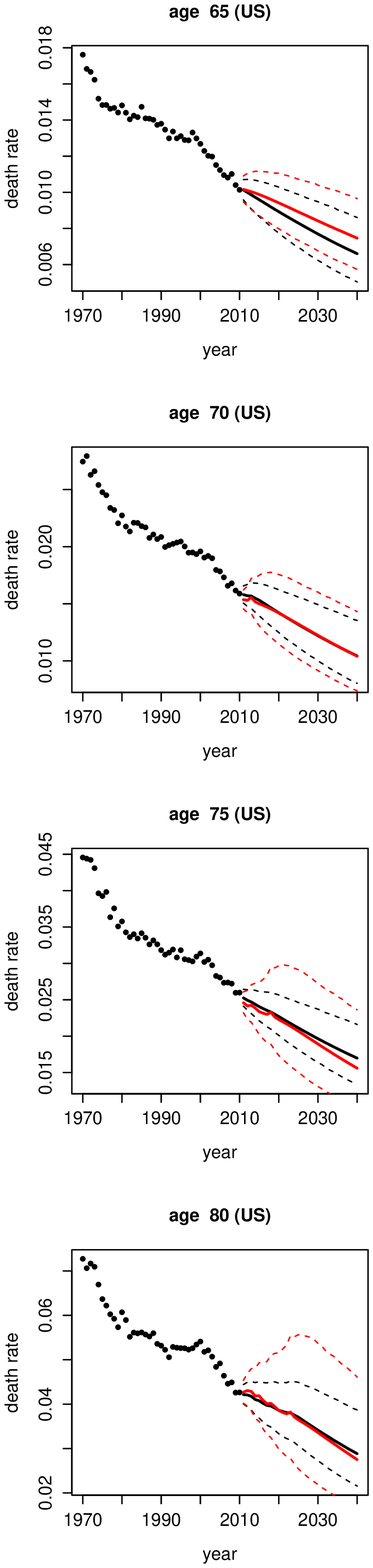}\includegraphics[width=5.5cm, height=20cm]{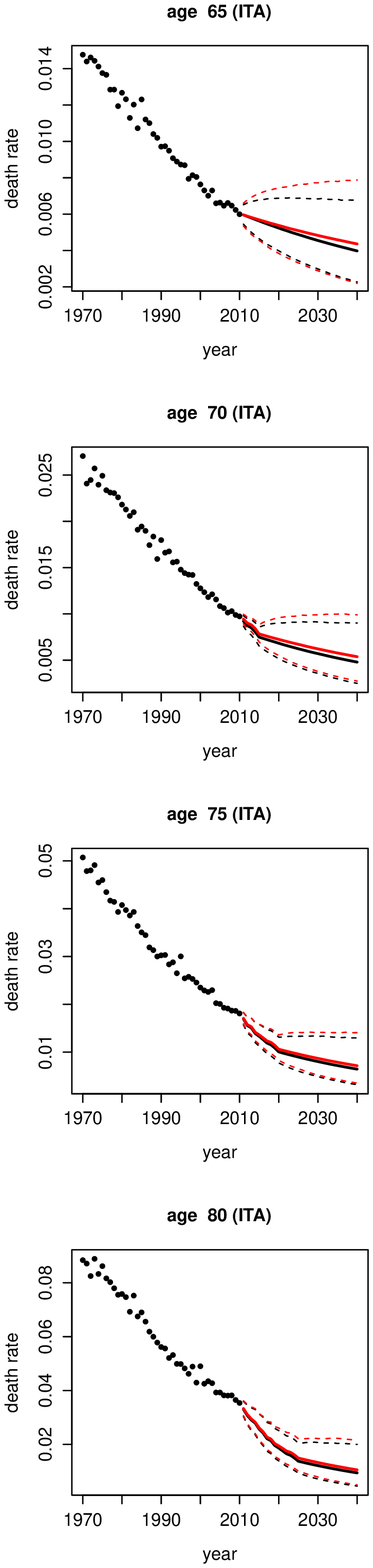}
\caption{\small{Forecasted death rates for the UK, US and Italy female populations from the full cohort model (black lines) and the simplified cohort model (red lines). Solid circles: observed data; solid lines: posterior mean; dash lines: 95\% forecasting interval.}}
\label{fig:forecastDeathRatesFullSimpCohortfemales}
\end{center}
\end{figure}

A comparison of forecasting from the full cohort model and the simplified cohort model is shown in Figure~\ref{fig:forecastDeathRatesFullSimpCohortmales} for the male populations and in Figure~\ref{fig:forecastDeathRatesFullSimpCohortfemales} for the female populations. It is clear that the full and simplified models generate similar forecasting distribution for the UK and Italy populations. However, the forecasting intervals produced by the simplified model are substantially wider than the intervals generated by the full cohort model for the US male population; the difference is even more pronounced for the female populations. These results suggest that if cohort patterns are strongly present in the data, then there is little difference between the full model and the simplified model in terms of model fitting and forecasting. On the contrary, if the data show only a weak presence of cohort patterns, the full cohort model can lead to significant different fitting and forecasting results to the simplified cohort model, which indicates that cohort models may not be required in this case.

\subsubsection{Projection of period and cohort factors}\label{sec:projectCohortFactors}
The regular occurrence of trend changing behaviour of the forecasted death rates at year-of-birth 1945, which was discussed in Section~\ref{sec:projectDeathrates}, suggest that this phenomenon may originate from the projection of cohort factor starting at year-of-birth 1945.

\begin{figure}[h]
\begin{center}
\includegraphics[width=5.5cm, height=9cm]{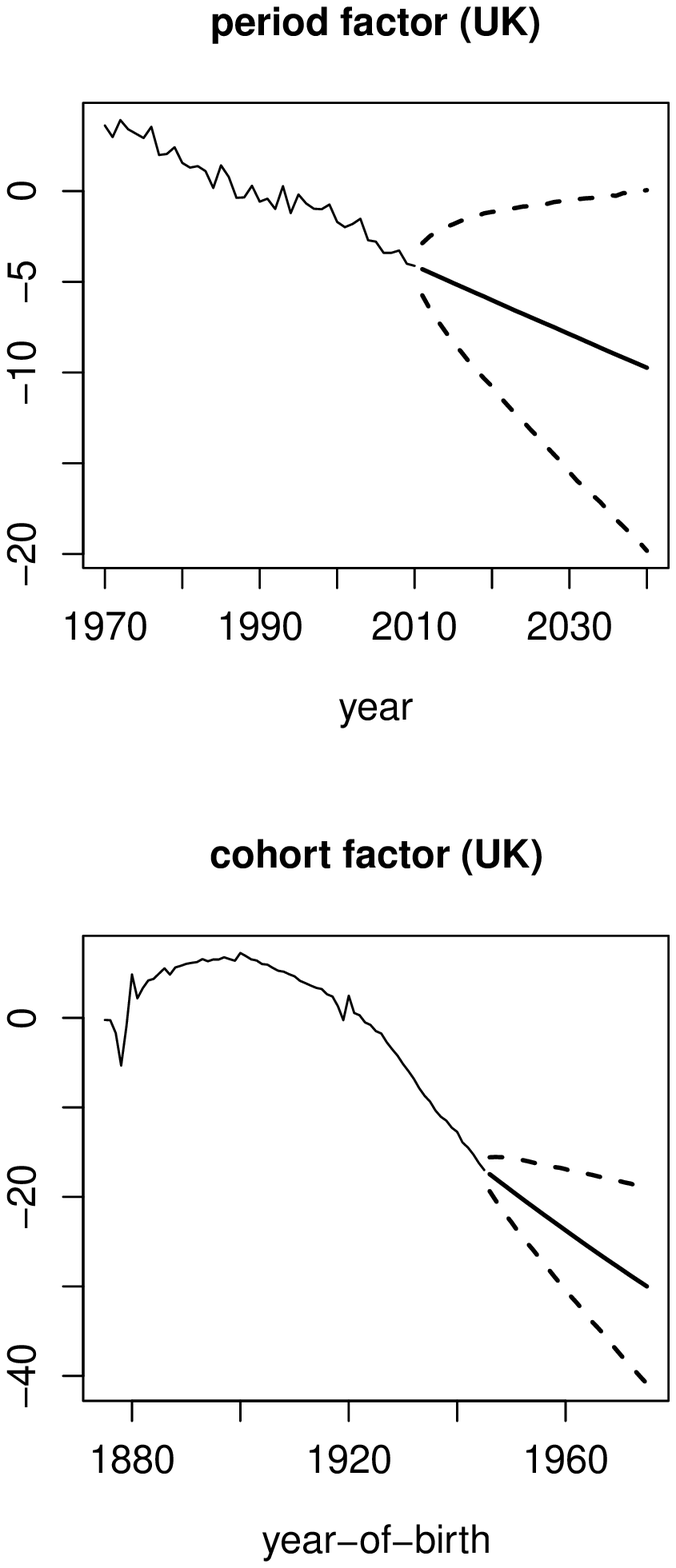}\includegraphics[width=5.5cm, height=9cm]{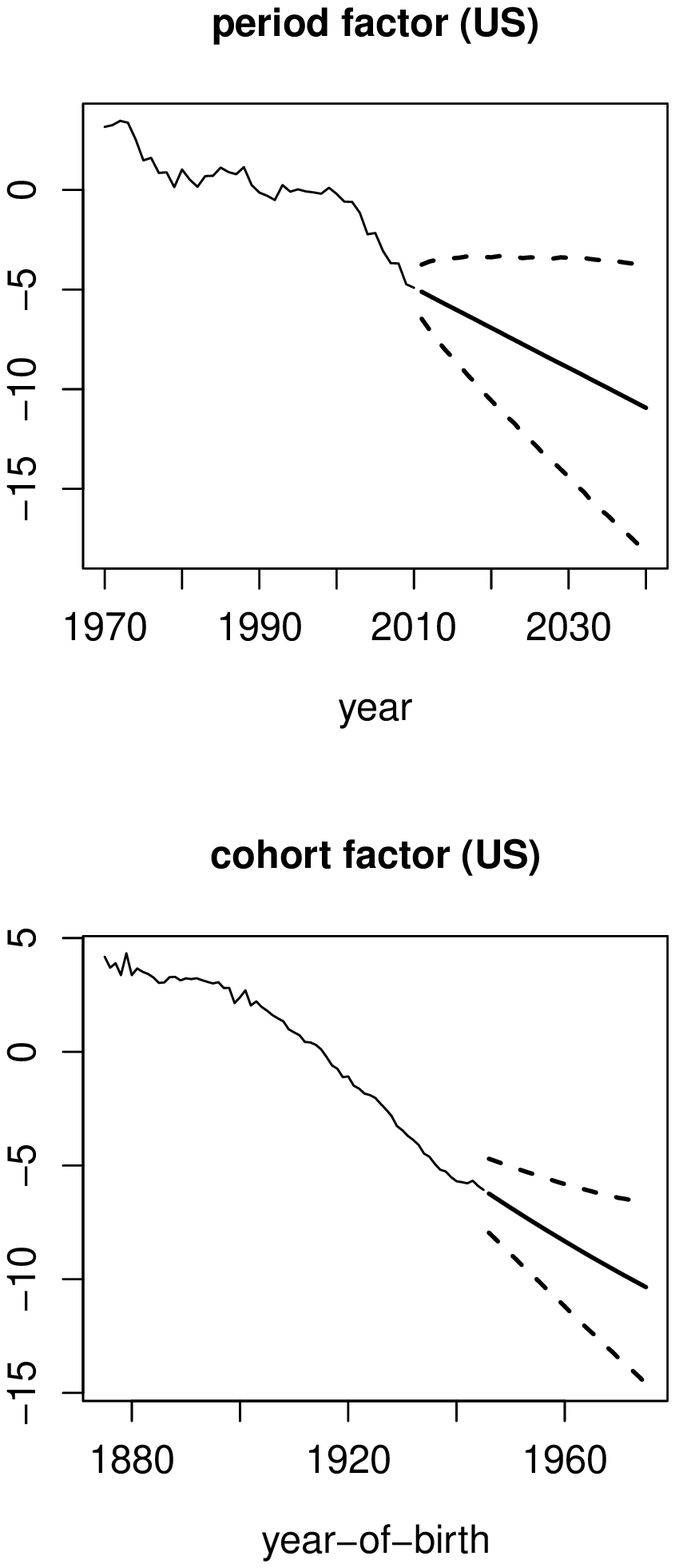}\includegraphics[width=5.5cm, height=9cm]{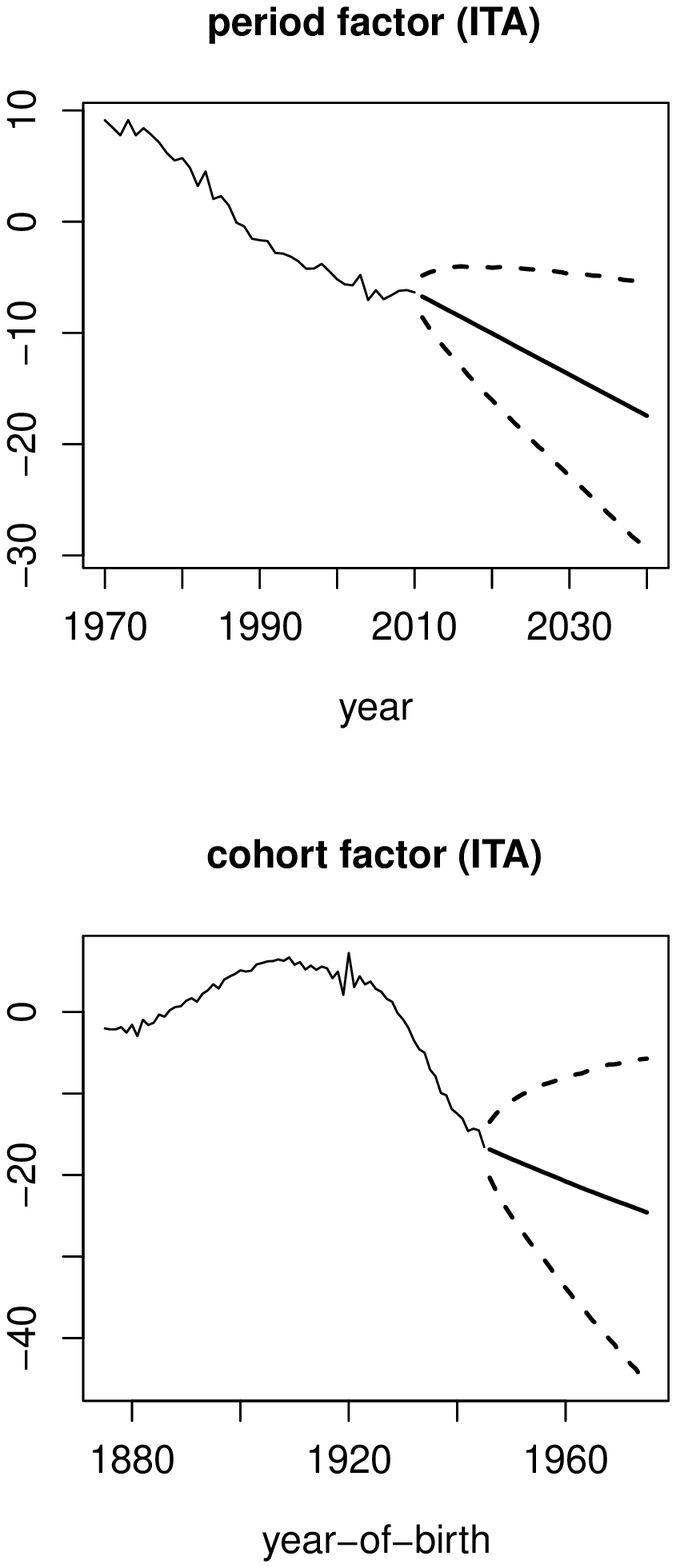}
\caption{\small{Projection of the period factor $\bm{\kappa}$ and cohort factor $\bm{\gamma}$ from the full cohort model for the UK, US and Italy male populations.}}
\label{fig:projectFactorsfullcohortmale}
\end{center}
\end{figure}

\begin{figure}[h]
\begin{center}
\includegraphics[width=5.5cm, height=9cm]{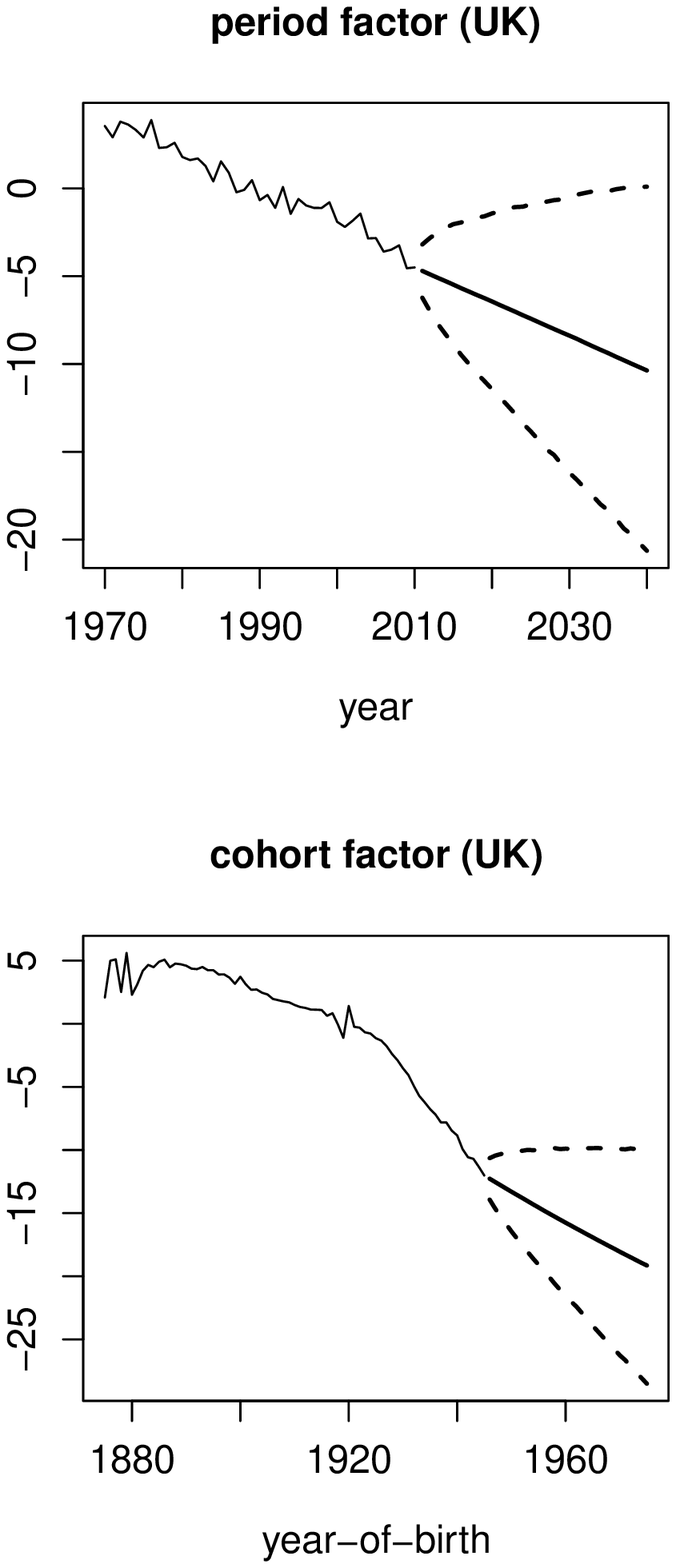}\includegraphics[width=5.5cm, height=9cm]{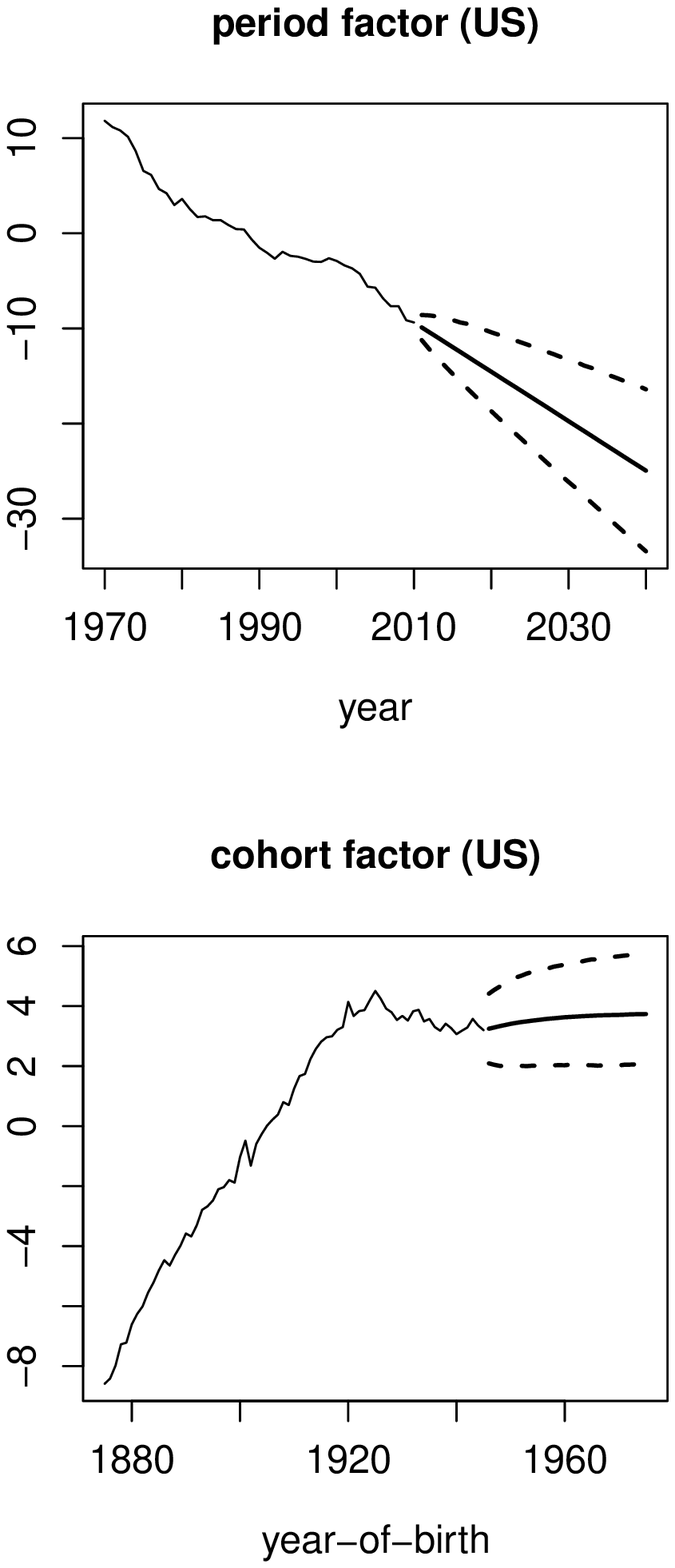}\includegraphics[width=5.5cm, height=9cm]{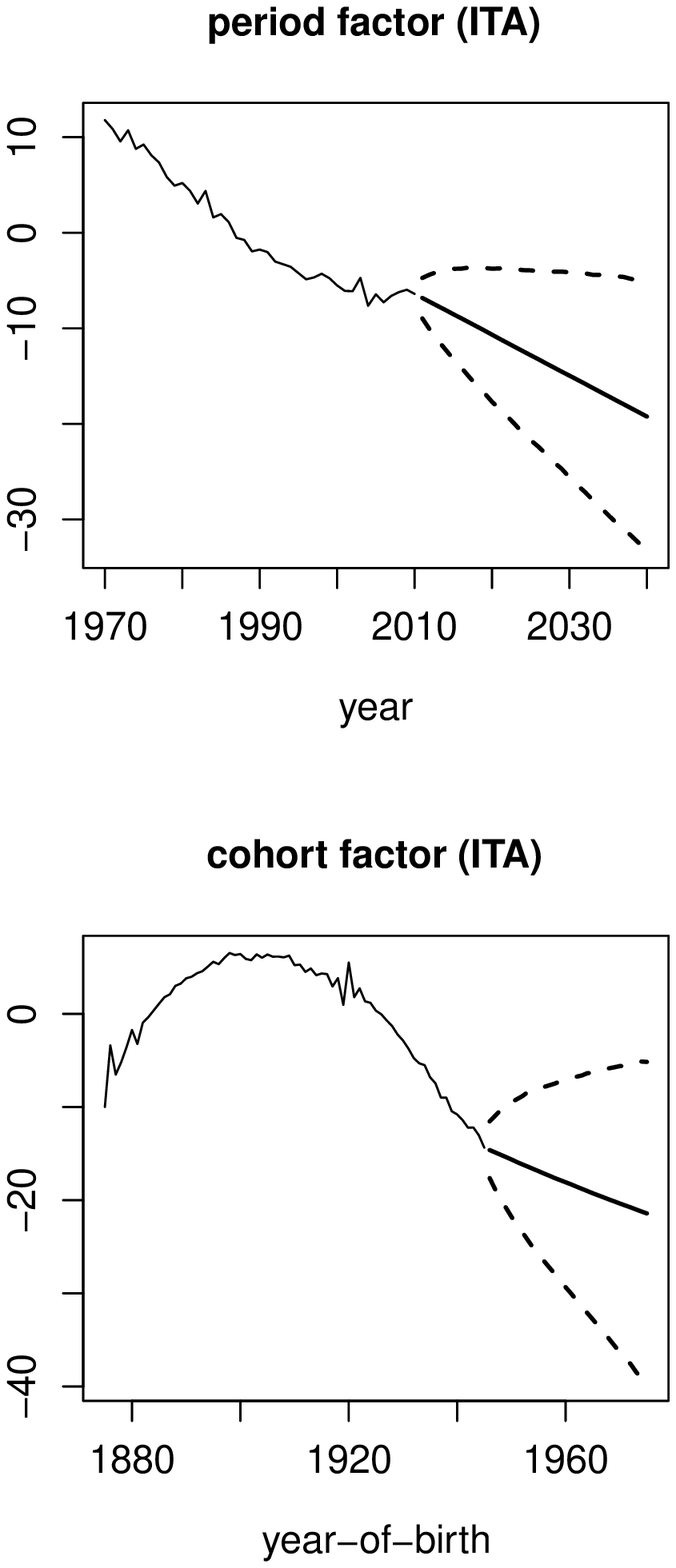}
\caption{\small{Projection of the period factor $\bm{\kappa}$ and cohort factor $\bm{\gamma}$ from the full cohort model for the UK, US and Italy female populations.}}
\label{fig:projectFactorsfullcohortfemale}
\end{center}
\end{figure}

To investigate this, we plot the projection of the period and cohort factors from the full cohort model in Figure~\ref{fig:projectFactorsfullcohortmale} for the male populations and Figure~\ref{fig:projectFactorsfullcohortfemale} for the female populations in the considered countries.

We first notice that the projected period factor continues the linear trend as expected for both genders in UK, US and Italy populations. It reassures that the projection of the period factor behaves properly and therefore we should focus on the projection of the cohort factor to explain the aforementioned trend-changing behaviour.

Interestingly, we observe from the male populations that there is a slight change of direction of the projected cohort factor starting from the year 1945 compared to the trend during the calibration period 1920-1945 in the UK population. We observe essentially no change of direction for the US population and a significant change of direction for the Italy population. The same observation also applies to the female mortality data.

These results strongly indicate that whether a change of trend will be observed for the forecasted death rates depend on the property of the projected cohort factor; namely it depends on whether there will be a change of direction from the estimated cohort factor when the projection starts. In the UK and Italy cases, the change of direction of the projected cohort factor at year 1945 is observed and hence results in the trend-changing behavior of the forecasted death rates shown in Figure~\ref{fig:forecastDeathRatesFullCohortLCmales} and \ref{fig:forecastDeathRatesFullCohortLCfemales}, in contrast to the case of US data. Note also that these results are consistent with our discussion in previous sections that cohort patterns are strongly presence in the UK and Italy data while for the US data cohort patterns are not clear.

The estimated cohort factors shown in Figure~\ref{fig:FullCohortEstMales}-\ref{fig:SimpCohortEstFemales} in fact raise questions about whether it is reasonable to assume that cohort factor dynamics can be captured by stationary ARIMA models which are commonly found in the literature. One may argue that cohort factors apparently exhibit trend changing behaviour. Recently there are growing interests in applying structural change dynamics for the period factor, see \cite{LiChCh11}, \cite{vanBerkumAnVe16} and \cite{LiuLi16b}. The results suggest that this new type of models may be equally suitable for the cohort factor dynamics. However the investigation of this issue is out-of-scope of the current paper and will be left for future research.

\section{Conclusion}\label{sec:conclusion}
In this paper we investigate the formulation, estimation and forecasting of cohort models under the state-space approach. The state-space framework provides a unified environment where estimation and forecasting of dynamics mortality models can be carried out in a statistically rigorous manner. Continuing development of advanced statistical techniques in numerical filtering and model estimations suggest that state-space method can be an essential tool to handle the modelling of human mortality. The paper contributes to the literature in this area by showing that cohort models are compatible with the state-space framework.

We demonstrate in this paper that the problem of cohort factors being indexed according to year-of-birth instead of year can be overcome by considering a multi-dimensional state-space system. The defining property of the cohort factor imposes a restriction on the observation and state equations of the resulting state-space mortality model. Dynamics of the period and cohort factors are specified in the state equation which allows for a range of time series models to be considered.

By treating the period and cohort factors as the state dynamics of a state-space model, Bayesian inference for cohort models can be performed based on filtering and MCMC method. We develop an efficient MCMC sampler for the resulting model involving a combination of conjugate Gibbs sampling steps for the static parameters and a forward-backward Kalman filtering for the latent state dynamics. The overall algorithm can be applied naturally to the full cohort model as well as its nested models including the simplified cohort model and the LC model.

We apply the cohort models in state-space formulation to analyse male and female mortality data of the UK, US and Italy population. The Bayesian approach allows us to present estimation of the cohort models where parameter uncertainty is properly quantified and accounted for. Our empirical studies show that UK and Italy populations exhibit strong cohort patterns while cohort patterns for the US population are weak. We also find that both genders share common mortality characteristics including cohort effects for the considered countries. Examination of residual heatmaps produced from the cohort models and LC model suggest that cohort models are able to capture cohort effects while the LC model fails to do so. Using DIC for model ranking, we find that the full cohort model is preferred over the simplified cohort model, which in turn outperform the LC model for the considered countries. We show that forecasted death rates from cohort models display trend-changing behaviour at the year where cohort factors are projected, for countries that show strong presence of cohort patterns.

Our estimation and forecasting studies of state-space cohort models suggest that more sophisticated models such as structural change models may be required to capture the dynamics of cohort factors adequately. Moreover, the observed mortality characteristics shared by both genders indicate that one can generalise the approach proposed in this paper to multi-population setting where common period and cohort factors can be assumed for male and female populations within a country. Recently there are also growing interests in applying state-space method to deal with the problem of pricing and hedging of longevity risk (\cite{LiuLi16a} and \cite{LiuLi16b}). The approach presented in this paper can significantly enhance the capability of the state-space methodology to solving practical problems involving longevity risk since cohort models are shown to be compatible with the state-space framework. Applications of the state-space cohort models studied in this paper to the key issues in longevity risk pricing and management will be left for future research.

\section*{Acknowledgements}
This research was supported by the CSIRO-Monash Superannuation Research
Cluster, a collaboration among CSIRO, Monash University, Griffith
University, the University of Western Australia, the University of
Warwick, and stakeholders of the retirement system in the interest
of better outcomes for all. This research was also partially supported under the
Australian Research Council's Discovery Projects funding scheme (project number:
DP160103489). We acknowledge the grant of the interdisciplinary project ``Research on Urban Intelligence"  from Research Organization of Information and Systems in Japan.


\clearpage
\bibliographystyle{elsart-harv}
\bibliography{mcf}

\end{document}